# Contact-driven super-resolution with dielectric microlenses at practical photon budgets

Kwang S. Kim*

Center for Superfunctional Materials, Department of Chemistry, Ulsan National Institute of Science and Technology (UNIST), Ulsan 44919, Republic of Korea

* Corresponding author: Kwang S. Kim, Center for Superfunctional Materials, Department of Chemistry, UNIST, 50 UNIST-gil, Ulsan 44919, Republic of Korea. E-mail: kimks@unist.ac.kr. Tel.: +82-10-5067-2110

## Abstract

A transparent microsphere resting on a specimen lets an ordinary light microscope see beyond its diffraction limit. Yet what sets its resolution, how to design better lenses and how much light finer detail requires have remained open. We answer by counting the independent pieces of information or channels that a lens delivers and pricing them in photons. Exact and full-wave calculations show that the finest detail is read only where the lens touches the specimen. A lens in contact acts as two instruments: a wide-field immersion imager up to the solid-immersion limit and, beyond it, a near-field pixel under its contact. Light that the lens returns onto the specimen doubles the reach of the imager; beyond it, only curved contacts read. Contacts can therefore be designed. In three-dimensional calculations a bump and three rings on a flat face pass nearly four times as many channels beyond the immersion limit as the best sphere, and in two-dimensional simulations 17 contacts resolve lines $\lambda_0/10$ (where $\lambda_0$ is the wavelength in vacuum) and, with more light, $\lambda_0/12.5$ apart (40 and 32 nm at 400 nm wavelength) everywhere across a four-wavelength field. For a weak object, $\lambda_0/10$ takes 46 s at 1 kW cm$^{-2}$ at the quantum limit, and an ordinary camera only three times longer. Contacts set the resolution; detectors set the speed.

## Introduction

Light leaving an object carries its details as ripples of different fineness. Coarse ripples travel freely and are collected by the microscope objective, but ripples with periods finer than about half a wavelength cling to the object as evanescent waves, which fade within a few tens of nanometres and never arrive (Fig. 1a). This is why a microscope cannot resolve a period finer than about $\lambda_0/(2\mathrm{NA})$, as shown by Abbe, where $\lambda_0$ is the wavelength in vacuum and NA the numerical aperture of the objective[1]. Each ripple is described by its spatial frequency $k_{\parallel}$, $2\pi$ divided by its period; ripples with $k_{\parallel}$ larger than the wavenumber of light, $k_0 = 2\pi/\lambda_0$, are the evanescent ones. Fluorescence nanoscopy overcomes this limit by switching labels on and off[2,3], near-field probes scan a tip through the evanescent waves[4], solid-immersion lenses of refractive index $n$ raise the collected band to $nk_0$ (ref.[5]), and hyperlenses carry evanescent waves through engineered media[6].

A strikingly simple alternative is to rest a transparent bead a few micrometres across on the specimen. Touching the object, the bead sits inside the evanescent waves and turns part of them into travelling light that the objective collects, much as a fingertip reads Braille dots that the eye cannot see from a distance. In 2009, self-assembled organic nanolenses magnified line patterns that the microscope alone could not resolve[7], and full-wave analysis showed that such wavelength-scale lenses focus beyond the corresponding macroscopic solid-immersion lens[8]. In 2011, Wang et al. extended the approach to white light with silica microspheres[9], opening it to many laboratories. The field now spans high-index and liquid-immersed beads[10], scanning instruments[11,12], living cells[13], on-chip lenses[14,15], semiconductor metrology[16] and analytes that act as their own lenses[17], and is the subject of a community roadmap[18], a review[19] and a perspective[20].

Four issues have kept this simple technique from becoming a designable one. First, there is no agreed mechanism. Photonic nanojets[21], whispering-gallery resonances[22], a larger effective numerical aperture[23] and, in simulations of complete imaging, coherent re-illumination of the object by light circulating in the bead[24] each capture part of the physics (Fig. 1b), but none states which effect sets the resolution. Second, there is no common yardstick: reported resolutions range from about $\lambda_0/2$ for resolved periods to below $\lambda_0/10$ for visible features, because studies measure different, complementary aspects of image quality[25,26]. Third, there is no design rule: a higher index and tighter focusing have been the main levers, yet larger beads do not necessarily see finer, and detail resolved at the best spot need not be resolved across the field of view. Fourth, the price in light is unknown. The finest predictions and measurements are quoted against a dynamic range, the ratio of the strongest signal to the noise floor, or against a contrast, rather than a photon count[25,27], so it is unknown how much light each step of resolution costs, or whether detectors that sort light into spatial modes[28,29] or shaped illumination[30,31] would help.

We address all four with one quantity that every imaging system has: the number of independent pieces of information about the object that it delivers above the noise[32]. We call each piece a channel: an object pattern that reaches the detector independently of all others and so, like a camera pixel, carries one number about the object. The channels of a lens can be pictured as bars of different heights: those above the noise floor can be read, those below are lost, and more light lowers the floor (Fig. 1c). Mathematically, the channels and their strengths are the singular values of the operator that maps the object's evanescent waves onto the light collected by the objective[20,32]; they had not been computed for a lens in contact. To price channels in photons we use the Fisher information[33], the currency of localization microscopy[34] and optical metrology[31,35], which states how precisely the detected photons pin down the object, and its quantum limit, the most that any physically allowed detector could extract.

Here we compute the channels exactly for spheres and with full-wave simulations for flat-faced and textured lenses, in two dimensions (2D) and, for rotationally symmetric lenses, in three (3D); simulate complete imaging, including light that the lens returns onto the object; and convert resolution into photons for a camera and for mode sorters. Five results emerge. (1) The contact sets the limit: a smooth lens in contact is a wide-field imager up to $nk_0$ and a single nanoscale pixel beyond it. (2) The competing pictures fit together: light reaching the object from or through the lens extends every lens towards $2nk_0$, and beyond $2nk_0$ only curved contacts read. (3) Contacts can be designed with a simple counting rule: a bump and three rings on a flat face pass nearly four times as many channels beyond $nk_0$ as the best sphere, and 17 contacts resolve lines $\lambda_0/10$ and $\lambda_0/12.5$ apart everywhere across a four-wavelength field. (4) The photon price is practical: for a weak test line 12 nm wide, $\lambda_0/10$ costs 46 s at 1 kW cm$^{-2}$ at the quantum limit and only three times more with an ordinary camera, whose reference wave is the lens background itself. (5) Contacts set the ceiling and detectors set the speed (Fig. 1d): no photon budget or detector resolves beyond what the contacts deliver, and a mode sorter pays off only against systematic errors. Together, these results turn microlens nanoscopy from an observed phenomenon into a design problem; Supplementary Note 8 retells the argument in questions and answers.

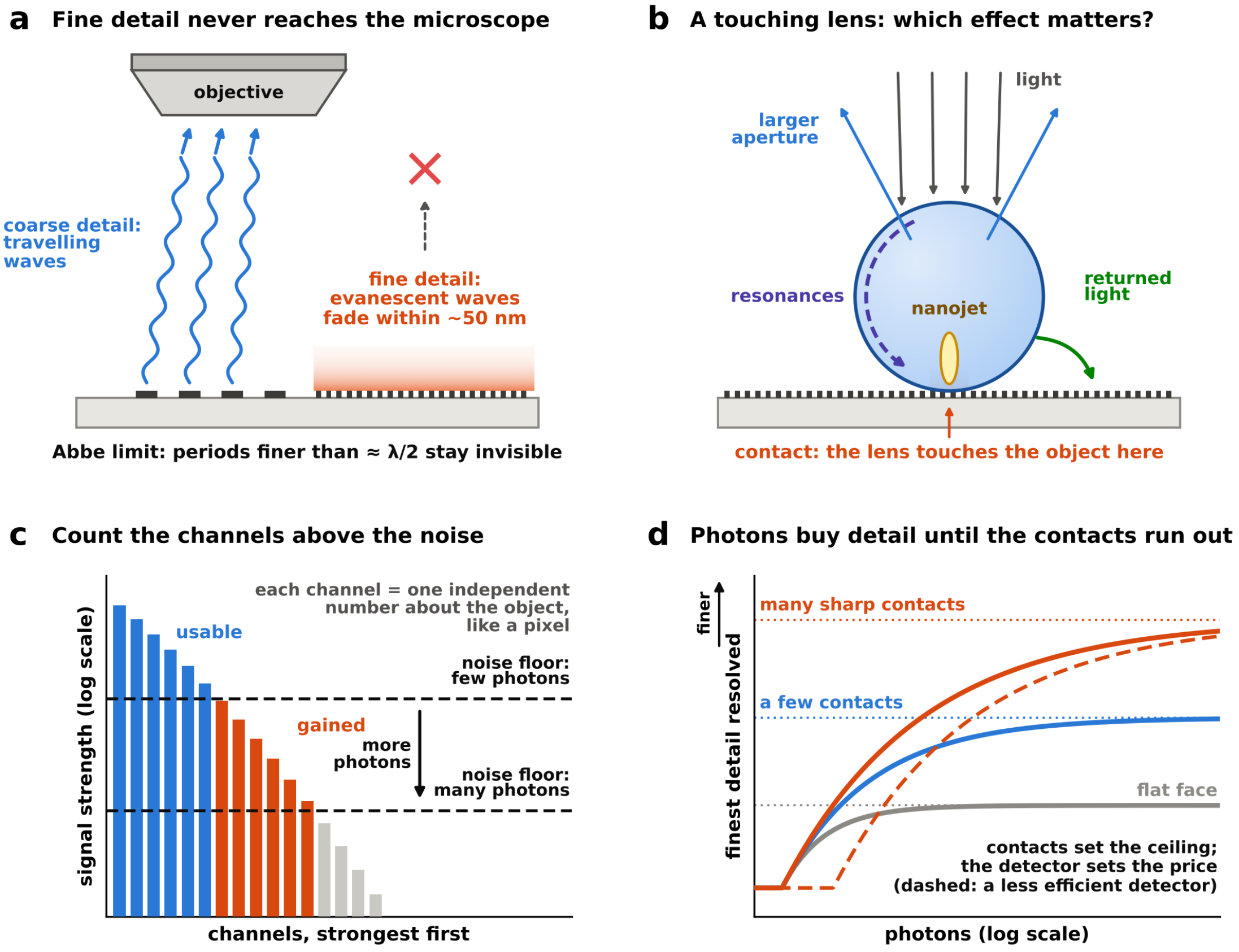

**Fig. 1 | Why a lens in contact sees finer detail, and what limits it.** Schematics, not to scale. **a**, An object sends its coarse detail into travelling waves that reach the objective (blue), but detail finer than about half a wavelength produces evanescent waves (orange) that fade within a few tens of nanometres and never arrive: the diffraction limit. **b**, A transparent lens resting on the object reaches into these waves. Several effects have been proposed to explain the resulting super-resolution: a larger effective aperture, resonances of light circulating in the lens, focusing into a photonic nanojet and light returned onto the object. The contact (red) is where the lens touches the object. **c**, Channels are the independent object patterns that a lens delivers to the microscope, ordered by strength; each carries one number about the object, like a pixel. Channels above the noise floor are usable (blue), and more photons lower the floor and add channels (red). **d**, Resolution improves with the photon budget until the channels that the lens delivers are exhausted. The contacts set this ceiling; the detector changes how many photons each step costs (dashed, a less efficient detector).

# Results

## Counting the channels of a lens

To compare lenses we must count what each delivers. We describe the object as a sum of gratings of different spatial frequency $k_\parallel$, as a musical chord is a sum of pure tones. A grating with $k_\parallel > k_0$ produces a field that decays away from the object as $\exp(-\alpha z)$, with $\alpha = (k_\parallel^2 - k_0^2)^{1/2}$: the finer the grating, the faster it fades. A dielectric lens of diameter $D$ and index $n$ resting on the object scatters part of this field into an NA 0.9 objective (Fig. 2a). The singular-value decomposition of this linear map, $H$, sorts all object patterns into independent channels[32]; the strength of a channel, the eigenvalue $\lambda_i$ of $H^\dagger H$, is the power that the objective collects from it. A channel counts if it is at most 1,000 times (30 dB) weaker than the brightest evanescent channel of the same lens and measurement, $\lambda_i \geq 10^{-3}\lambda_{max}$; 20 and 40 dB test the sensitivity (Methods). A count is not a resolution, which also depends on noise, object and reconstruction. For spheres $H$ follows exactly from Mie theory[36,37], and rotational symmetry sorts the patterns by their azimuthal order $l$, the number of times a pattern repeats around the contact point, like the petals of a flower (Methods).

We divide fine detail into two bands at the solid-immersion limit $nk_0$. Gratings in the immersion band, $k_0 < k_\parallel < nk_0$, can enter a flat glass face as travelling light, so any immersion lens collects them. Finer gratings, with $k_\parallel > nk_0$, cannot enter a flat face at all; to deliver them, a lens must redirect light by its shape. This band beyond the immersion limit is the real prize of microlens nanoscopy.

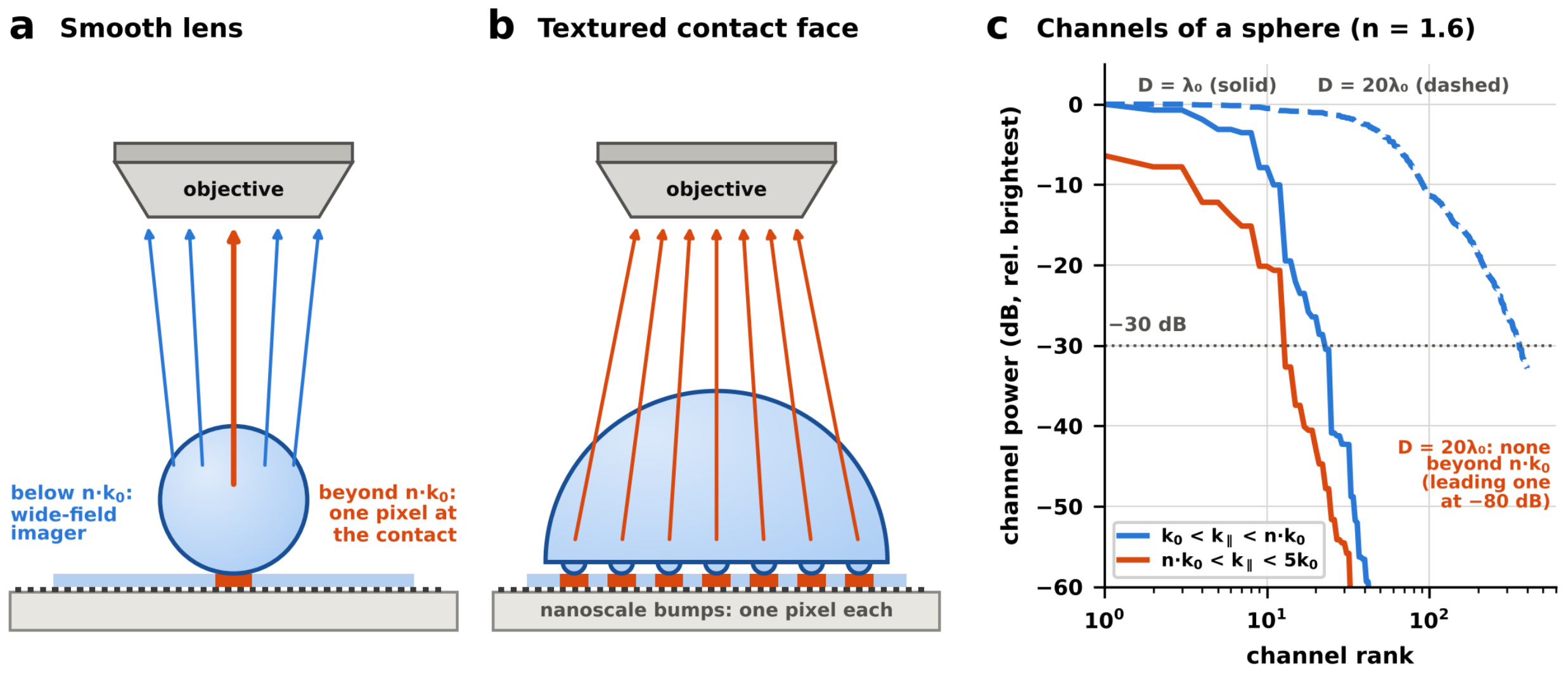

**Fig. 2 | A dielectric lens in contact is two instruments. a**, A smooth lens couples the object's evanescent components below the solid-immersion limit ($k_0 < k_\parallel < nk_0$, blue) from a wide region and forms a wide-field image; components beyond $nk_0$ (red) are read only from a small patch under the contact. **b**, Nanoscale bumps on a flat contact face each read their own patch, so one lens reads many patches at once. Schematics, not to scale. **c**, Computed channel spectra of spheres of index 1.6 in contact with an NA 0.9 objective (exact vector Mie theory; general 2D objects): collected power $\lambda_i$ of each independent object pattern relative to $\lambda_{max}$ (the brightest channel in either band), in rank order, for the band below $nk_0$ (blue) and the band from $nk_0$ to $5k_0$ (red), for spheres $\lambda_0$ (solid) and $20\lambda_0$ (dashed) across. Dotted line, the 30 dB threshold used to count channels ($\lambda_i = 10^{-3}\lambda_{max}$).

## A lens in contact is an imager and a pixel

A simple example shows the principle before any detailed analysis (Supplementary Note 1 and Supplementary Figs. 11 and 12). We placed three pieces of glass ($n$ = 1.6) on a plane of sources and counted, with independent solvers and no assumed contact size, the channels beyond $nk_0$ that radiate into the air. A flat block passes none, because a flat face leaves $k_\parallel$ unchanged. A half-disk one wavelength in radius resting on its flat face passes 5 in 2D, read at its two corners, and 24–30 in 3D, where the corners become a circular rim that reads azimuthal orders up to $|l| = 7$. Turned over, so that only its dome touches, it passes 3 in 2D and 5–7 in 3D. Doubling the gap between glass and sources to $0.1\lambda_0$ removes more than half of the 2D channels. Fine detail is read only where the glass leaves the object within a small fraction of a wavelength, at an edge or a curved tip.

The channel counts of spheres show that a bead in contact is two instruments at once. Without a lens the count is zero in both bands. In the immersion band the count grows almost in proportion to diameter, from 22–28 for spheres one wavelength across to 310–380 for spheres twenty wavelengths across ($n$ = 1.46–2.3; Figs. 2c and 3a): here the lens is a wide-field immersion imager whose field of view grows with its size. Beyond the immersion limit the count first rises and then falls (Fig. 3b). It peaks at 36 channels for $n = 1.46$ and 14 for $n = 2.3$, at diameters of $4.0\lambda_0$ and $1.0\lambda_0$, and drops below the threshold beyond $13\lambda_0$ and $2.7\lambda_0$, respectively (Table 1).

The structure of these channels shows why. A sphere passes one dominant channel beyond $nk_0$ for each azimuthal order and polarization; the next channel of the same order is 25–40 dB weaker (Fig. 3c and Supplementary Fig. 3c). Mapped back onto the object, the leading channel is a spot centred on the contact point (Fig. 3d). Beyond the immersion limit a sphere is therefore a single nanoscale pixel under its contact: it reads one number per azimuthal order from that patch, and a wider patch resolves more orders, up to $|l| = 3$ at $D = \lambda_0$ and 8 at $3.3\lambda_0$. Moving the lens moves the pixel: for a sphere one wavelength across, each new position at least half a wavelength from the others adds three to four channels (Fig. 4c), which helps to explain the success of scanning microsphere microscopes[11,12].

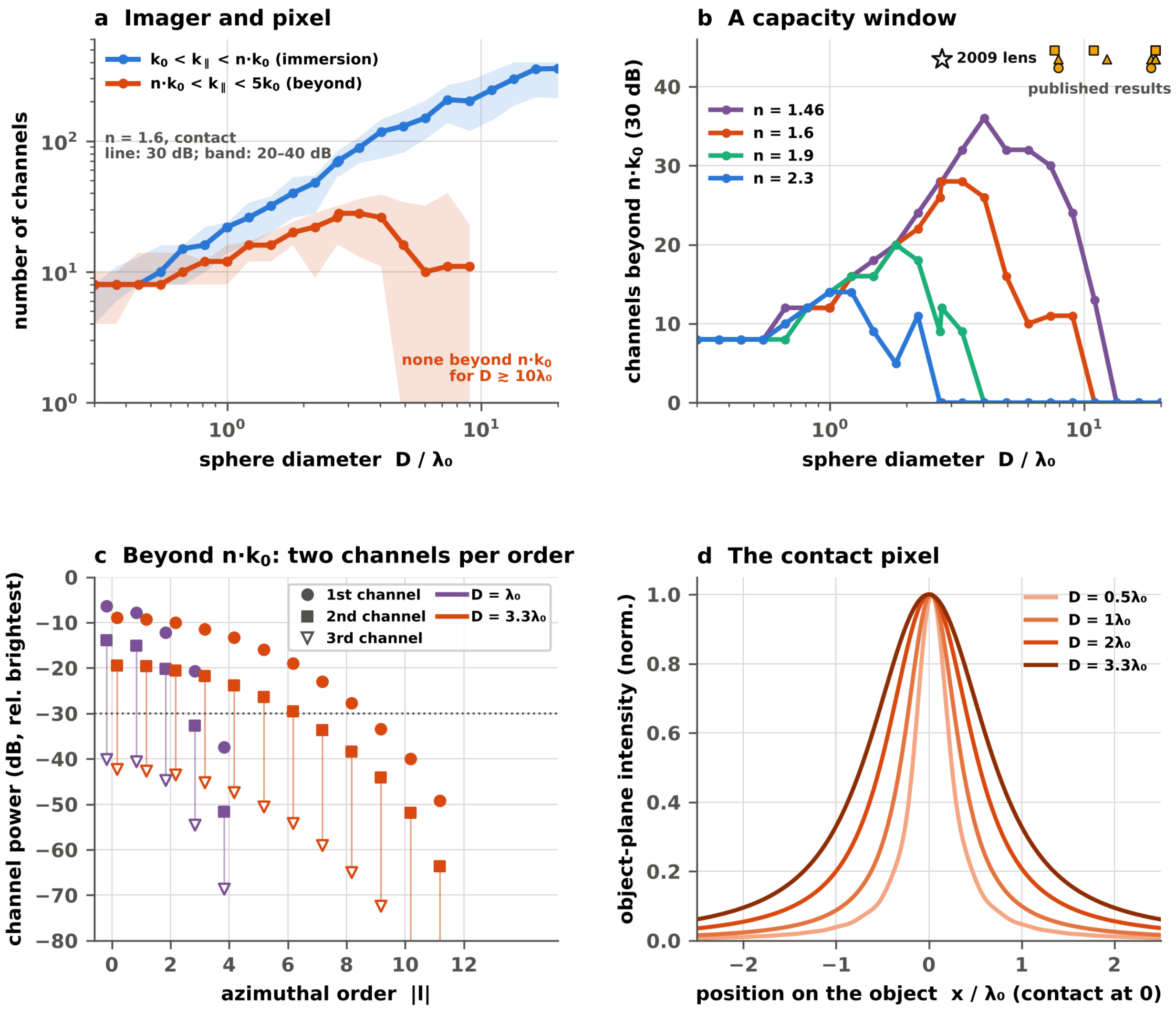

**Fig. 3 | Information channels of dielectric spheres in contact.** Exact vector Mie theory, single pass; air, NA 0.9; channels counted within 30 dB ($\lambda_i \geq 10^{-3}\lambda_{max}$; $\lambda_{max}$, the brightest channel of either evanescent band of the same lens and measurement). **a**, Number of channels below $nk_0$ (blue) and beyond it (red) versus diameter for $n = 1.6$, for general 2D objects; shading, 20–40 dB thresholds. **b**, Channels beyond $nk_0$ for four indices. Symbols above the curves mark the diameters of the lenses in published results (circles, resolved periods; triangles, visible gaps or features; squares, point spread function (PSF) or convolution widths; results beyond $19\lambda_0$ are drawn at $19\lambda_0$; Supplementary Table 1); star, the 2009 self-assembled lens ($D \approx 2.8\lambda_0$, $n \approx 1.5$). **c**, Collected power of the three strongest object patterns beyond $nk_0$ in each azimuthal order $|l|$ for spheres $\lambda_0$ (purple) and $3.3\lambda_0$ (red) across ($n = 1.6$), relative to $\lambda_{max}$; the first two belong to the two polarizations, and the third, a second radial pattern, lies far below them. Dotted line, −30 dB. **d**, Object-plane intensity of the leading channel beyond $nk_0$ for a line object (s polarization, $n = 1.6$): the patch read by the lens, centred on the contact.

**Table 1 | Channel budget of dielectric spheres in contact** (exact vector Mie theory; air; NA 0.9; general 2D objects; single pass; counts within 30 dB of the brightest channel of the same sphere, with 20–40 dB in parentheses).

| **Index $n$** | **Channels below $nk_0$: $D = \lambda_0$ / $20\lambda_0$** | **Channels beyond $nk_0$ at $D = \lambda_0$** | **Largest count beyond $nk_0$ (at $D$)** | **No single-pass channel beyond $nk_0$ from $D$ =** | **Beyond $nk_0$ at $D = \lambda_0$: $g = 0.05\lambda_0$ / $0.1\lambda_0$** |
|---|---|---|---|---|---|
| 1.46 | 22 / 382 | 12 (10–20) | 36 ($4.0\lambda_0$) | $13.4\lambda_0$ | 12 / 12 |
| 1.6 | 22 / 358 | 12 (8–16) | 28 ($2.8\lambda_0$) | $11.0\lambda_0$ | 12 / 12 |
| 1.9 | 22 / 357 | 14 (8–17) | 20 ($1.8\lambda_0$) | $4.0\lambda_0$ | 12 / 8 |
| 2.3 | 28 / 312 | 14 (8–19) | 14 ($1.0\lambda_0$) | $2.7\lambda_0$ | 12 / 5 |

Channels below $nk_0$ span $k_0 < k_\parallel < nk_0$; channels beyond $nk_0$ span $nk_0 < k_\parallel < 5k_0$. Diameters were sampled at 22 logarithmically spaced values from $0.3\lambda_0$ to $20\lambda_0$, plus $\lambda_0$ and $2.75\lambda_0$; the fifth column gives the smallest sampled diameter from which no larger sphere has a single-pass channel beyond $nk_0$ within 30 dB (light returned onto the object extends the reach of such lenses), and the last column refers counts to the brightest channel in contact, as for a fixed detection floor.

## Why the contact sets the limit

Why does only the contact matter, and why do large beads not see finer? Two simple effects answer both. The first is geometry. A wave decaying as $\exp(-\alpha z)$ is felt only where the lens surface lies within about $1/\alpha$ of the object, a few tens of nanometres for the finest detail. Near the contact a sphere's surface rises as $h \approx r^2/D$ at a distance $r$ from the contact point, so the lens meets the wave only over a patch of width $w \sim (D/\alpha)^{1/2}$. Like a slit, this patch spreads the transverse momentum of the light by about $1/w$: a wider patch reads a larger area but a narrower band of frequencies, and the product of the two is fixed by the contact, much as the number of degrees of freedom of an optical system is fixed[32]. The computed values bear this out (Fig. 4a): from $D = 0.45\lambda_0$ to $3.3\lambda_0$ ($n = 1.6$) the patch widens 3.6-fold while the bandwidth beyond $nk_0$ narrows 3.7-fold.

The second is trapping. Light coupled beyond $nk_0$ enters the lens near grazing incidence. In a small sphere, diffraction turns it towards the objective; in a large one, total internal reflection guides it around the surface, and it escapes only through narrow resonances[22,24] (Supplementary Fig. 1d). Relative to the brightest channel, the leading channel beyond $nk_0$ falls from −6 dB at $D \approx \lambda_0$ to −80 dB, a hundred-millionth, at $20\lambda_0$ ($n = 1.6$; Fig. 4b). A higher index raises the immersion limit but traps the light more strongly, closing the size window earlier, and makes the pixel more sensitive to a gap between lens and object (Fig. 4d and Table 1). This explains why neither a larger bead nor a higher index guarantees finer detail: what matters is a sharp, close contact from which the light can escape.

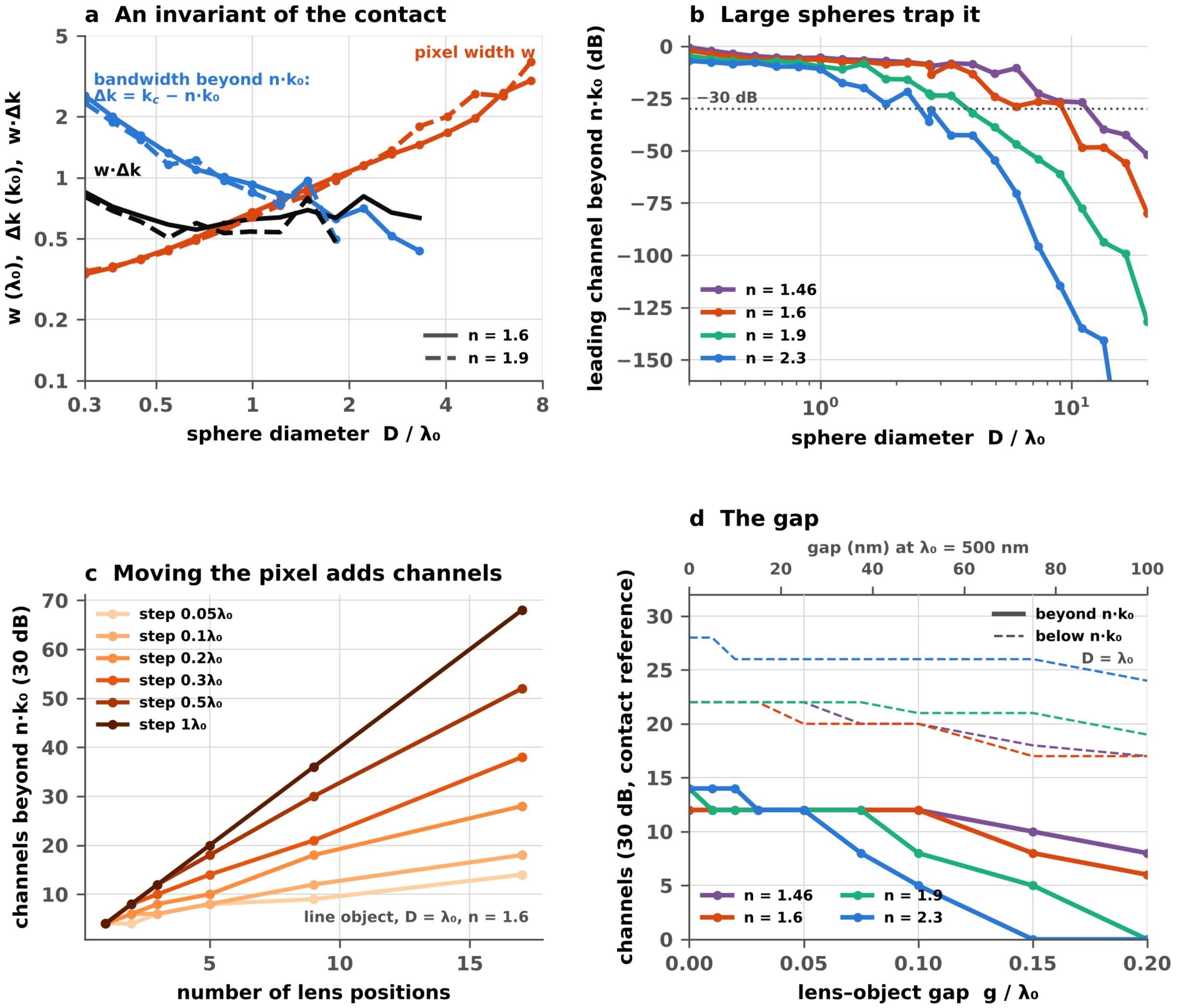

**Fig. 4 | Why the contact sets the limit.** Spheres in contact, NA 0.9, exact vector Mie theory, single pass. **a**, Width $w$ of the patch read beyond $nk_0$ (full width at half maximum of the leading line-object channel; red), bandwidth beyond $nk_0$ of a single object component ($\Delta k = k_c - nk_0$, 30 dB; blue) and their product (black) versus diameter, for $n = 1.6$ (solid) and 1.9 (dashed); bandwidth and product are shown up to the diameter of largest capacity beyond $nk_0$. **b**, Collected power of the leading channel beyond $nk_0$ relative to $\lambda_{max}$ of the same sphere. **c**, Channels beyond $nk_0$ for a line object read by a sphere ($D = \lambda_0$, $n = 1.6$) at M positions spaced by the step shown, counted against the brightest channel of a single position. **d**, Channels beyond $nk_0$ (solid) and below it (dashed) versus the lens–object gap for spheres one wavelength across, counted against the brightest channel in contact; upper axis, gap in nanometres at $\lambda_0 = 500$ nm.

## Many contacts read many pixels

If the limit belongs to the contact, a lens with many contacts should read many pixels at once. We tested this with 2D full-wave calculations, which reproduce the exact channel operators of cylinders within 1% (Methods and Supplementary Fig. 2). A smooth cylinder of index 1.6 passes two channels beyond $nk_0$ up to $7\lambda_0$ and none from $9\lambda_0$, mirroring the sphere. A plano-convex lens $6\lambda_0$ wide resting on its flat face also passes two (Fig. 5a), read at the edges of the face, where the surface leaves the object abruptly; the flat face itself conserves $k_\parallel$, as a solid-immersion lens does[5,8]. Small hemicylindrical bumps of radius $0.1\lambda_0$ on the flat face change this (Fig. 5a,b). Each bump acts as a pixel: the channels beyond $nk_0$ concentrate under the bumps, and their number rises from 2 to 12 for 9 bumps $0.6\lambda_0$ apart. Randomly placed bumps give 11, so the gain does not come from grating diffraction. A textured face flattens the channel spectrum beyond $nk_0$ while keeping the immersion band (Fig. 5c,d). The two jobs of a sphere are thereby separated, which gives a design rule: the contacts define the pixels and supply the momentum, and the body collects the light.

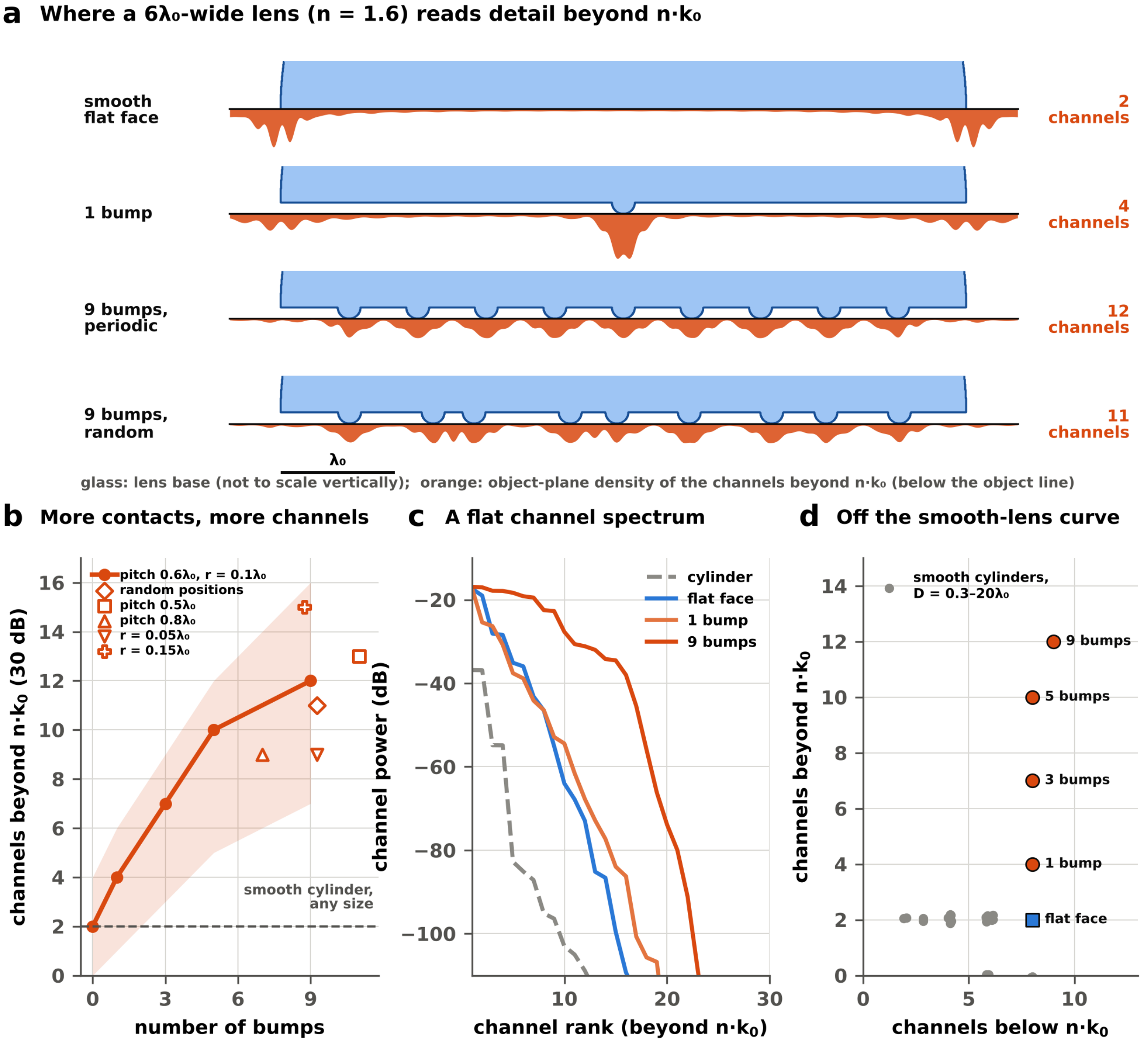

**Fig. 5 | Textured contact faces multiply the channels beyond $nk_0$.** Full-wave 2D finite-difference frequency-domain calculation; transverse-magnetic polarization, index 1.6, air, NA 0.9, line objects with $k_\parallel$ of either sign; channels counted within 30 dB ($\lambda_i \geq 10^{-3}\lambda_{max}$; $\lambda_{max}$, the brightest channel of either evanescent band of the same lens and measurement). **a**, Plano-convex lenses $6\lambda_0$ wide resting on a smooth flat face, or on hemicylindrical bumps of radius $0.1\lambda_0$ (pitch $0.6\lambda_0$, or displaced randomly by up to $\pm 0.15\lambda_0$). Glass, lens base (vertical scale exaggerated); red, object-plane density of the channels beyond $nk_0$ (sum of their intensities, drawn below the object line), showing where the lens reads; right, number of channels. **b**, Channels beyond $nk_0$ versus the number of bumps (pitch $0.6\lambda_0$, radius $0.1\lambda_0$; shading, 20–40 dB), with variants of position, pitch and radius (open symbols). Dashed line, smooth cylinders of any diameter up to $7\lambda_0$. **c**, Collected power of the channels beyond $nk_0$, in rank order, for a $6\lambda_0$ cylinder, the smooth plano-convex lens and lenses with one and nine bumps. **d**, Channels beyond versus below $nk_0$: smooth cylinders of diameter 0.3–$20\lambda_0$ (grey, slightly jittered for visibility) and the flat-faced lenses of a,b.

## Contacts multiply in 3D

In 3D contacts multiply even more strongly (3D full-wave model for rotationally symmetric lenses, validated against exact Mie theory; Methods and Supplementary Fig. 5a). A plano-convex lens $4\lambda_0$ across ($n = 1.6$) resting on its flat face passes 51 channels beyond $nk_0$: its rim is a circular contact line, which reads about one channel per polarization in each azimuthal order up to $|l| = 13$. Held $0.1\lambda_0$ above the object, the same face passes 33, read through the gap at its rim. A central hemispherical bump of radius $0.1\lambda_0$ touching the object adds one channel, a ring of the same cross-section at a radius of $0.6\lambda_0$ raises the count to 58, and three rings $0.6\lambda_0$ apart raise it to 108 (Fig. 6a,b), or 167 at 40 dB. That is twice the smooth face in contact and nearly four times the 28 channels of the best smooth sphere of the same index.

These numbers follow a simple counting rule (Supplementary Note 2, Supplementary Fig. 13 and Supplementary Table 2). A lens delivers azimuthal orders up to about the number of wavelengths that fit around it, $|l| \lesssim \pi D/\lambda_0$: 13 for this lens and 7 for the best sphere. A contact at radius ρ meets a pattern of order $l$ only where that pattern is not dark, $|l| \lesssim k_{\parallel}\rho$. A bump on the axis therefore meets only $l = 0$ and adds one channel, whereas each ring meets many orders and adds about 25. The count is the number of orders, times the contacts that meet each order, times the polarizations above the threshold. In 3D a ring is the natural contact, and channels can be counted before they are computed.

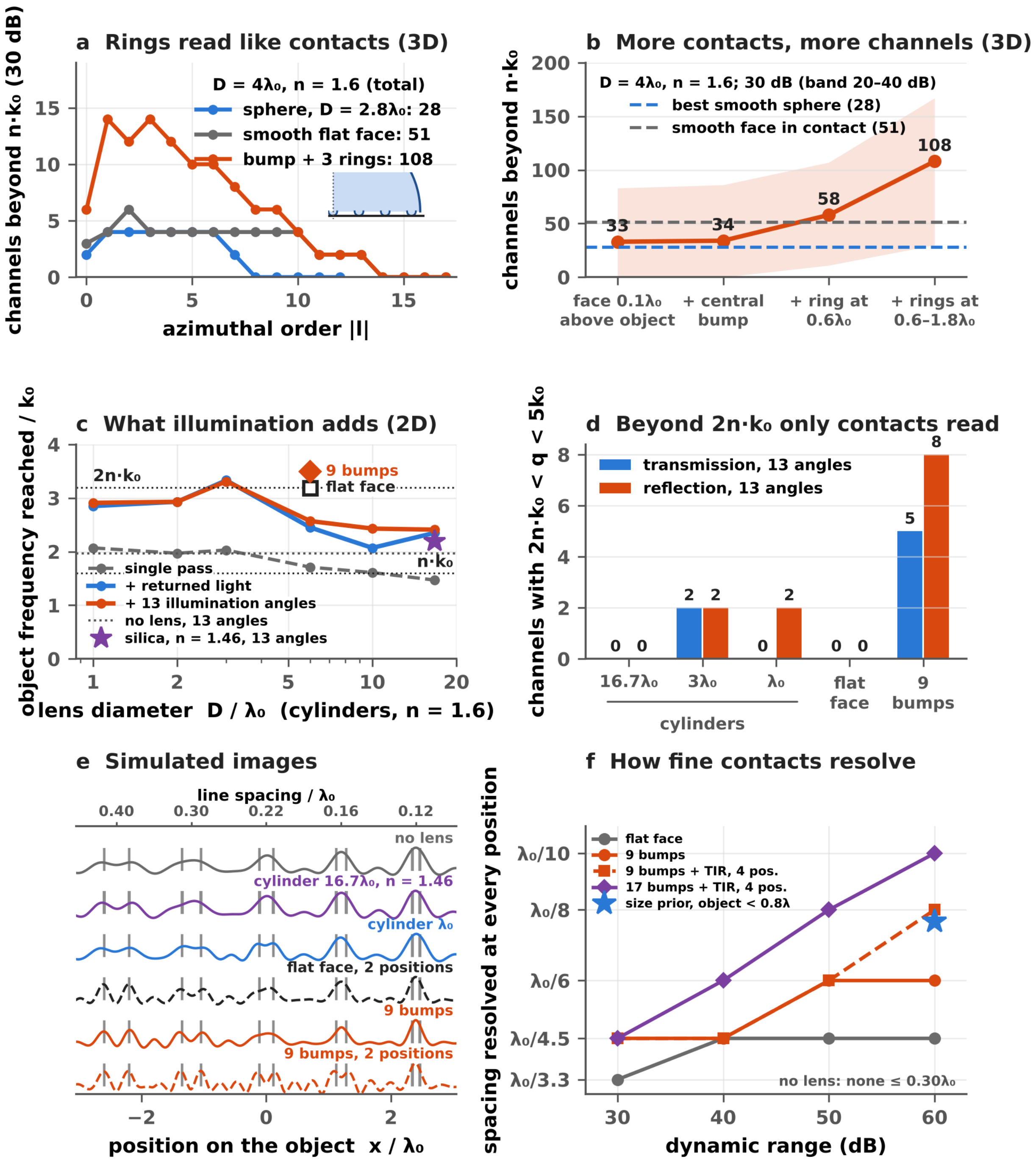

**Fig. 6 | Contacts in 3D and in complete imaging. a,b**, 3D full-wave model for rotationally symmetric lenses, single pass; **c–f**, 2D complete-imaging model (object on glass, $n = 1.6$). **a**, Channels beyond $nk_0$ per azimuthal order $|l|$ (both signs and polarizations; 30 dB) of plano-convex lenses $4\lambda_0$ across on a smooth flat face (grey) or on a face with a central bump and three rings of radius $0.1\lambda_0$ at a pitch of $0.6\lambda_0$ (red; inset, half cross-section, not to scale), and of a sphere $2.8\lambda_0$ across (blue); totals in the legend. **b**, Channels beyond $nk_0$ of a smooth face held $0.1\lambda_0$ above the object as contacts are added; band, 20–40 dB; dashed lines, smooth face in contact (grey) and best smooth sphere (blue). **c**, Largest object frequency read within 30 dB: single pass under normal illumination (grey), with returned light (blue) and with 13 illumination angles within NA 0.9 (red); open square and filled diamond, flat-faced and nine-bump lenses $6\lambda_0$ wide; star, silica cylinder $16.7\lambda_0$ across (geometry of ref.[24]); dotted lines, $nk_0$, $2nk_0$ and the bare microscope. **d**, Channels between $2nk_0$ and $5k_0$ (30 dB) under 13 transmission (blue) or reflection (red) angles. **e**, Simulated images (reconstructed $\Delta\varepsilon$) of five line pairs (grey bars; spacings above) from exact far fields with noise 30 dB below the signal, 13 transmission angles; dashed, two lens positions $0.3\lambda_0$ apart. **f**, Finest spacing resolved at all seven test positions in all 20 noise realizations (Methods) versus dynamic range; TIR, evanescent illumination by total internal reflection in the substrate. Star, $\lambda/8$ point-spread width of limited-size-object microscopy for isolated objects smaller than $0.8\lambda$ (60 dB)[27], a different metric shown for orientation.

## Returned light doubles the reach; contacts read beyond it

In a microscope the lens also lights the object, and light bouncing between lens and object can carry extra detail; Maslov and Astratov identified such coherent re-illumination, in first-principles simulations of complete imaging, as a key to the fine features seen through large spheres[24]. To see how returned light and contacts share the work, we simulated complete 2D imaging. The object is a thin patterned layer on glass under the lens, illuminated from below or through the lens at up to 13 angles within NA 0.9; the model includes every reflection (Methods and Supplementary Fig. 5b). The reach of a lens is the finest object frequency whose collected power stays within 30 dB of that of coarse, propagating detail.

Returned light matters, as Maslov and Astratov found[24]. Under normal illumination (Fig. 6c), a cylinder one to three wavelengths across reads frequencies up to 2.9–3.3$k_0$ instead of 2.0$k_0$ in a single pass, and one 16.7$\lambda_0$ across up to 2.35$k_0$ instead of 1.5$k_0$. With 13 angles large lenses reach 2.4–2.6$k_0$, close to the $(n + \mathrm{NA})k_0 = 2.5k_0$ of an immersion lens under oblique light. In the 10 μm silica geometry (16.7$\lambda_0$, $n = 1.46$) the reach is 2.2–2.5$k_0$, close to the resolution of $\lambda/2.8$ that Fourier analysis of their first-principles images gives[24]. A flat face in contact returns light across its whole width and reaches $3.2k_0 = 2nk_0$, the doubling familiar from solid-immersion microscopy[5].

Beyond $2nk_0$, returned light and contacts part ways (Fig. 6d). In the geometries tested, the large lenses and the flat face pass no channels there, whereas curved contacts do: cylinders 1–3 wavelengths across pass up to 2, and the face with 9 bumps passes 5 under transmission and 8 under reflection illumination (Supplementary Fig. 6c). This reconciles the competing pictures: the nanojet, resonance, effective-aperture and re-illumination pictures all describe light that reaches the object from or through the lens, which extends every lens towards $2nk_0$, whereas detail beyond $2nk_0$ is read only by curved contacts.

## Channels become images

Do the extra channels turn into sharper images? We simulated images of five pairs of thin lines 0.40–0.12$\lambda_0$ apart under 13 transmission angles, from exact far fields that include all multiple scattering, with noise 30 dB below the signal, and reconstructed them with the known linear operator of each lens (Methods). Without a lens only the pair 0.40$\lambda_0$ apart is resolved (Fig. 6e). A silica cylinder 16.7$\lambda_0$ across, in the geometry of ref.[24], narrows a single line from 0.33$\lambda_0$ to 0.27–0.29$\lambda_0$ but resolves no closer pair. A flat face within 0.05$\lambda_0$ of the object resolves pairs 0.30$\lambda_0$ apart. The face with nine bumps, read at two positions 0.3$\lambda_0$ apart, also resolves the pairs 0.22$\lambda_0$ and 0.16$\lambda_0$ ($\lambda_0$/6) apart, in the order predicted by the channel counts and in all 20 noise realizations (Supplementary Fig. 8). Phase need not be measured: bright-field intensity images alone resolve the 0.22$\lambda_0$ pair, and through-focus stacks the $\lambda_0$/6 pair (Supplementary Figs. 7 and 8).

## A tenth of a wavelength across the whole field

A useful imager must resolve detail everywhere in its field, not only at the best spot. We therefore count a spacing as resolved only if pairs of lines placed at seven positions across the contact array all show two clear peaks, with a dip of at least 25% between them, in every one of 20 noise realizations (Methods and Supplementary Note 3). By this field-wide test the bare microscope resolves no spacing of 0.30$\lambda_0$ or less, the flat face 0.22$\lambda_0$ at best and the nine-bump face, read at four positions, 0.12$\lambda_0$ at a dynamic range of 60 dB, a signal-to-noise ratio of $10^6$ (Fig. 6f and Supplementary Fig. 9). Seventeen bumps of radius 0.1$\lambda_0$, 0.3$\lambda_0$ apart, read at four positions with evanescent illumination by total internal reflection in the substrate, as in chip-based label-free nanoscopy[38], resolve $\lambda_0/6$ at 40 dB, $\lambda_0/8$ at 50 dB and $\lambda_0/10$ at 60 dB at all seven positions, which span $4\lambda_0$; complete full-wave data confirm this at every dynamic range. With a reconstruction matched to the expected signal, the same data resolve $\lambda_0/12.5$ at 60 dB in the linear model (Methods). At $\lambda_0$ = 400 nm these spacings are 40 and 32 nm, across 1.6 µm.

Limited-size-object microscopy, introduced by Chang et al., has resolved a $\lambda/8$ point-spread width in 1D ($\lambda/7$ in 2D) for isolated objects smaller than $0.8\lambda$, with evanescent illumination and channels read 50–60 dB below the strongest[27]. It extracts fine detail from the far field by extrapolation, using the prior knowledge that the object is confined; each further mode arrives weaker than the last, so the field of view must stay below a wavelength and the object must be isolated. Contacts work differently: they convert evanescent detail into travelling light where the detail lies, need no prior on the size of the object, and read a field that grows with the number of contacts, here $\lambda_0/8$ to $\lambda_0/10$ across four wavelengths at the same 50–60 dB. Both approaches need an operator calibrated to the noise level, and our results reflect an idealized 2D model. A dynamic range, however, is not a knob on a microscope; the number of photons is.

**The photon price of field-wide resolution**

Detecting light is inherently noisy, like counting raindrops in a bucket: a count of $N$ photons fluctuates by about $\sqrt{N}$, so every tenfold gain in precision costs a hundredfold more photons. The Fisher information makes this rigorous: its inverse bounds the variance of any unbiased estimate of the object[33]. We computed it for the most capable configuration of Fig. 6f: 17 contacts of radius $0.1\lambda_0$ at a pitch of $0.3\lambda_0$, 13 transmitted and 10 totally internally reflected plane waves and four lens positions, 92 exposures in all (Fig. 7a). The image-plane field is $E = B + Gc$, where $B$, the background, is the light of the lens alone and $Gc$ the weak contribution of the object $c$. The photon budget $N$ is the number of photons per $\lambda_0^2$ summed over all exposures, and the photon threshold of a spacing is the smallest budget from which it, and every coarser spacing, passes the field-wide test (Methods). At $\lambda_0$ = 500 nm, 1 kW cm$^{-2}$ delivers $6.3 \times 10^{12}$ photons per $\lambda_0^2$ per second; the test lines are 12 nm wide and 6 nm thick, with a weak permittivity contrast $\Delta\varepsilon = 0.05$.

At the quantum limit, $\lambda_0/6$ costs $2.9 \times 10^{12}$ photons per $\lambda_0^2$, $\lambda_0/10$ costs $2.9 \times 10^{14}$ and $\lambda_0/12.5$ costs $9.2 \times 10^{15}$ (Fig. 7b and Supplementary Table 6): 46 s and 24 min at 1 kW cm$^{-2}$ for the last two. Each further step of resolution costs roughly ten times more photons than the one before, because the channels that carry finer detail leave the contacts weaker (Fig. 7f), much as superoscillatory hotspots are paid for with sideband energy[39]. The reconstruction rule moves the finest prices by up to tenfold but not the comparisons below (Supplementary Fig. 10b). The dynamic ranges of Fig. 6f, 40–60 dB, correspond to $6.9 \times 10^{14}$–$6.9 \times 10^{16}$ photons per $\lambda_0^2$ (triangles in Fig. 7b). To our knowledge this is the first photon budget for field-wide microlens super-resolution: it turns dynamic-range claims into an exposure that an experimenter can plan.

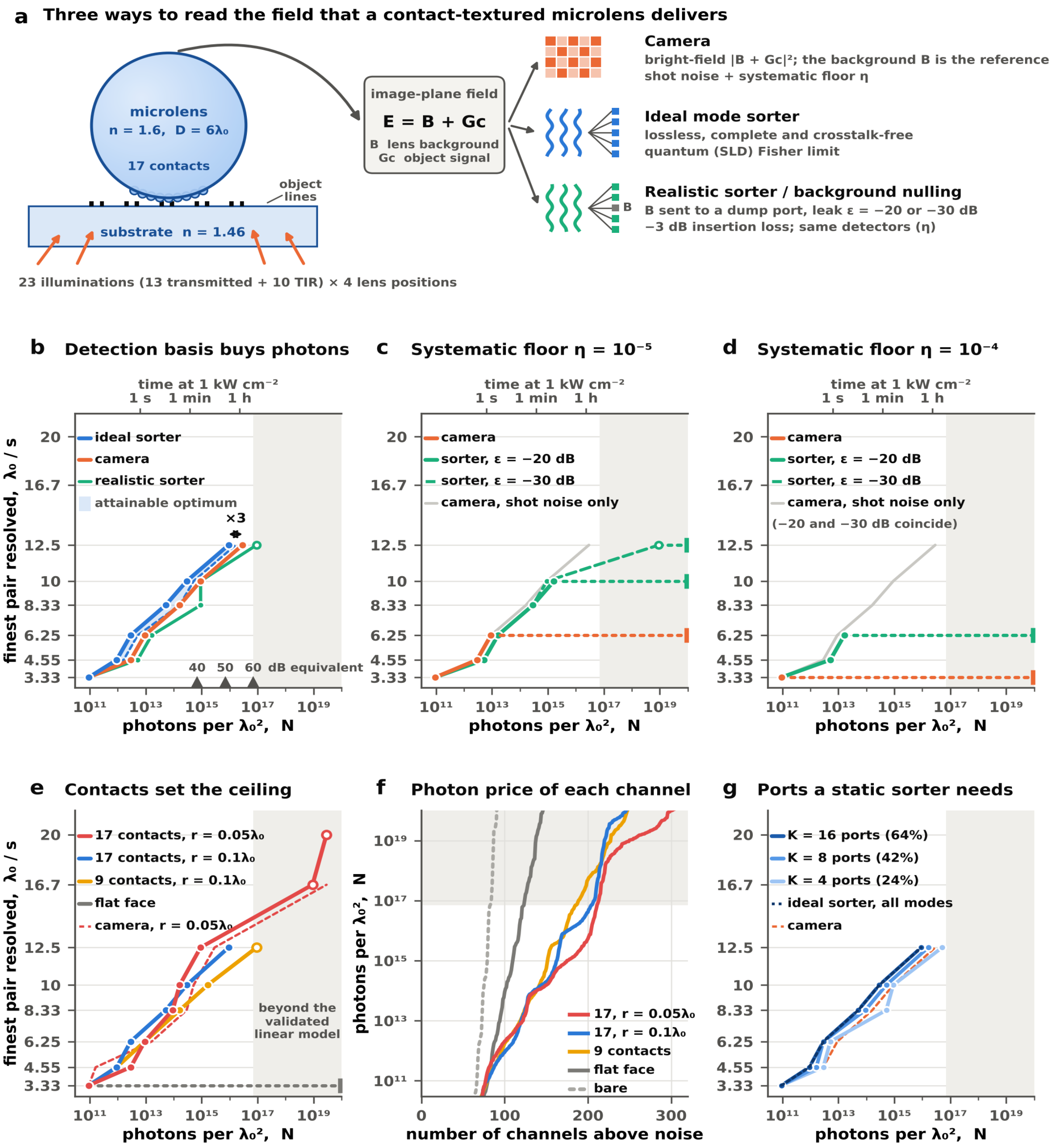

**Fig. 7 | The photon price of field-wide resolution. a**, The lens of Fig. 6f (17 contacts of radius $0.1\lambda_0$, pitch $0.3\lambda_0$) on glass, illuminated by 13 transmitted and 10 totally internally reflected plane waves and read at four positions; the image-plane field $E = B + Gc$ ($B$, lens background; $Gc$, object signal) is detected by a bright-field camera, an ideal mode sorter or a realistic sorter that sends $B$ to a dump port. **b–e,g**, Finest line-pair spacing $s$, as $\lambda_0/s$, resolved at all seven positions under the contacts in all 20 noise realizations, versus the photon budget $N$ (photons per $\lambda_0^2$ summed over the 92 exposures); points, smallest $N$ from which that spacing and all coarser ones stay resolved; upper axis, time at 1 kW cm$^{-2}$ for $\lambda_0 = 500$ nm. Grey regions, budgets above the 60 dB equivalent, beyond the validated linear model; open symbols, points there. **b**, Shot-noise-limited detection: ideal sorter (quantum bound), camera and realistic sorter (3 dB loss); dashed line, off-axis holography (half the quantum information); band, attainable optimum; triangles, budgets equivalent to the dynamic ranges of Fig. 6f; ×3, camera-to-quantum photon ratio at $\lambda_0/12.5$. **c,d**, Systematic floors $\eta = 10^{-5}$ (**c**) and $10^{-4}$ (**d**): camera, and realistic sorter with background leakage $\varepsilon = -20$ or $-30$ dB; dotted lines ending in bars, ceilings that no budget overcomes; grey, shot-noise camera. **e**, Contact faces at the quantum bound: 17 contacts of radius $0.05\lambda_0$ or $0.1\lambda_0$, nine contacts (radius $0.1\lambda_0$, pitch $0.6\lambda_0$) and a flat face; dashed, camera with the $0.05\lambda_0$ contacts. The bare microscope resolves no spacing of $0.30\lambda_0$ or less. **f**, Photon budget at which each Fisher channel of the quantum bound rises above the noise of the two-line test; the number of usable channels at a given $N$ is read horizontally. **g**, Static sorters with $K$ fixed ports, quantum-limited in their ports, with the share of the total quantum Fisher information they collect; dotted, ideal sorter; dashed, camera.

**An ordinary camera is nearly as good as an ideal detector**

Is an advanced detector needed? In the shot-noise limit, no. A bright-field camera needs $9.2 \times 10^{14}$ photons per $\lambda_0^2$ for $\lambda_0/10$ and $2.9 \times 10^{16}$ for $\lambda_0/12.5$, about three times the quantum price, and reaches the same finest spacing (Fig. 7b). The reason is the background. The camera records the intensity $|B + Gc|^2$, whose object-dependent part, $2\mathrm{Re}(B^*Gc)$, is the interference of the weak object light with the bright light of the lens. As in holography, the bright wave amplifies the weak one: its shot noise grows with $|B|$ exactly as the signal does, so the information does not depend on how bright the background is. The lens thus supplies a free, perfectly aligned reference wave, as transmitted light does in Fourier ptychotomography[40]. The loss is exact and small: the quantum Fisher information is the sum of what the camera records and what lies in the quadrature 90° out of phase with the background, which it cannot see, $F_Q = F_{cam} + F_{quad}$ (Supplementary Note 4). Across the channels that set resolution the camera keeps between 1/2.6 and 1/2.1 of the quantum information (Supplementary Fig. 10c).

A mode sorter, which separates light by its spatial pattern as a prism separates it by colour[28,29,41], could in principle reach the quantum limit, but not quite here. When many object features are estimated at once, the ideal measurements for different features can be incompatible, and here they are almost maximally so: the incompatibility measure $R$, which runs from 0 for compatible to 1 for maximally incompatible measurements[42], is at least 0.999 (Methods). The attainable optimum therefore lies between the quantum limit and off-axis holography, which always reaches half of it (shaded band in Fig. 7b). For $\lambda_0/10$ the camera needs 1.8 times the photons of holography and 3.2 times those of the quantum limit, and a realistic sorter with 3 dB insertion loss as many as the camera.

**Rejecting the background defeats systematic errors**

Real detectors also make errors that more photons cannot average away. We model fixed-pattern noise, drift and imperfect knowledge of the lens background as a systematic floor $\eta$, the residual error relative to the detected background (Methods and Supplementary Note 5). The object signal is a tiny part of that background ($|Gc|/|B| \approx 10^{-4}$ in median), so the floor stops the camera: at $\eta = 10^{-5}$ it resolves $\lambda_0/6$ and no finer at any photon budget, and at $\eta = 10^{-4}$ only $\lambda_0/3.3$ (Fig. 7c,d). Here a sorter, or any interferometric nulling, earns its place. Routing the background into a dump port with leakage $\varepsilon$ lowers the floor error, which scales with the detected background power, faster than the signal, which scales with its amplitude: the effective floor becomes $\eta\sqrt{\varepsilon}$. With a realistic sorter (3 dB loss, $\varepsilon = -20$ dB), $\lambda_0/10$ returns at $\eta = 10^{-5}$ ($1.6 \times 10^{15}$ photons per $\lambda_0^2$). Rejection also prevents camera saturation, cutting the detector-limited time for $\lambda_0/10$ from about 3 h to about 4 min (Methods). Interferometric scattering microscopy reached the same conclusion experimentally[43].

**Contacts set the ceiling**

Can more photons or a better detector make up for a plain lens face? No: every face has a resolution ceiling that no photon budget overcomes, and denser, sharper contacts raise it (Fig. 7e). At the quantum limit a flat face resolves $\lambda_0/3.3$ and no finer at any budget. Nine contacts $0.6\lambda_0$ apart reach $\lambda_0/10$ at $1.6 \times 10^{15}$ photons per $\lambda_0^2$. Seventeen contacts of radius $0.1\lambda_0$ reach $\lambda_0/12.5$ at $9.2 \times 10^{15}$, and sharpening them to radius $0.05\lambda_0$ reaches it at $9.2 \times 10^{14}$, ten times cheaper: 2.4 min at 1 kW cm$^{-2}$ at the quantum limit, and 8 min ($2.9 \times 10^{15}$) with a camera.

Figure 7f shows why. Each channel becomes usable when the photon budget lifts it above the noise. For the bare microscope each tenfold increase in photons adds about 3 usable channels; a flat face adds 7 and textured faces add 19–25. At the budget equivalent to 60 dB the bare microscope has 82 usable channels, the flat face 122 and the 17-contact faces 203–210. Photons cannot recover detail that the contacts do not deliver.

**Few ports and many views suffice**

A mode sorter need not separate every pattern: a static sorter, one fixed set of $K$ output modes shared by all 92 exposures and designed from the image-plane patterns of fine detail, performs like the camera with four ports, like off-axis holography with eight, and reaches the full quantum-limited curve to $\lambda_0/12.5$ with 16 (Fig. 7g), although 16 ports collect only 64% of the total Fisher information. The contacts concentrate fine detail into a few far-field patterns, and multi-plane light converters already separate more than 200 spatial modes[41]. Shaping the illumination, by contrast, does not lower the price: spending all photons on the single wavefront that couples most strongly to fine detail needs about thirty times more photons for $\lambda_0/10$ than the 23 plane waves used in turn (Supplementary Fig. 10a). Field-wide resolution needs many independent views rather than one bright view, unlike the estimation of a single parameter, for which an optimal wavefront exists[30].

## Discussion

Counting channels and pricing them in photons settles the four issues raised at the outset. Mechanism: only contacts read beyond the immersion limit in a single pass, and the nanojet[21], resonance[22], effective-aperture[23] and re-illumination[24] pictures all describe light that extends every lens towards $2nk_0$. In terms of Fisher-information flow[31,35], contacts release into the far field what evanescent waves would keep at the object. Yardstick: the channel count measures transferred information above a threshold and complements image simulations[24], resolution criteria[25,26] and information limits for finite objects[27]. Design: channels can be counted before they are computed, and multiplied by adding contacts, from 28 for the best sphere to 108 for a ring-textured face. Price: field-wide $\lambda_0/10$ costs minutes with an ordinary camera, and $\lambda_0/12.5$ is within reach of sharper contacts.

These rules place published measurements in context (Fig. 3b and Supplementary Table 1). The pioneering experiments with self-assembled nanolenses[7], white-light silica spheres[9] and liquid-immersed high-index spheres[10] used lenses inside the single-pass window, where 28–40 channels pass beyond the immersion limit. Later experiments mostly used larger, easier-to-handle lenses, for which returned light[24], focusing through the lens[21], scanning[11,12] and prior knowledge of the object[27] supply what a single pass cannot, consistent with resolved periods of about $\lambda_0/2$ for spheres 7.9–18$\lambda_0$ across[9,44] and with the reach of about $(n + 1)k_0$ computed here.

The results turn design around. Rather than a higher index or a tighter focus, one should engineer the contact region: keep the lens within a few tens of nanometres of the object across the field, so that returned light doubles the immersion band, and present many sharp contacts there to read beyond $2nk_0$. Arrays of plano-convex nanolenses[7], flat-faced hyper-hemispheres[45] and lenses assembled from high-index nanoparticles[46] already exploit these routes, phase-ring microspheres[47] show the value of shaping the lens, and textured faces and printed or condensed lens arrays[14,15,48,49] are natural next steps. Detection follows from the photon analysis: a camera suffices in the shot-noise limit; where systematic errors or saturation dominate, the background should be rejected by 20–30 dB; and 16 well-chosen ports reach the quantum limit. Joint inversion across lens positions and illumination angles assembles the channels into images, as in Fourier ptychography[50].

The finest images need 50–60 dB, as for limited-size objects[27], or $3 \times 10^{14}$ to $10^{16}$ photons per $\lambda_0^2$ for the weak test lines used here. In the linear regime these photon numbers scale as $(\Delta\varepsilon \cdot w \cdot t)^{-2}$, with $w$ and $t$ the width and thickness of a line, so an object that scatters ten times more strongly needs a hundred times fewer photons. Absorption, roughness, strong objects and photon budgets in 3D remain for future study. The results suggest direct tests: large spheres should resolve periods down to about $\lambda_0/(n + 1)$; a textured face should show high-frequency contrast under each contact, measurable with Siemens-star targets[25]; and at $\lambda_0 = 400$ nm, contacts 40 nm in radius and 120 nm apart, read at four positions under evanescent illumination, should resolve line pairs 40 nm apart across 1.6 μm within minutes at 1 kW $\text{cm}^{-2}$ with a camera, and contacts 20 nm in radius should resolve pairs 32 nm apart within about a quarter of an hour. Beyond the immersion limit, a dielectric lens sees its finest detail through its contacts; to see more, we must give it more of them, and detectors only change how quickly we see it.

## Materials and methods

### Channel operator for spheres

An object component $\mathbf{E} = \hat{\mathbf{e}} \exp[\mathrm{i}\mathbf{k}_\parallel \cdot \mathbf{r} - \alpha(z + g + a)]$, $a = D/2$, with complex unit propagation vector and unit object-plane field magnitude for s ($\hat{\mathbf{e}} = \hat{\mathbf{y}}$) and p polarization, is expanded about the sphere centre in vector spherical harmonics continued to complex polar angle[36]; scattered coefficients follow from the Mie coefficients[37]. For $\mathbf{k}_\parallel$ along $x$, the far field is evaluated on Gauss–Legendre nodes $\theta_j$ (48 points) in the objective cone, $\sin\theta \le \mathrm{NA} = 0.9$, as azimuthal Fourier components $F_m(\theta_j)$ of its $\theta$ and $\varphi$ components. For a line object, the columns of $H$ are these far fields for s and p components at $k_\parallel = k_1$, $k_2$, … (step $\Delta k = 0.01k_0$), weighted by $(2\pi w_j \Delta k)^{1/2}$ with quadrature weights $w_j$, so that $\|Hc\|^2$ is the collected power for an object spectrum normalized as $\Sigma|c_i|^2\Delta k = 1$. For a general 2D object, components with in-plane direction $\psi$ are rotations of those along $x$; an object pattern varying as $\exp(\mathrm{i}l\psi)$ therefore couples only to far-field order $m = l$, so $H$ is block-diagonal in $l$, and each block is assembled from $F_l$ with weights $2\pi(w_j k\Delta k)^{1/2}$ (object normalization $\Sigma|c|^2 k\Delta k\Delta\psi = 1$). Channels are the eigenvalues $\lambda_i$ of $H^\dagger H$, computed separately for the immersion band ($k_0 < k_\parallel < nk_0$) and the band beyond it ($nk_0 < k_\parallel < 5k_0$). The $\lambda_i$ are powers, so a threshold of $X$ dB counts channels with $\lambda_i \ge 10^{-X/10}\lambda_{\max}$: $10^{-2}$, $10^{-3}$ and $10^{-4}$ for the 20, 30 and 40 dB used here; absolute floors of $10^{-4}$–$10^{-6}$ of the object spectral energy are a separate check. A threshold of $X$ dB corresponds to a measurement whose noise floor lies $X$ dB below the strongest channel, the role played by the photon budget for limited-size objects[27]. In every model in this paper (spheres, 2D and 3D full-wave lenses, complete imaging and intensity detection), $\lambda_{\max}$ is the brightest channel of the same lens and measurement in either evanescent band, that is, the larger of the leading eigenvalues of the two bands; the only exceptions, stated where they occur, are scanning counts, referred to the brightest channel of a single position, and gap counts, referred to the brightest channel in contact, as for a fixed detection floor. The input norm is the norm of the object-plane field spectrum; another source norm, such as a fixed permittivity contrast in the complete-imaging model, reweights the singular values and can shift individual counts. Diameters are 22 values spaced logarithmically from $0.3\lambda_0$ to $20\lambda_0$, plus $\lambda_0$ and $2.75\lambda_0$; indices are 1.46, 1.6, 1.9 and 2.3, with 1.35 and 1.40 added for the relative indices of immersed lenses in Supplementary Table 1; the lens is in air and in contact. Halving $\Delta k$ changes the counts beyond $nk_0$ by at most one channel and those below it by at most two channels or 5% (Supplementary Fig. 3), and raising the upper band limit from $5k_0$ to $7.5k_0$ and $10k_0$ leaves the counts beyond $nk_0$ unchanged at 20, 30 and 40 dB ($D = \lambda_0$, $2.75\lambda_0$ and $4\lambda_0$; $n = 1.6$).

**Pixel width, scanning and gap**

The object-plane pattern of a channel is $\int v(k_\parallel)\exp(\mathrm{i}k_\parallel x)\mathrm{d}k_\parallel$, where $v$ is its right singular vector; the pixel width is the full width at half maximum of its intensity for the leading line-object channel beyond $nk_0$. The bandwidth beyond $nk_0$ in Fig. 4a is the largest $k_\parallel$ at which the collected response of a single component stays within 30 dB of its value at the light line, minus $nk_0$ (Supplementary Fig. 1). For line objects with $k_\parallel$ of either sign read at $M$ lens positions $x_0$, the Gram matrices $H^\dagger H$ are summed after multiplying each column by $\exp(\mathrm{i}k_\parallel x_0)$, and channels are counted against the brightest channel of a single position. A gap $g$ multiplies each column by $\exp(-\alpha g)$ exactly; counts are referred to the brightest channel in contact, as for a fixed detection floor.

**Full-wave 2D model**

Flat-based and textured lenses were computed with a finite-difference frequency-domain solution of the transverse-magnetic wave equation in a scattered-field formulation, in which the object component drives a source inside the lens; the object is a fixed source as in the multipole model. The grid is $0.004\lambda_0$ near the object plane and $0.008\lambda_0$ laterally across the lens ($0.016\lambda_0$ elsewhere), with area-averaged permittivity, stretched-coordinate perfectly matched layers and one sparse factorization for all components. The collected far field follows from the polarization current. Lenses are full cylinders, plano-convex half-cylinders (flat face down) and plano-convex lenses whose flat face, raised by $r_b$, carries $N$ hemicylindrical bumps of radius $r_b$ whose tips touch the object plane, at a fixed pitch or at positions displaced randomly by up to $\pm 0.15\lambda_0$. Columns of $H$ are computed for $k_\parallel$ of both signs in steps of $0.02k_0$ and weighted as for line objects, with trapezoidal weights over the 241 far-field angles. For full cylinders the operator reproduces the exact multipole operator (Gram matrices within 0.2% at $D = \lambda_0$ and 1% at $6\lambda_0$), its collected response agrees with the exact one within 0.4% for $D \leq \lambda_0$, and refining the grid by 25% leaves every channel count of the smooth and textured lenses unchanged (Supplementary Figs. 2 and 4).

**Full-wave imaging model**

In complete imaging the object is a layer of thickness $0.012\lambda_0$ at the top of a substrate of index 1.46 whose permittivity is changed by $\Delta\varepsilon(x)$; the lens (index 1.6, or 1.46 for the geometry of ref.[24]) rests on it with its lowest point at the surface, and the objective (NA 0.9) collects from the air side. To first order in $\Delta\varepsilon$, the far field scattered by the object is $F(\varphi) = (\mathrm{i}/4)k_0^2\int\Delta\varepsilon(\mathbf{r})E_{\mathrm{rec}}(\mathbf{r};\varphi)E_{\mathrm{ill}}(\mathbf{r})\mathrm{d}^2\mathbf{r}$, where $E_{\mathrm{ill}}$ is the field that illuminates the object with the lens present (incident and substrate-reflected light, and everything the lens returns onto the object) and $E_{\mathrm{rec}}(\mathbf{r};\varphi)$ is, by reciprocity, the field at $\mathbf{r}$ of a unit plane wave arriving from the detection direction $\varphi$, again with the lens present. Both fields are computed with the model above for the lens on the substrate, with one factorization per lens, 96 Gauss–Legendre detection directions across the objective cone, and 13 illumination directions with in-plane wavenumbers from $-0.9k_0$ to $0.9k_0$, from below through the substrate (transmission) or from above through the lens (reflection). In the single-pass variant $E_{\mathrm{ill}}$ is the illumination without the lens. The operator maps windowed object gratings $w(x)\exp(iqx)$ ($w$ flat for $|x| < 3\lambda_0$ with Gaussian edges of width $\lambda_0$; steps of $0.02k_0$) onto the far fields of all illuminations; channels are counted as for spheres, against the brightest channel of either evanescent band ($k_0 < |q| < nk_0$ or $nk_0 < |q| < 5k_0$) of the same lens and set of illuminations, and the reach is the largest $|q|$ whose collected power, summed over all measurements, lies within 30 dB of the strongest component with $|q| < 0.9k_0$. In this model the source is the permittivity pattern, each grating normalized to unit norm of $\Delta\varepsilon$ over the window, the measurement is the complex far field in the objective cone (power summed over detection directions and illuminations), and the illumination set and detector are stated with each result; these norms differ from the single-pass field-spectrum norm, so single-pass and imaging counts are compared only within each model. The operator reproduces complete full-wave solutions with the object included with a relative error of about $0.01\Delta\varepsilon$, that is about 1% for the line objects used here ($\Delta\varepsilon = 1$) and 10% for $\Delta\varepsilon = 10$ (Supplementary Fig. 5b). Lenses $10\lambda_0$ across and larger used a grid of $0.010\lambda_0$ laterally and $0.020\lambda_0$ elsewhere; for $D = 10\lambda_0$ the two grids give the same channel counts and reaches within $0.02k_0$, and diameters of 16.65, 16.70 and $16.75\lambda_0$ and a gap of $0.033\lambda_0$ bound the effect of individual resonances (Supplementary Fig. 5c). Extending the object band to $7.5k_0$ and $10k_0$ leaves every reach unchanged and changes the counts beyond $2nk_0$ by at most one channel.

**Intensity detection**

For weak, lossless objects the image intensity in a plane at height $y_f$ along the optical axis ($y$; $y_f = 0$ is the object plane and negative values lie below it) is $I = |D + E_0 + \delta E|^2$, where $D$ is the directly transmitted wave, $E_0$ the light scattered by the lens alone and $\delta E$ the contribution of the object, so that $\delta I = 2\mathrm{Re}[(D + E_0)^*\delta E]$ to first order. Image-plane fields follow from the far fields within the objective aperture, $E(x, y_f) = (1/\pi)\Sigma_j w_j F(\varphi_j)\exp[ik_0(x \cos \varphi_j + y_f \sin \varphi_j)]$, at unit magnification, where $\varphi_j$ is the detection angle measured from the object plane and $w_j$ are the quadrature weights. We evaluated images in the object plane for each illumination separately and with the intensities of the 13 illuminations added in one partially coherent (Köhler) image, and a through-focus stack from $30\lambda_0$ below to $2\lambda_0$ above the object plane, and compared reach and channel counts with complex-field detection for the same real cosine and sine object gratings (Supplementary Fig. 7).

**Full-wave 3D model: rotationally symmetric lenses**

3D lenses of revolution were computed with a full-wave vector body-of-revolution finite-integration method, which solves Maxwell's equations for a rotationally symmetric lens and object patterns of any form: fields are expanded in $\exp(im\varphi)$, and each order is a vector problem in ($\rho$, $z$) with line-integral unknowns $E_\rho$, $\rho E_\varphi$ and $E_z$, stretched-coordinate perfectly matched layers at the outer radius and at both ends, and sub-cell averaging of the permittivity. With the reflection $\varphi \rightarrow -\varphi$ the system matrix becomes complex symmetric and is factorized once per order. The grid is $0.005\lambda_0$ in $z$ near the object plane, $0.01\lambda_0$ in $\rho$ across the lens and $0.02\lambda_0$ elsewhere. The object field of azimuthal order $l$, $\int d\psi \exp(il\psi)\hat{\mathbf{e}} \exp(i\mathbf{k}_\parallel\cdot\mathbf{r} - \alpha z)$, is written in closed form with Bessel functions for s and p polarization, the collected far field follows from an exact azimuthal transformation of the polarization current, and channels are counted as for spheres ($\Delta k = 0.02k_0$; orders until the strongest eigenvalue falls at least 45 dB below the brightest channel of the lens; order $-l$ has the eigenvalues of $l$). For spheres one and 2.8 wavelengths across, the eigenvalues within 30 dB of the brightest agree with exact Mie theory within 0.5 dB (median deviation 2%) and give the same channel counts (Supplementary Fig. 5a). Textured lenses are plano-convex lenses whose flat face, raised by $0.1\lambda_0$, carries a central hemispherical bump and toroidal rings of semicircular cross-section of radius $0.1\lambda_0$ at radii of 0.6, 1.2 and $1.8\lambda_0$, all touching the object plane, and are compared with the same smooth face raised by $0.1\lambda_0$; refining the grid by 25% leaves every channel count of the textured lens unchanged at 20, 30 and 40 dB (Supplementary Fig. 5d).

**Simulated imaging**

The test object consists of five pairs of lines $0.04\lambda_0$ wide with $\Delta\varepsilon = 1$, separated by 0.40, 0.30, 0.22, 0.16 and $0.12\lambda_0$ and centred at $x = -2.4, -1.2, 0, 1.2$ and $2.4\lambda_0$. Its far fields under the 13 transmission illuminations were computed with the object included in the full-wave operator, for the object centred under the lens and displaced by $0.3\lambda_0$ (second lens position); the far fields of the lens alone were subtracted, and complex Gaussian noise was added at 30 dB below the root-mean-square signal. Images were reconstructed with the linear operator of the nominal lens by truncated singular-value inversion, keeping singular values within 30 dB of the largest, and for two lens positions by joint inversion of both data sets. The width of a reconstructed line is the full width at half maximum of the corresponding resolution kernel, and a pair of spacing $s$ counts as resolved when the two strongest peaks within $s/2 + 0.08\lambda_0$ of its centre lie on either side of it, the reconstruction dips by at least 25% between them, and both reach at least half the largest $|\Delta\varepsilon|$ within $0.5\lambda_0$; the last condition rejects ringing that a dip test alone would accept. Robustness to other noise levels, truncation thresholds, dip criteria and mismatched lenses (gap, index and position errors) was tested with 20 noise realizations each (Supplementary Fig. 8 and Supplementary Note 6). For intensity-only reconstruction, exact bright-field images $|D + E|^2$ were formed from the same far fields as in 'Intensity detection', the image of the lens alone was subtracted, real Gaussian noise was added at 30 dB below the root-mean-square intensity signal, and the object, taken as real, was reconstructed with the linearized intensity operator from one image per illumination direction in the object plane, from one partially coherent image per lens position, or from stacks of 14 partially coherent images between $30\lambda_0$ below and $2\lambda_0$ above the object plane. As ref.[25] recommends, we state the prior knowledge used: the operator of the lens; a thin object layer of known height, akin to the finite-thickness prior of analytic Fourier ptychotomography[40]; confinement of the object to the $13\lambda_0$ computational window; and, for intensity data, a real permittivity change. No sparsity, positivity or shape prior is imposed, and truncation of the singular-value expansion is the only regularization. The window is too wide for support-based extrapolation[27] to contribute: without a lens the reach is $1.97k_0$, close to the $1.8k_0$ of the bare microscope under 13 illumination angles.

**Resolution study**

Faces with 17 bumps of radius $0.1\lambda_0$ or $0.05\lambda_0$ at a pitch of $0.3\lambda_0$ were added to the flat and nine-bump faces (lenses $6\lambda_0$ wide, $n = 1.6$; Supplementary Fig. 9a). Besides the 13 transmission angles, ten evanescent illuminations with in-plane wavenumbers ±1.0, 1.1, 1.2, 1.3 and $1.4k_0$ were applied from below by total internal reflection in the glass substrate. The object was read at 1–10 lens positions spaced evenly over one bump pitch. Each test object consists of nine pairs of lines $0.024\lambda_0$ wide ($\Delta\varepsilon = 0.05$) with spacing $s$ between $0.30\lambda_0$ and $0.06\lambda_0$, placed so that together they sample every position relative to the contacts; the seven inner pairs lie under the contact array. Data were generated with the linear operator, with complex Gaussian noise 30–60 dB below the root-mean-square signal (20 realizations), and reconstructed by truncated singular-value inversion at the noise level. A spacing counts as resolved when all seven inner pairs pass the test above in every realization, as do all larger spacings. For these weak objects the linear operator reproduces complete full-wave solutions to 0.04% (−68 dB); for the 17-bump face of radius $0.1\lambda_0$ with evanescent illumination and four positions, the test was repeated with complete full-wave data, including all multiple scattering, for every spacing from $0.30\lambda_0$ to $0.08\lambda_0$ at 30, 40, 50 and 60 dB, and gives the same finest spacings as the linear operator: $0.22\lambda_0$, $\lambda_0/6$, $\lambda_0/8$ and $\lambda_0/10$ (Supplementary Fig. 9d,e). The transfer support is the largest object frequency, up to $10k_0$, whose collected power lies within the stated dynamic range of the strongest propagating component; its inverse is the smallest full-pitch period transferred[25]. Details are given in Supplementary Note 3. Twenty realizations make the test strict at a manageable cost: a spacing that passes all of them fails in fewer than 15% of realizations with 95% confidence, while every spacing is still tested at every noise level and photon budget.

**Photon budgets and Fisher information**

For the photon analysis the far fields of the resolution study (13 transmission and 10 evanescent illuminations; four positions; 92 exposures) were mapped to image-plane fields at unit magnification, sampled every $dx = 0.1\lambda_0$ for $|x| \le 9\lambda_0$ (181 pixels). In exposure $(p, m)$ the field is $B_m + G_{pm}c$, with $c = \Delta\varepsilon\sqrt{w}$ on the $0.008\lambda_0$ object grid (1,625 real values), $B_m$ the image of the lens alone plus the directly transmitted wave, and fields referred to a unit-amplitude plane wave in the substrate. A fluence of $\Phi$ photons per $\lambda_0^2$ per exposure delivers $\Phi dx|E|^2$ photons to a pixel, and the budget is $N = 92\Phi$. At $\lambda_0 = 500$ nm, 1 kW cm$^{-2}$ corresponds to $6.29 \times 10^{12}$ photons per $\lambda_0^2$ per second. Image-plane and far-field normalizations differ by about 25% in the trace of the Fisher matrix, which sets the accuracy of the absolute photon numbers.

Each exposure is a coherent state of amplitude $(\Phi dx)^{1/2}(B + Gc)$ in the pixel modes. The quantum (symmetric-logarithmic-derivative) Fisher matrix[30] is $F_Q = 4\Phi dx\ \Sigma_{p,m}\ \mathrm{Re}(G_{pm}^{\dagger}G_{pm})$; off-axis holography and heterodyne detection reach $F_Q/2$. A bright-field camera with shot noise and a systematic floor $\eta$ has pixel variance $n_b + \eta^2 n_b^2$, with $n_b = \Phi dx|B|^2$, so that $F_{cam} = \Sigma\ 4\Phi dx(1 + \eta^2 n_b)^{-1} r^T r$, with $r = \mathrm{Re}(e^{-i\varphi}g)$, $\varphi$ the phase of $B$ and $g$ the row of $G_{pm}$ for that pixel; for $\eta = 0$ it does not depend on $|B|$. The realistic sorter attenuates the background power by $\varepsilon$ before the same detectors and transmits half of all light, the residual background serving as the reference. The static $K$-port sorter projects every exposure onto the $K$ leading eigenvectors of $\Sigma\ G_{pm}UU^T G_{pm}^{\dagger}$, where the columns of $U$ span windowed object gratings with $|q| = 3.2$–$10k_0$ confined to $|x| \le 2.4\lambda_0$, and is quantum-limited in its ports; these modes collect 24%, 42%, 64% and 88% of the trace of $F_Q$ for $K = 4$, 8, 16 and 32. With $D = 4\Phi dx\ \Sigma\ G_{pm}^{\dagger}G_{pm}$, the incompatibility $R = ||F_Q^{-1}\ \mathrm{Im}\ D||_\infty$ bounds the gap between the attainable (Holevo) bound and the quantum bound by a factor $1 + R$ (ref.[42]); on the 133–206 channels retained for $N = 9.2 \times 10^{13}$–$9.2 \times 10^{16}$, $R \ge 0.9996$. Derivations are given in Supplementary Note 4 and summarized in Supplementary Table 5.

**Reconstruction, photon thresholds and validity**

For every scheme the object is estimated by Fisher-weighted eigen-inversion: with $F = V\Lambda V^T$, $\hat{c} = V_k\Lambda_k^{-1}V_k^T(Fc + n)$, $n \sim \mathrm{N}(0, F)$, keeping the $k$ eigenvalues $\lambda_i \ge N_c/||c||^2$ ($N_c = 1{,}625$) whose expected signal exceeds the noise; the rule uses the expected object energy as prior knowledge[25]. Scaling its threshold by 0.1 or 10 changes the photon price of $\lambda_0/10$ and $\lambda_0/12.5$ by up to an order of magnitude, while the camera stays within a factor of 1–3.3 of the quantum bound (Supplementary Fig. 10b). $\Phi$ was stepped on a quarter-decade grid from $10^9$ to $10^{18}$, and every spacing was tested at every $\Phi$ with fresh noise. The threshold of a spacing is the smallest Φ from which it stays resolved at every larger Φ, an isolated failure between two successes being treated as a noise dropout; thresholds are uncertain by one grid step (a factor of 1.8). Noise $X$ dB below the root-mean-square far-field signal of the object with $s = 0.12\lambda_0$ corresponds to $N = 6.9 \times 10^{14}$, $6.9 \times 10^{15}$ and $6.9 \times 10^{16}$ for 40, 50 and 60 dB; with the far-field Fisher matrix of that noise model, the signal-matched rule resolves $\lambda_0/10$, $\lambda_0/10$ and $\lambda_0/12.5$ at 40, 50 and 60 dB, against $\lambda_0/6$, $\lambda_0/8$ and $\lambda_0/10$ with truncation at the noise level (Fig. 6f). The 60 dB-equivalent budget is the largest at which the linear model was validated against complete full-wave data; results above it (open symbols in Fig. 7, asterisks in Supplementary Table 6) indicate trends that require full-wave inversion (Supplementary Note 3).

### Systematic floor, detector load and shaped illumination

The floor η represents errors that do not decrease with more photons, such as fixed-pattern noise, drift and imperfect knowledge of $B$, expressed relative to the detected background in each pixel; it was set to $10^{-4}$ and $10^{-5}$. Detector-limited times follow from the brightest background pixel ($|B|^2 = 4.13$), a well depth of $3 \times 10^4$ electrons and a 4.2-megapixel camera at 100 frames per second, and illumination-limited times from $N/\Gamma$ with $\Gamma = 6.29 \times 10^{12}$ s$^{-1}$ per $\lambda_0^2$ at 1 kW cm$^{-2}$. For shaped illumination, the $K$ leading eigenvectors of the coupling matrix $Q_{mm'} = \mathrm{tr}(U^{\mathrm{T}} G_m^{\dagger} G_{m'} U)$ give $K$ wavefronts that share the photon budget of the 23 plane waves used in turn ($K$ = 1, 2, 4, 8 and 23). Expressions are given in Supplementary Note 5.

### Simple example: half-disk and half-sphere

The example of Supplementary Note 1 was computed with separate frequency-domain finite-difference solvers written for it: a 2D solver for transverse-electric and transverse-magnetic polarization and a 3D vector solver for bodies of revolution, solved order by order in $l$. Glass ($n = 1.6$, radius $\lambda_0$) stands in air above a layer of electric current sources $0.05\lambda_0$ below it, sampled every $0.1\lambda_0$; the detector is all light radiated into air, expanded in cylindrical or vector spherical multipoles; and channels are the singular values of the response table restricted to source patterns with at least 99.9% (or 99.99%) of their spectral weight beyond $nk_0$, counted within 30 dB of the strongest channel of the system. Because source, detector and normalization differ from those of the channel operator above, these counts are compared with one another and not with the counts of the main calculations (Supplementary Tables 3 and 4).

### Literature comparison

Results were included when lens material, surrounding medium, illumination wavelength and type of reported length could be identified; 14 were taken from the full text of the primary paper and 5 from the primary paper together with a published compilation[47]. Lengths were normalized by the stated free-space wavelength, or the stated spectral centre for broadband sources, and particle indices are recorded as stated (Supplementary Table 1). The sample is a selected comparison, not a systematic survey.

### Use of AI tools

Large language model assistants were used to help write and check code and to cross-check calculations. The author designed the study, checked the code and results, and takes full responsibility for the content.

## Code availability

All calculations, figures and quoted numbers were produced with the Python package nanolens-transfer, which contains the exact multipole models for spheres and cylinders, the 2D and 3D full-wave models, the complete-imaging model, intensity detection, image robustness, the resolution study and the photon-budget analysis, together with all computed data. The package is deposited in Zenodo (https://doi.org/10.5281/zenodo.24252657); its files are under restricted access during peer review and will be made openly available on publication.

## Data availability

Source data are provided with this paper. The computed data underlying Figs. 2–7, Table 1, Supplementary Figs. 1–13 and Supplementary Tables 2–4 and 6 are deposited in Zenodo together with the code (https://doi.org/10.5281/zenodo.24252657); the literature compilation is provided in Supplementary Table 1.

## Acknowledgements

K.S.K. thanks the support from the Basic Science Research Program through the National Research Foundation of Korea, funded by the Korean government (MSIT) (No. RS-2024-00346451).

## Author contributions

K.S.K. conceived the study, performed the analysis and wrote the paper.

## Conflict of interest

The author declares no conflict of interest.

# Supplementary Information

## Contact-driven super-resolution with dielectric microlenses at practical photon budgets

Kwang S. Kim (kimks@unist.ac.kr)

**Contents.** Supplementary Figs. 1–13; Supplementary Tables 1–6 (Supplementary Table 1 with its references S1–S16); Supplementary Note 1 (a simple example: half-disk and half-sphere contacts); Supplementary Note 2 (a counting rule for contacts in three dimensions); Supplementary Note 3 (resolution study, reconstruction and photon thresholds); Supplementary Note 4 (Fisher information of the detection schemes); Supplementary Note 5 (systematic floor, detector load and shaped illumination); Supplementary Note 6 (robustness of the simulated images); Supplementary Note 7 (code, data and numerical checks); Supplementary Note 8 (fundamental questions and answers). Reference numbers without S refer to the main text.

## Supplementary Figures

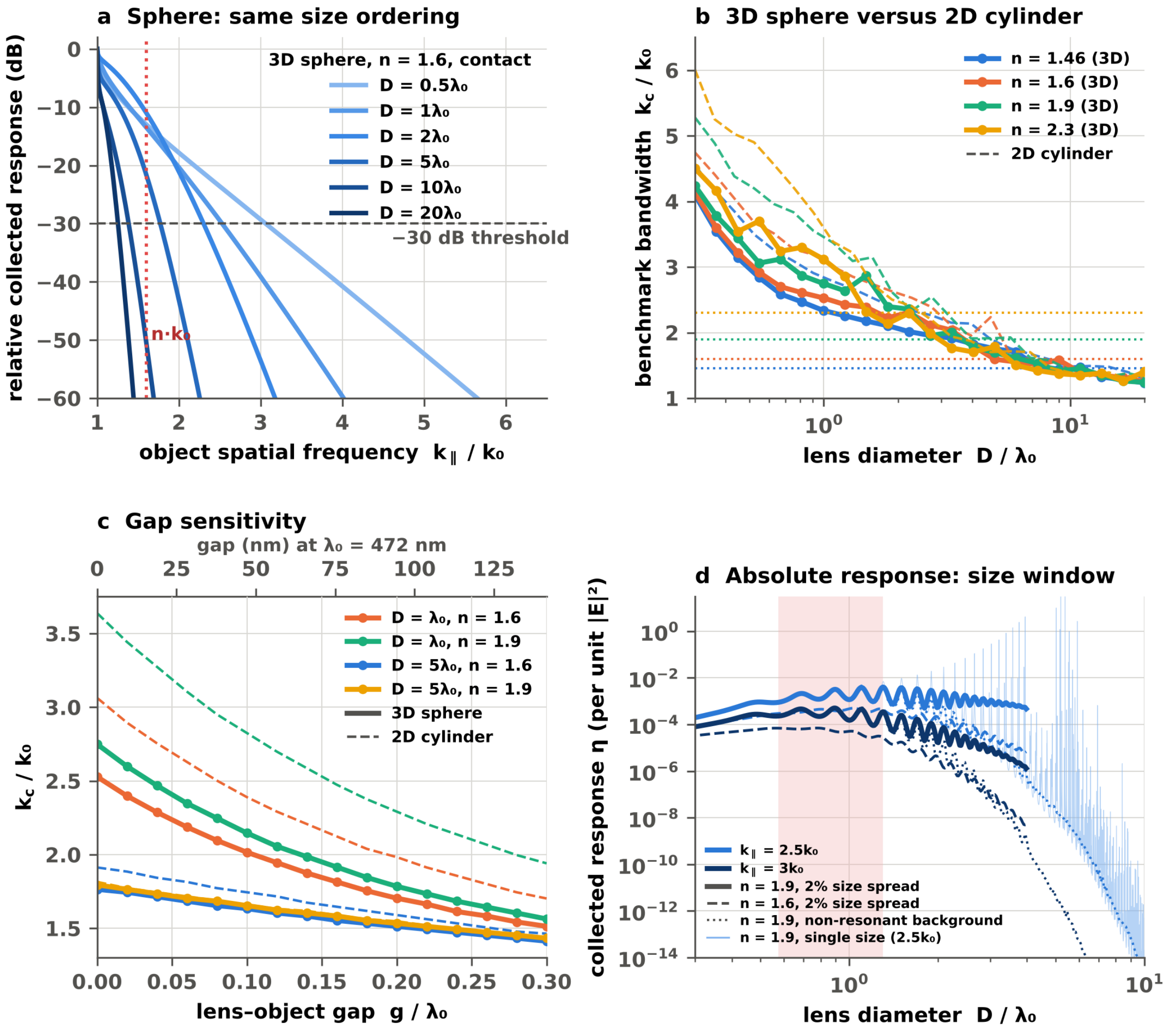

**Supplementary Fig. 1 | Transfer of single object components by spheres.** Exact vector Mie theory; air, NA 0.9; s and p polarizations each normalized to unit object-plane field and averaged. **a**, Relative collected response of one object component versus $k_{\parallel}$ for spheres of index 1.6 in contact; dotted line, $nk_0$; dashed line, −30 dB. **b**, Bandwidth $k_c$ (largest $k_{\parallel}$ within 30 dB of the response at the light line) versus diameter for four indices (solid, sphere) and for cylinders (dashed); dotted lines, $nk_0$. **c**, $k_c$ versus gap for spheres one and five wavelengths across (solid, sphere; dashed, cylinder); upper axis, gap in nanometres at $\lambda_0$ = 472 nm. **d**, Collected response at $k_{\parallel} = 2.5k_0$ and $3k_0$ versus diameter: bold lines, average over a Gaussian 2% spread of diameters (solid, $n$ = 1.9; dashed, $n$ = 1.6); dotted, non-resonant background ($n$ = 1.9); thin line, single-size spheres ($n$ = 1.9, $2.5k_0$) sampled down to $10^{-6}\lambda_0$, showing that above a few wavelengths the response beyond $nk_0$ survives only in narrow resonances, whose weight in practice is reduced by finite spectral bandwidth, size dispersion and absorption.

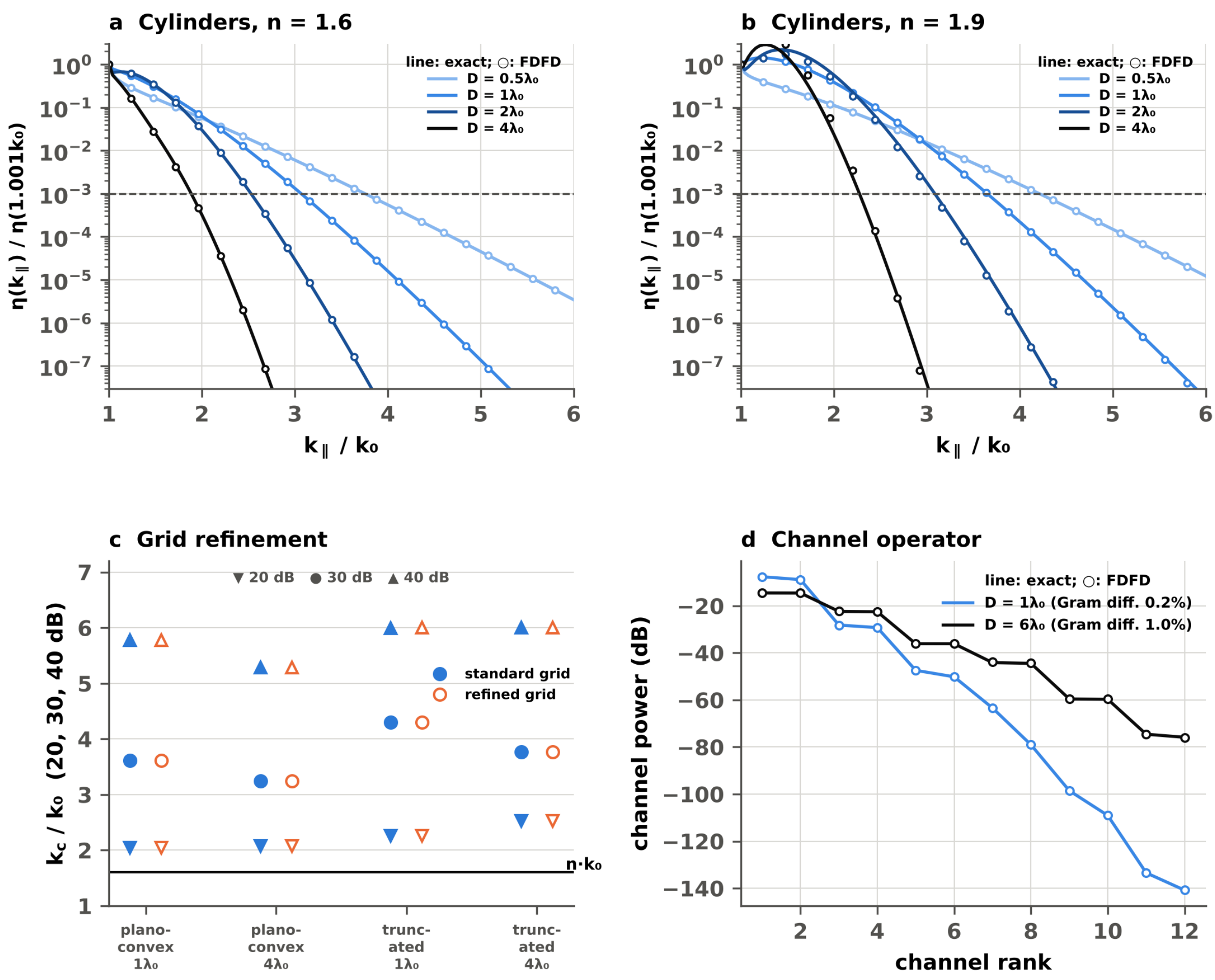

**Supplementary Fig. 2 | Validation of the full-wave model. a,b**, Relative collected response of full cylinders in air from the finite-difference frequency-domain model (circles) and the exact multipole solution (lines), for $n$ = 1.6 (a) and 1.9 (b) and $D$ = 0.5–4$\lambda_0$; dashed line, −30 dB. **c**, Bandwidth at 20, 30 and 40 dB for plano-convex and truncated lenses ($n$ = 1.6) on the standard grid (filled) and on a grid refined by 25–33% (open). **d**, Channel spectra of cylinders ($n$ = 1.6, $D = \lambda_0$ and $6\lambda_0$, both bands, line objects with $k_\parallel$ of either sign) from the full-wave operator (circles) and the exact operator (lines); their Gram matrices agree within 0.2% and 1%. In **a**,**b** responses below $3 \times 10^{-8}$, where the finite grid limits the accuracy of the finite-difference model, are not shown.

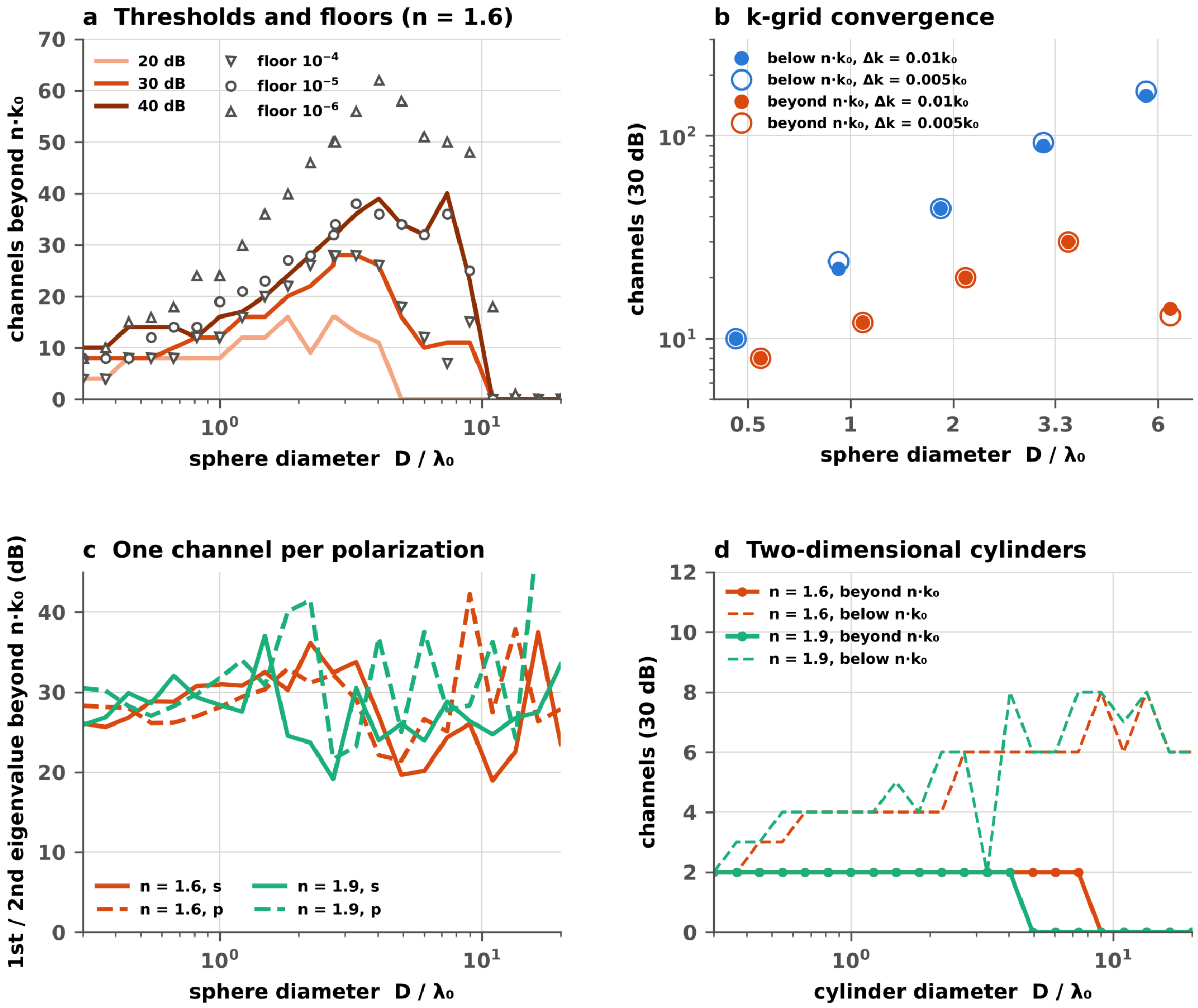

**Supplementary Fig. 3 | Robustness of the channel counts.** **a**, Channels beyond $nk_0$ for spheres of index 1.6 counted at 20, 30 and 40 dB below $\lambda_{max}$ (lines; $10^{-2}$, $10^{-3}$ and $10^{-4}$ in power) and above absolute floors of $10^{-4}$, $10^{-5}$ and $10^{-6}$ of the object spectral energy (symbols). **b**, Counts for $k$-grid steps of $0.01k_0$ (filled) and $0.005k_0$ (open), both bands, $n = 1.6$. **c**, Ratio of the first to the second eigenvalue beyond $nk_0$ for line objects, s (solid) and p (dashed) polarization, $n = 1.6$ and 1.9: inside the capacity window (up to the diameter of largest capacity) one channel per polarization dominates by 25–40 dB. **d**, Exact channel counts of 2D cylinders ($n = 1.6$ and 1.9, line objects with $k_\parallel$ of either sign) below and beyond $nk_0$ versus diameter.

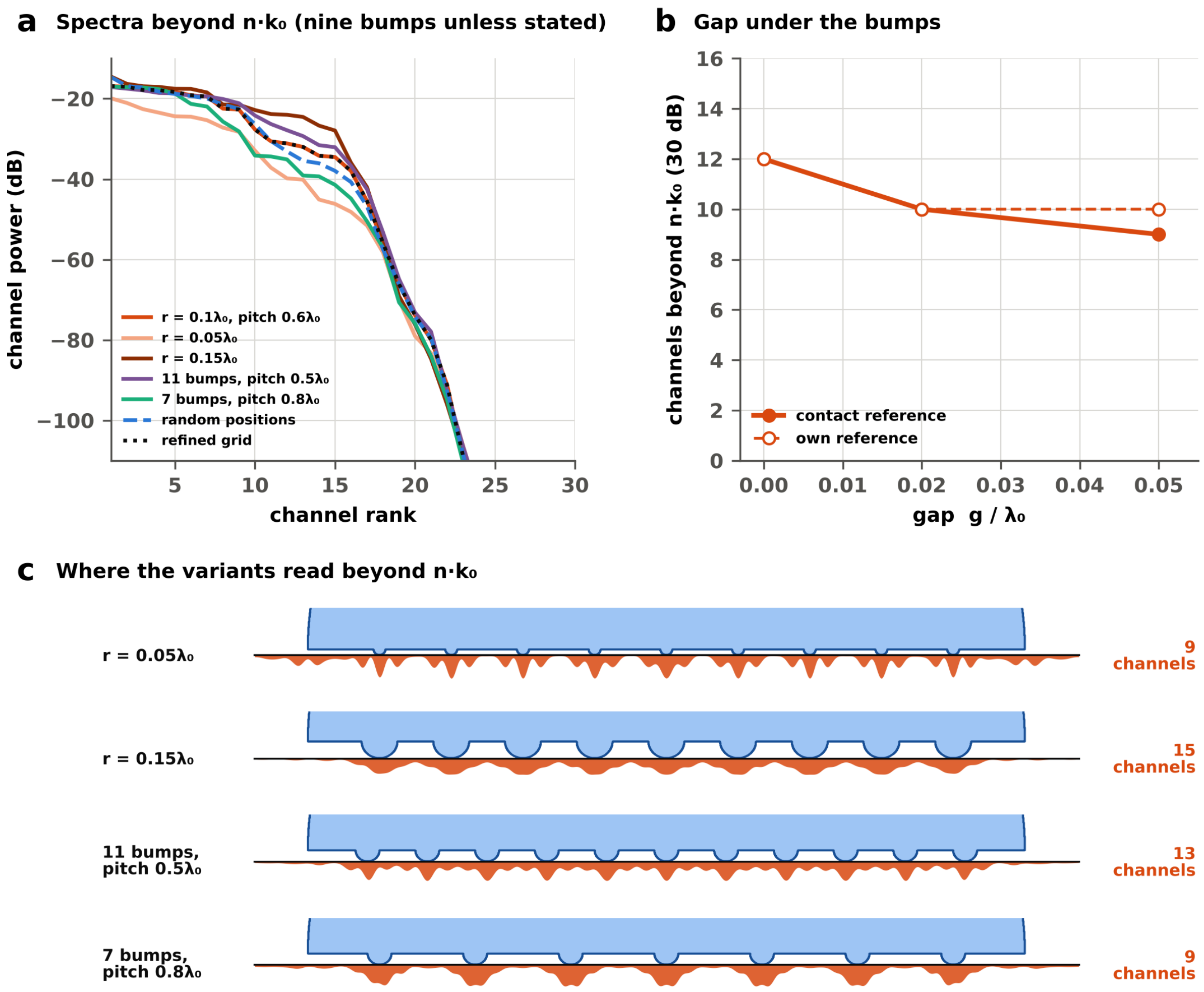

**Supplementary Fig. 4 | Textured lenses: variants, gap and convergence.** Full-wave calculation as in Fig. 5. **a**, Channel spectra beyond $nk_0$ for nine bumps of radius 0.05, 0.1 and 0.15$\lambda_0$, for 11 bumps at a pitch of 0.5$\lambda_0$ and 7 at 0.8$\lambda_0$, for randomly displaced bumps and for the standard lens on a grid refined by 25%. **b**, Channels beyond $nk_0$ versus a uniform gap under nine bumps, counted against the brightest channel in contact (filled) or of the same lens (open). **c**, Object-plane density of the channels beyond $nk_0$ for four variants; right, number of channels.

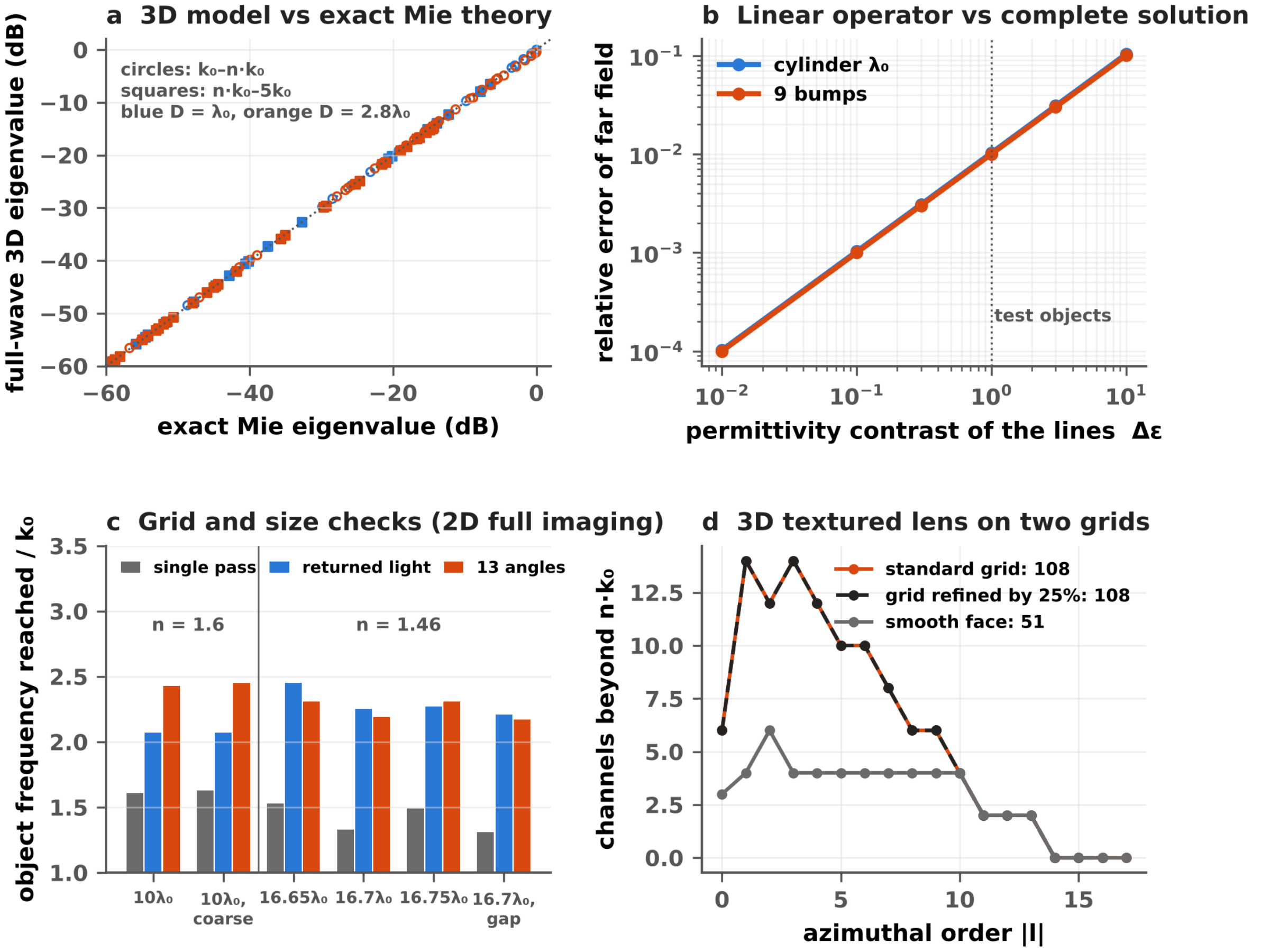

**Supplementary Fig. 5 | Checks of the 3D and complete-imaging models.** **a**, Eigenvalues of the channel operator of spheres ($n$ = 1.6; $D = \lambda_0$, blue; $2.8\lambda_0$, red) in each azimuthal order from the full-wave 3D model versus exact Mie theory, relative to $\lambda_{max}$ (open circles, immersion band; filled squares, beyond $nk_0$). **b**, Relative error of the far fields predicted by the linear imaging operator, against complete full-wave solutions with the object included, versus the permittivity contrast of the line-pair object, for a cylinder $\lambda_0$ across and for the nine-bump lens. **c**, Largest object frequency read (30 dB) for a $10\lambda_0$ cylinder ($n$ = 1.6) on the standard and the coarse grid, and for silica cylinders 16.65–16.75$\lambda_0$ across in contact and $16.7\lambda_0$ across at a gap of $0.033\lambda_0$ (20 nm at 600 nm): single pass under normal illumination (grey), with returned light (blue) and with 13 angles (red). **d**, Channels beyond $nk_0$ per azimuthal order for the 3D smooth and textured lenses of Fig. 6a on the standard grid and on a grid refined by 25%.

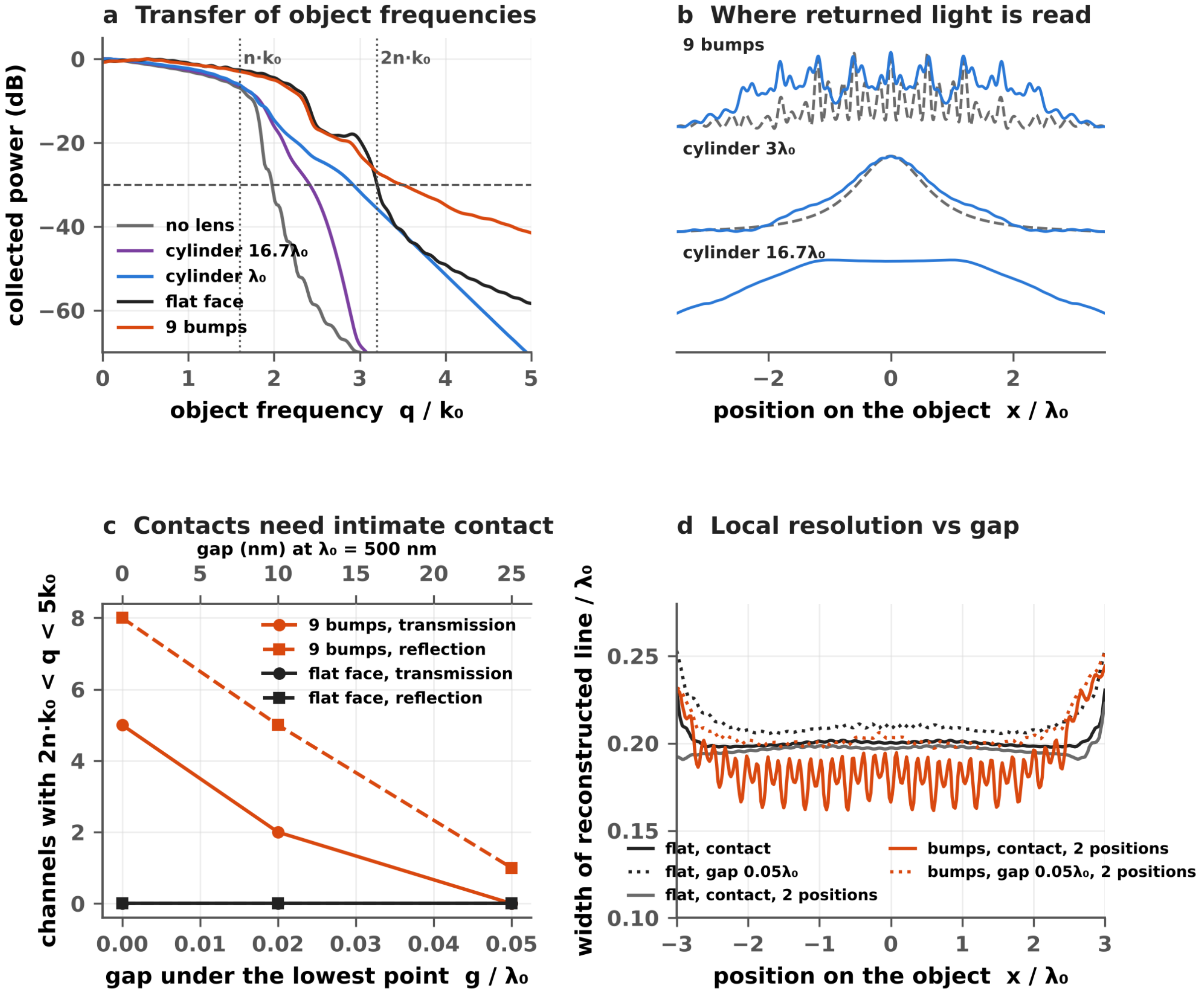

**Supplementary Fig. 6 | Complete imaging in detail.** 2D model, $n$ = 1.6, object on glass, 13 transmission angles unless stated. **a**, Collected power per object frequency $q$, summed over illuminations and detection directions, relative to the strongest propagating component (the transfer-function support recommended in ref. 29); dashed line, −30 dB. **b**, Object-plane density of the channels beyond $nk_0$ under normal illumination, single pass (dashed) and with returned light (solid): for large lenses the returned light reads a wide region around the contact. **c**, Channels with object frequencies between $2nk_0$ and $5k_0$ versus the gap under the lowest point of flat-faced and nine-bump lenses $6\lambda_0$ wide, under transmission (circles) and reflection (squares) illumination; upper axis, gap in nanometres at $\lambda_0$ = 500 nm. **d**, Width of a reconstructed line versus position for flat and nine-bump faces in contact and at a gap of $0.05\lambda_0$.

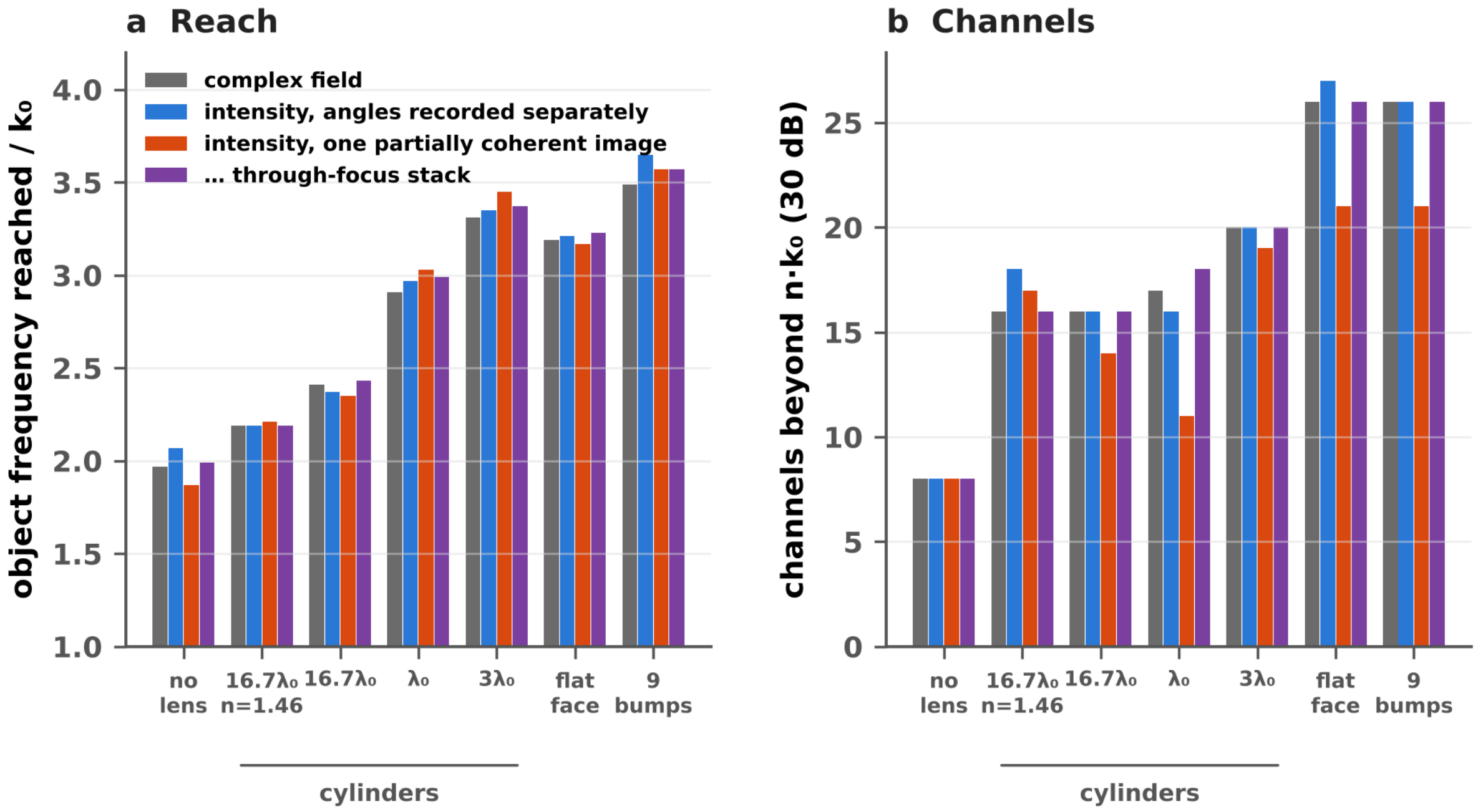

**Supplementary Fig. 7 | Intensity detection.** Complete 2D imaging model (object on glass, 13 transmission angles within NA 0.9, weak lossless object; lenses of index 1.6 unless stated). **a**, Largest object frequency read within 30 dB, and **b**, channels beyond $nk_0$ (30 dB), for complex-field detection (grey), for bright-field intensity images in the object plane recorded for each illumination separately (blue) or added in one partially coherent (Köhler) image (red), and for a through-focus stack of partially coherent images from $30\lambda_0$ below to $2\lambda_0$ above the object plane (purple). Channels counted against $\lambda_{max}$ of the same lens and detection scheme. The directly transmitted light acts as the reference that makes the intensity images linear in the object.

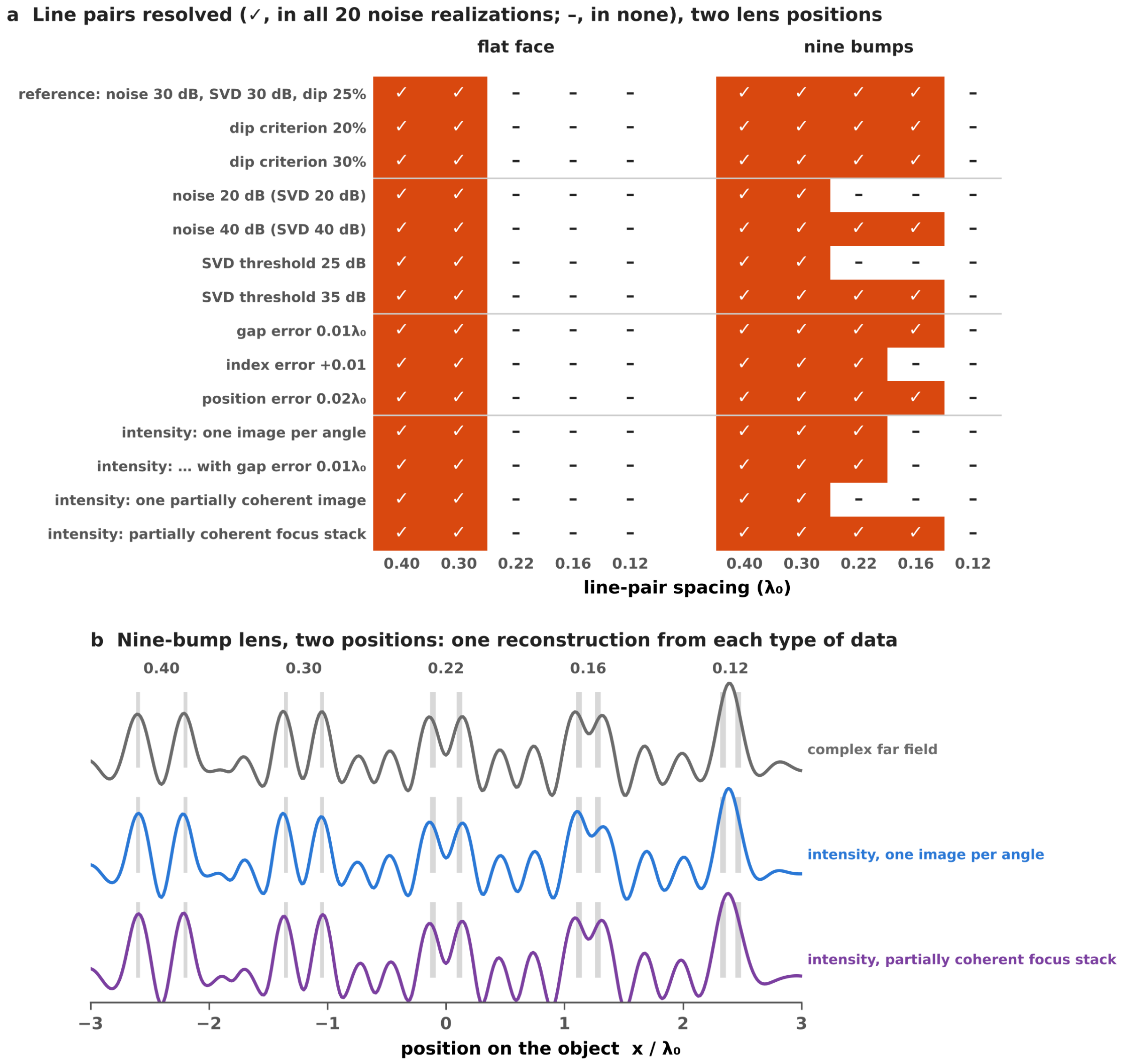

**Supplementary Fig. 8 | Robustness of the simulated images and reconstruction from intensity alone.** 2D complete-imaging model; object of Fig. 6e (five pairs of lines $0.04\lambda_0$ wide, $\Delta\varepsilon = 1$) on glass; 13 transmission angles within NA 0.9; flat-faced and nine-bump lenses $6\lambda_0$ wide ($n = 1.6$) read at two positions $0.3\lambda_0$ apart. **a**, Pairs resolved (two dominant peaks with a dip of at least 25% between them unless stated; Methods) in 20 noise realizations: ✓, in all; –, in none; numbers, fraction. Rows: reference reconstruction (complex far fields, noise 30 dB below the signal, truncated singular-value inversion at 30 dB); other dip criteria; other noise levels and truncation thresholds; data from a lens with a gap of $0.01\lambda_0$, an index of 1.61 or a second position displaced by $0.32\lambda_0$, inverted with the operator of the nominal lens; and bright-field intensity data alone (noise 30 dB below the intensity signal), recorded as one image per illumination angle in the object plane, one partially coherent image per position, or a stack of 14 partially coherent images from $30\lambda_0$ below to $2\lambda_0$ above the object plane. **b**, One reconstruction of the nine-bump lens from each type of data (grey bars, true lines; spacings in $\lambda_0$ above each pair).

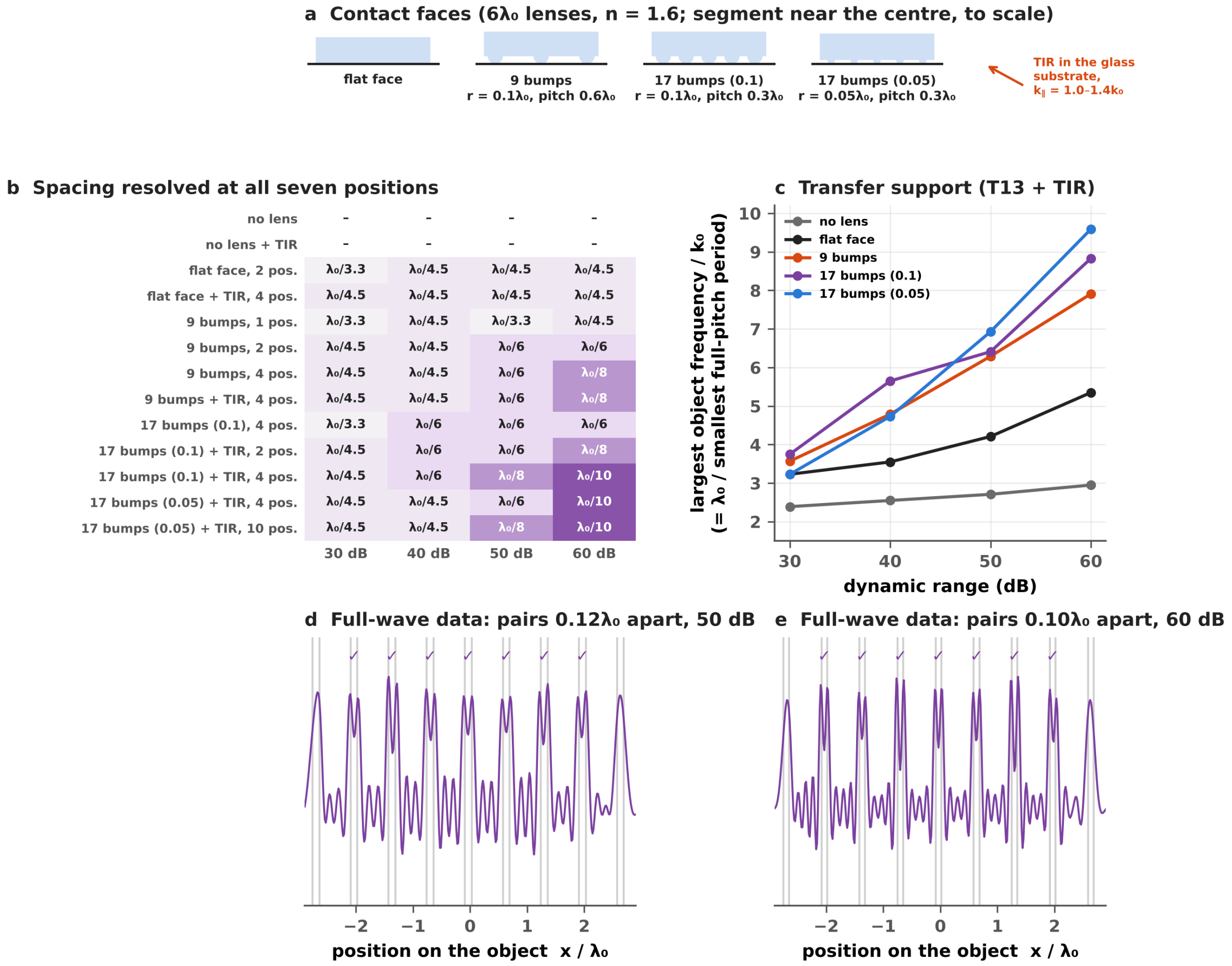

**Supplementary Fig. 9 | How fine contacts resolve across a field.** 2D complete-imaging model; lenses 6$\lambda_0$ wide ($n$ = 1.6) on glass. **a**, Contact faces near the centre (to scale): flat; nine bumps (radius 0.1$\lambda_0$, pitch 0.6$\lambda_0$); 17 bumps of radius 0.1$\lambda_0$ or 0.05$\lambda_0$ at a pitch of 0.3$\lambda_0$. Evanescent illumination by total internal reflection (TIR) in the glass, in-plane wavenumbers 1.0–1.4$k_0$, adds to the 13 transmission angles within NA 0.9. **b**, Finest spacing of line pairs resolved at all seven positions across the contact array, in every one of 20 noise realizations, for lens, illumination, number of lens positions and dynamic range; –, no spacing of 0.30$\lambda_0$ or less. **c**, Largest object frequency transferred within the stated dynamic range (13 transmission angles and TIR), the inverse of the smallest full-pitch period. **d**,**e**, One reconstruction from complete full-wave data (all multiple scattering) for the 17-bump face of radius 0.1$\lambda_0$ with TIR and four positions: pairs 0.12$\lambda_0$ apart at 50 dB (d) and 0.10$\lambda_0$ apart at 60 dB (e); grey lines, true lines; ✓, pair resolved in all 20 realizations. For this face, complete full-wave data give the same finest spacings as in **b** at 30–60 dB (spacings 0.30–0.08$\lambda_0$ tested; Methods).

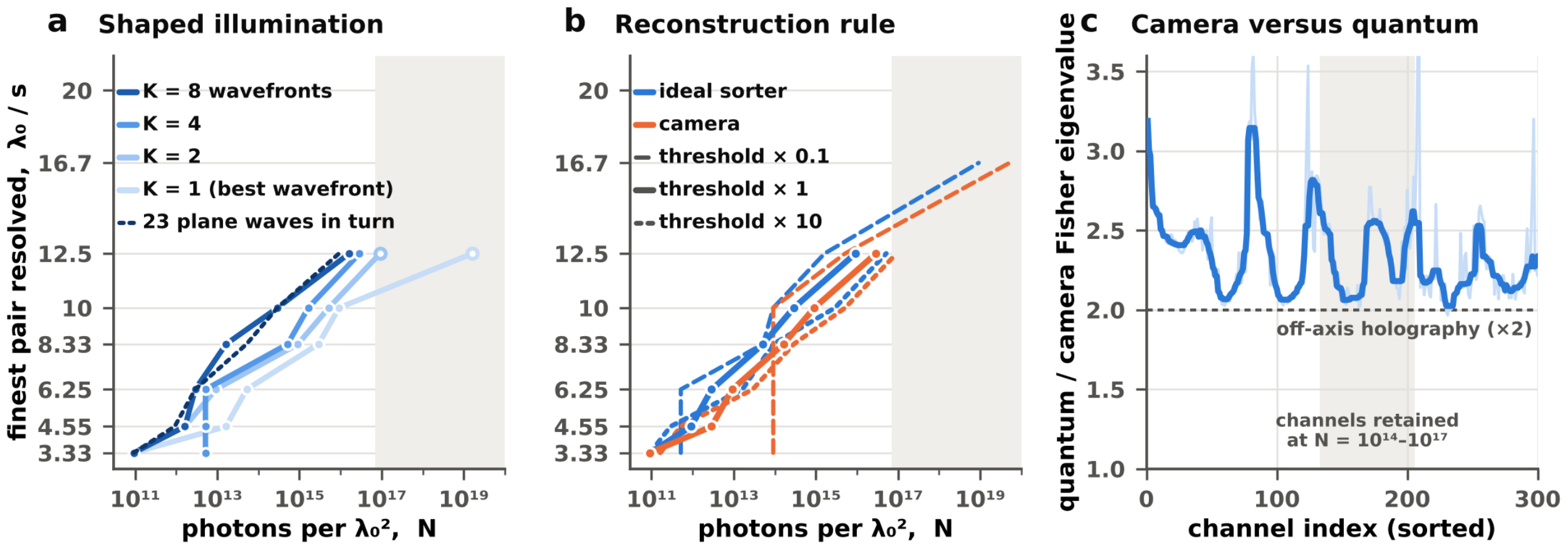

**Supplementary Fig. 10 | Robustness of the photon prices.** Lens with 17 contacts of radius $0.1\lambda_0$ read at four positions under 13 transmitted and 10 evanescent illuminations, as in Fig. 7; finest line-pair spacing resolved at all seven positions in all 20 noise realizations versus the photon budget *N*; grey regions and open symbols as in Fig. 7. **a**, Shaped illumination at the quantum bound: the *K* wavefronts that couple most strongly to fine detail ($K = 1$, 2, 4 and 8), sharing the photon budget of the 23 plane waves used in turn (dotted). The best single wavefront couples 10.9 times more fine-detail power than an average plane wave but needs about thirty times more photons for $\lambda_0/10$; $K = 23$ shaped wavefronts reproduce the plane-wave thresholds exactly. **b**, Retention threshold of the reconstruction scaled by 0.1 (dashed), 1 (solid) and 10 (dotted) for the ideal sorter (blue) and the camera (orange): the photon prices of $\lambda_0/10$ and $\lambda_0/12.5$ move by up to an order of magnitude, while at these steps the camera stays within a factor of 1–3.3 of the quantum bound; the lowest threshold fails at coarse steps for low budgets. **c**, Ratio of the sorted Fisher eigenvalues of the quantum bound and of the camera (thin line; thick line, running median over 11 channels); dashed line, off-axis holography (factor 2); grey band, channels retained at $N = 10^{14}$–$10^{17}$.

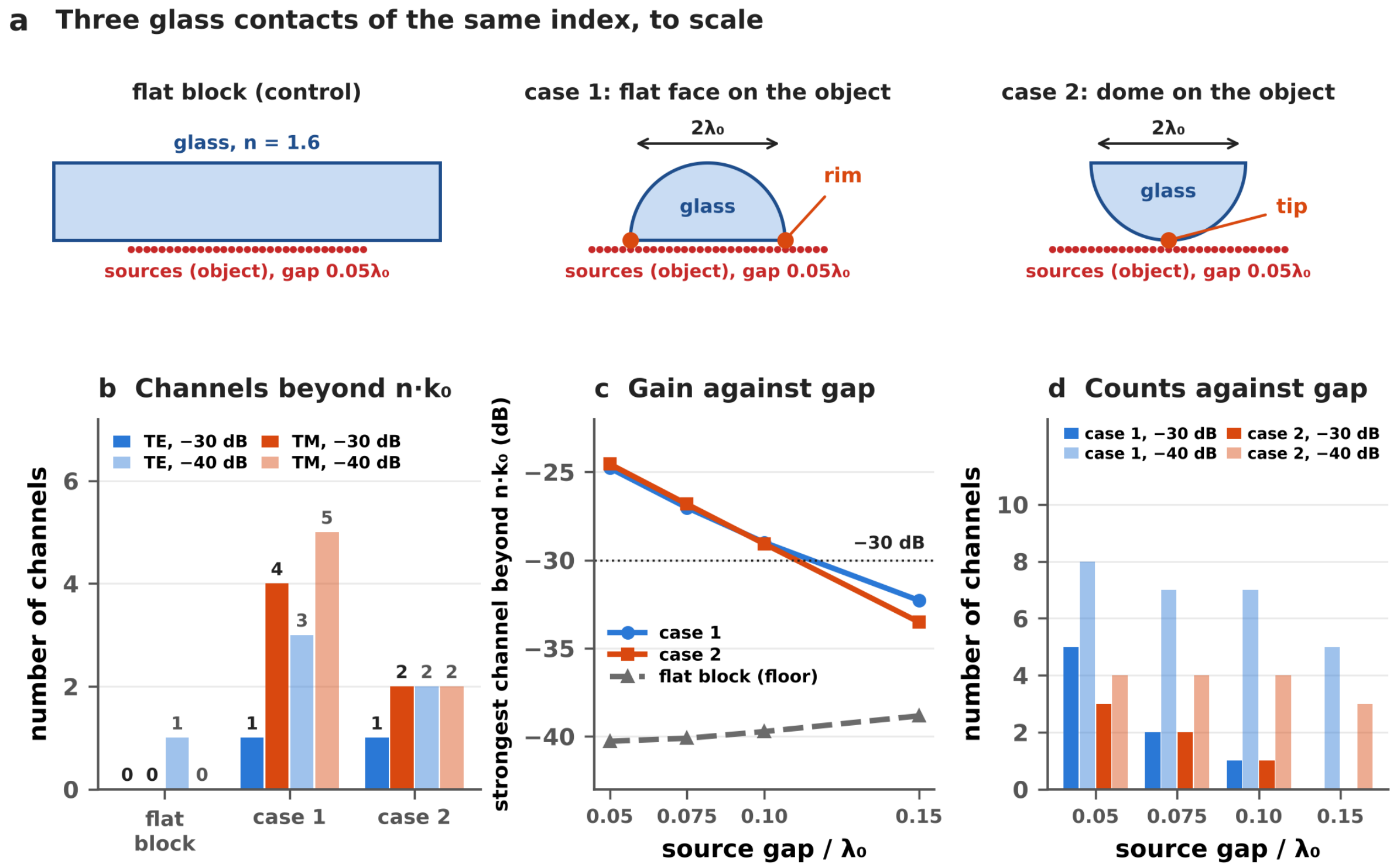

**Supplementary Fig. 11 | A simple example in 2D: three pieces of glass on a row of sources. a**, The three systems, to scale: a flat block (control), a half-disk of radius $\lambda_0$ resting on its flat face (case 1) and the same half-disk turned over, so that only its dome touches (case 2). Glass, $n$ = 1.6, in air; red dots, the 31 sources of the object, $0.05\lambda_0$ below the glass; orange dots, where the glass leaves the source plane. **b**, Channels beyond $nk_0$ above −30 dB (dark) and −40 dB (light) for sources polarized along the edge (transverse electric, TE) and in the plane (transverse magnetic, TM), each counted against the strongest channel of its polarization; the single TE channel of the flat block at −40 dB lies at the leakage floor. **c**, Strongest channel beyond $nk_0$ against the gap between sources and glass (TE and TM together); dashed, leakage floor of the flat block; dotted, −30 dB. **d**, Channels above −30 dB (dark) and −40 dB (light) against the gap. Frequency-domain finite differences; all light radiated into air is detected; object patterns with at least 99.9% of their spectral weight beyond $nk_0$; values in Supplementary Table 3.

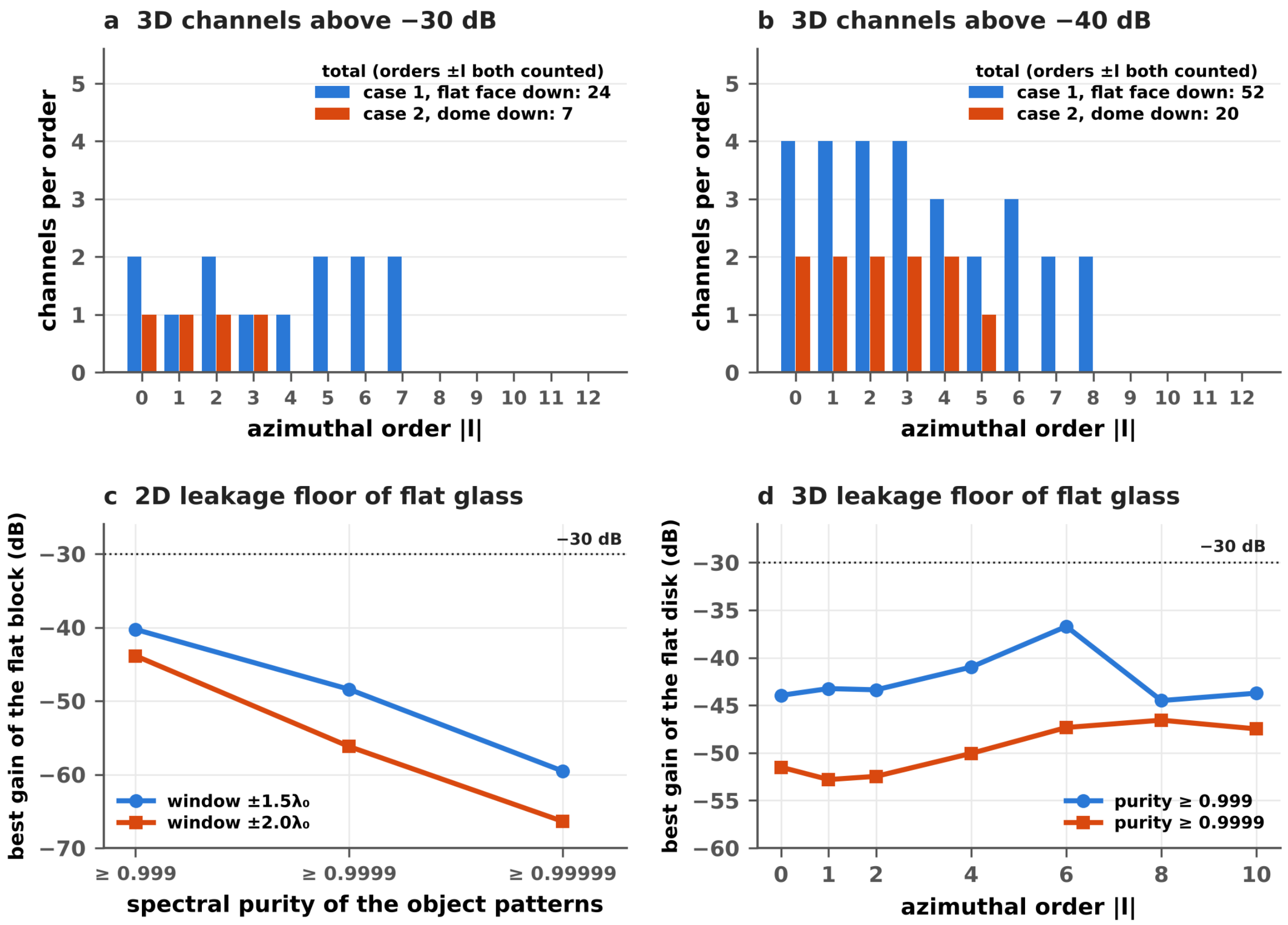

**Supplementary Fig. 12 | The simple example in 3D and its leakage floor.** **a**,**b**, Channels beyond $nk_0$ per azimuthal order $|l|$ for a half-sphere of radius $\lambda_0$ ($n$ = 1.6) resting on its flat face (case 1, blue) and turned over (case 2, orange), above −30 dB (**a**) and −40 dB (**b**). Source rings out to $1.95\lambda_0$, $0.05\lambda_0$ below the glass, three source orientations; object patterns with at least 99.99% of their spectral weight beyond $nk_0$; the totals in the legends count orders $l$ and $-l$. **c**, Strongest spurious channel beyond $nk_0$ of the flat block in 2D against the spectral purity required of the object patterns, for two source windows. **d**, The same for a flat disk in 3D, per azimuthal order (rings to $1.95\lambda_0$). Dotted lines, −30 dB: the counts of **a**,**b** lie well clear of the leakage. Values in Supplementary Table 4.

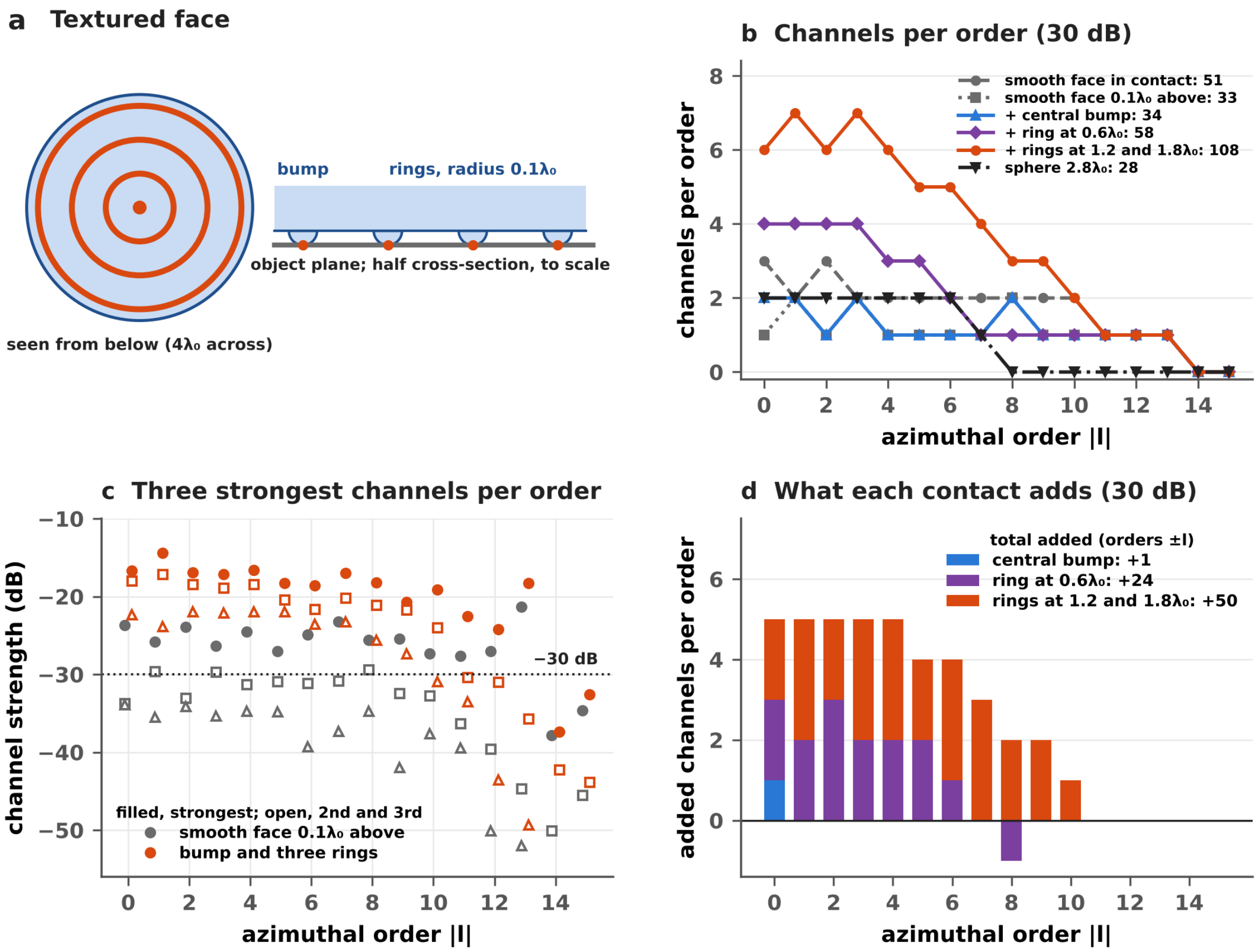

**Supplementary Fig. 13 | Channel accounting for the 3D textured face.** 3D full-wave model, single pass, NA 0.9; lens $4\lambda_0$ across, $n = 1.6$. **a**, The textured face seen from below and in half cross-section, to scale: a central half-sphere bump and three half-round rings (each half of a torus) of radius $0.1\lambda_0$, at radii of 0.6, 1.2 and $1.8\lambda_0$, touch the object; the flat face between them stays $0.1\lambda_0$ above it. **b**, Channels beyond $nk_0$ per azimuthal order (one sign of $l$, both polarizations; 30 dB) as contacts are added, and for the best smooth sphere; the totals in the legend count both signs. **c**, The three strongest channels beyond $nk_0$ in each order, relative to the brightest channel of the lens, for the smooth face $0.1\lambda_0$ above the object (grey) and for the textured face (orange); dotted line, −30 dB. **d**, Channels added per order by each contact, from the difference of counts. Values in Supplementary Table 2.

## Supplementary Table 1 | Literature compilation

Type (metric class): P, resolved period or pitch, as stated by the source (ref. 25 of the main text recommends the full period); G, smallest visible gap, slit or feature; T, PSF, line-spread, convolution or two-point-criterion width. These classes measure different properties of an image and are not compared with one another. λ/X is the value reported by the source when no length in nanometres was given; where a source reports a range, the range is listed. Provenance: "primary", taken from the full text of the original paper; "primary + compilation", from the original paper together with the tabulation of Zhong and Hong (ref. S16), which supplied the normalized length or a parameter not stated in the sources consulted. Particle indices are given as stated in the source: a single value where one is given (with its wavelength where stated), the stated range where only a range is given, and n.s. where none is stated. No index is assigned or averaged. For entry 1, the diameter of the lens that rendered the 220 nm pitch visible was measured on the scanning electron micrograph of Fig. 2f of ref. S1: 1.4 µm by 1.2 µm along the two axes (≈1.3 µm), using the 1 µm scale bar, which reproduces the 220 nm stripe period of the same image to within a few per cent; that lens is plano-convex. The last column gives, for orientation, a model estimate of how many channels a sphere of the same size and relative index passes beyond the immersion limit in a single pass (see the final note below the table); it is not a reanalysis of the experiment and does not bound what the experiment could resolve.

| # | Year | Lens | $n_p$ (as stated) | Medium | $D$ (µm) | $\lambda_0$ (nm) | Type | Value | Source | Single-pass channels beyond $n_{rel}k_0$ (model estimate) |
|---|---|---|---|---|---|---|---|---|---|---|
| 1 | 2009 | CHQ self-assembled plano-convex lens | 1.5 | air | 1.3 | 472 | P | 220 nm | S1 (primary) | 28† |
| 2 | 2011 | $SiO_2$ microsphere | 1.46 | air | 4.74 | 600 | G | 50 nm | S2 (primary) | 28 |
| 3 | 2011 | $SiO_2$ microsphere | 1.46 | air | 4.74 | 600 | P | 300 nm | S2 (primary) | 28 |
| 4 | 2012 | $BaTiO_3$ glass microsphere, fully immersed in IPA | ~1.9 | IPA | 4.2 | 550 | T | 75 nm | S3 (primary) | 40 |
| 5 | 2013 | $BaTiO_3$ microsphere | 1.9 | water | 100 | 550 | G | 75 nm | S4 (primary) | 0 |
| 6 | 2014 | fused-silica microsphere + confocal scanning | 1.47 (408 nm) | air | 5 | 408 | G | 25 nm | S5 (primary) | 5 |
| 7 | 2014 | $BaTiO_3$ glass microsphere (~13 µm) embedded in a PDMS film | ~1.9–2.2 | PDMS | 13 | 405 | T | λ/7 | S6 (primary) | 0 |
| 8 | 2016 | $BaTiO_3$ glass microsphere | 1.92 | water | 6 | 550 | T | λ/4–λ/7 | S7 (primary + compilation) | 1 |
| 9 | 2016 | $BaTiO_3$ glass microsphere (scanning superlens) | n.s. | water | 57 | 550 | T | λ/6.3–λ/8.4 | S8 (primary + compilation) | 0 |
| 10 | 2017 | melamine-formaldehyde microsphere + Mirau | 1.68 | air | 11 | 600 | G | 112 nm | S9 (primary) | 0 |
| 11 | 2017 | melamine-formaldehyde microsphere + Mirau | 1.68 | air | 11 | 600 | P | 323 nm | S9 (primary) | 0 |
| 12 | 2021 | $BaTiO_3$ glass two-sphere compound lens (30 µm bottom, 80 µm upper sphere) in oil | n.s. | oil | 30 | n.s. | G | λ/5.4 | S10 (primary + compilation) | 0 |
| 13 | 2021 | patchy $BaTiO_3$ glass microsphere | n.s. | air | 31 | 550 | G | 100 nm | S11 (primary) | 0 |
| 14 | 2022 | bilayer-coated $BaTiO_3$ glass microsphere | 1.9 | PDMS | 50 | 420 | G | λ/4.2 | S12 (primary + compilation) | 0 |
| 15 | 2023 | FIB-milled $BaTiO_3$ glass hyper-hemisphere | ~2.34 (420 nm) | air | 23 | 420 | G | 48 nm | S13 (primary) | flat contact face |

| # | Year | Lens | $n_p$ (as stated) | Medium | *D* (µm) | $\lambda_0$ (nm) | Type | Value | Source | Single-pass channels beyond $n_{rel}k_0$ (model estimate) |
|---|---|---|---|---|---|---|---|---|---|---|
| 16 | 2024 | Al-coated Janus $BaTiO_3$ glass microsphere on a PDMS support | ~1.90 | air | 30 | 550 | G | 100 nm | S14 (primary + compilation) | 0 |
| 17 | 2025 | microsphere on etched fibre (OCT) | – | air | – | 840 | T | 247 nm | S15 (primary) | – |
| 18 | 2026 | soda-lime glass microsphere (pristine) | n.s. | air | 25 | 450 | T | 112 nm | S16 (primary) | 0 |
| 19 | 2026 | phase-ring soda-lime glass microsphere | n.s. | air | 25 | 450 | T | 79 nm | S16 (primary) | 0 |

Entry 8: reported range λ/4-λ/7 (1D convolution); geometric midpoint plotted.

Entry 9: reported range λ/6.3-λ/8.4 (2D convolution); geometric midpoint plotted.

Entry 12: 76 nm gap from primary (scanned and stitched, non-contact); wavelength not stated in the sources consulted; normalized value λ/5.4 from the compilation.

Entry 13: 27-35 µm spheres (31 µm midpoint), air, 20x/0.4 objective, 100 nm Blu-ray gaps; index not stated (1.9 used in the paper's model).

Entry 14: 50 µm sphere fully embedded in PDMS; 420 nm; feature length λ/4.2 from the compilation.

Entry 16: imaging in air with the sphere held by PDMS; 100 nm gaps of 300 nm-period tracks.

Last column: number of single-pass channels beyond the immersion limit of the relative index $n_{rel} = n_p/n_m$ (30 dB, general two-dimensional objects) for a sphere of the same relative index and of diameter $Dn_m/\lambda_0$ in medium wavelengths, interpolated from the exact calculations of the main text (relative indices 1.35–2.3, diameters up to 20 medium wavelengths; larger spheres lie beyond every computed single-pass window and are entered as 0). The estimate assumes contact and a collection cone of sin θ ≤ 0.9 in the surrounding medium, which is generous for dry objectives viewing immersed lenses; † marks a plano-convex lens, estimated as a sphere; – marks an index that is not stated or lies outside the computed range. A zero does not mean that the experiment could not record detail beyond the immersion limit: light returned onto the object or focused through the lens, scanning and prior knowledge of the object extend the reach of large lenses to about $(n + 1)k_0$ (main text), and visible gaps and fitted widths (classes G and T) quantify the detectability of known structures rather than the number of independent channels.

## References for Supplementary Table 1

## Supplementary Table 2 | Channel accounting of the 3D textured face

| $\|l\|$ | **Smooth face in contact** | **Smooth face $0.1\lambda_0$ above** | **+ central bump** | **+ ring at $0.6\lambda_0$** | **+ rings at 1.2 and $1.8\lambda_0$** | **Best smooth sphere ($2.8\lambda_0$)** |
|---|---|---|---|---|---|---|
| Contacts | rim (circle of radius $2\lambda_0$) | rim, across the gap | adds a point on the axis | adds a circle, $\rho = 0.6\lambda_0$ | adds circles, $\rho$ = 1.2 and $1.8\lambda_0$ | one patch about $0.5\lambda_0$ wide |
| 0 | 3 | 1 | 2 | 4 | 6 | 2 |
| 1 | 2 | 2 | 2 | 4 | 7 | 2 |
| 2 | 3 | 1 | 1 | 4 | 6 | 2 |
| 3 | 2 | 2 | 2 | 4 | 7 | 2 |
| 4 | 2 | 1 | 1 | 3 | 6 | 2 |
| 5 | 2 | 1 | 1 | 3 | 5 | 2 |
| 6 | 2 | 1 | 1 | 2 | 5 | 2 |
| 7 | 2 | 1 | 1 | 1 | 4 | 1 |
| 8 | 2 | 2 | 2 | 1 | 3 | 0 |
| 9 | 2 | 1 | 1 | 1 | 3 | 0 |
| 10 | 2 | 1 | 1 | 1 | 2 | 0 |
| 11 | 1 | 1 | 1 | 1 | 1 | 0 |
| 12 | 1 | 1 | 1 | 1 | 1 | 0 |
| 13 | 1 | 1 | 1 | 1 | 1 | 0 |
| **Total, 30 dB** | **51** | **33** | **34** | **58** | **108** | **28** |
| Total, 40 dB | 103 | 83 | 86 | 107 | 167 | 32 |
| Highest $\|l\|$ read | 13 | 13 | 13 | 13 | 13 | 7 |

Channels beyond $nk_0$ in each azimuthal order (one sign of $l$, both polarizations) within 30 dB of the brightest channel of the same lens; 3D full-wave model, single pass, NA 0.9, lens $4\lambda_0$ across, $n = 1.6$ (sphere: exact Mie theory and the same model). Totals count $l = 0$ once and every $|l| \geq 1$ twice. No channel is counted at $|l| \geq 14$ ($|l| \geq 8$ for the sphere). Explanation in Supplementary Note 2.

## Supplementary Table 3 | The simple example in 2D

**a**

| System | $J_y$ (TE) | $J_x$ | $J_z$ | $J_x$ and $J_z$ (TM) | All three |
|---|---|---|---|---|---|
| Flat block (control) | 0 / 1† | 0 / 1† | 0 / 1† | 0 / 0 | 0 / 0 |
| Case 1, flat face on the object | 1 / 3 | 4 / 4 | 3 / 5 | 4 / 5 | 5 / 8 |
| Case 2, dome on the object | 1 / 2 | 2 / 2 | 2 / 2 | 2 / 2 | 3 / 4 |

**b**

| Gap ($\lambda_0$) | Case 1: gain (dB); counts | Case 2: gain (dB); counts | Flat block: strongest spurious channel (dB) |
|---|---|---|---|
| 0.050 | −24.8; 5 / 8 | −24.6; 3 / 4 | −40.3 |
| 0.075 | −27.0; 2 / 7 | −26.8; 2 / 4 | −40.1 |
| 0.100 | −29.0; 1 / 7 | −29.1; 1 / 4 | −39.7 |
| 0.150 | −32.3; 0 / 5 | −33.5; 0 / 3 | −38.8 |

Channels beyond $nk_0$ above −30 dB / −40 dB. Glass $n = 1.6$, radius $\lambda_0$; 31 sources at $|x| \leq 1.5\lambda_0$; object patterns with purity $\mu \geq 0.999$; all light radiated into air is detected. **a**, Gap $0.05\lambda_0$, grid $\lambda_0/60$; each column is counted against its own strongest channel, the last against the strongest channel of all source orientations at equal current. † At the leakage floor (−37 to −40 dB), not a channel of the glass. **b**, All source orientations together, grid $\lambda_0/40$: gain of the strongest channel beyond $nk_0$, and counts. The fifth channel of case 1 at a gap of $0.05\lambda_0$ lies on the threshold (−29.9 dB; −30.0 dB on the grid $\lambda_0/60$). Explanation in Supplementary Note 1.

## Supplementary Table 4 | The simple example in 3D

**a**

| $|l|$ | 0 | 1 | 2 | 3 | 4 | 5 | 6 | 7 | 8 | Total |
|---|---|---|---|---|---|---|---|---|---|---|
| Case 1, flat face on the object | 2 / 4 | 1 / 4 | 2 / 4 | 1 / 4 | 1 / 3 | 2 / 2 | 2 / 3 | 2 / 2 | 0 / 2 | 24 / 52 |
| Case 2, dome on the object | 1 / 2 | 1 / 2 | 1 / 2 | 1 / 2 | 0 / 2 | 0 / 1 | 0 / 0 | 0 / 0 | 0 / 0 | 7 / 20 |

**b**

| Source rings; purity | Case 1 | Case 2 | Flat disk: strongest spurious channel (dB) |
|---|---|---|---|
| to $1.45\lambda_0$; $\mu \geq 0.999$ | 26 / 49‡ | 5 / 18‡ | −36.1 |
| to $1.45\lambda_0$; $\mu \geq 0.9999$ | 17 / 37 | 3 / 14 | −46.2 |
| to $1.95\lambda_0$; $\mu \geq 0.999$ | 30 / 56‡ | 7 / 24‡ | −36.7 |
| to $1.95\lambda_0$; $\mu \geq 0.9999$ | 24 / 52 | 7 / 20 | −46.6 |

Channels beyond $nk_0$ above −30 dB / −40 dB for a half-sphere of radius $\lambda_0$ ($n = 1.6$); source rings $0.05\lambda_0$ below the glass, three source orientations; all light radiated into air is detected; grid $\lambda_0/40$. **a**, Per azimuthal order (one sign of $l$), rings to $1.95\lambda_0$, $\mu \geq 0.9999$; totals count $l = 0$ once and every $|l| \geq 1$ twice; no channel is counted at $|l| \geq 9$ (orders up to 14 were computed). **b**, Totals for two source windows and two purities, with the leakage floor measured on a flat disk (orders 0, 1, 2, 4, 6, 8 and 10). ‡ Counts at −40 dB lie near the leakage floor at this purity and are less reliable. Explanation in Supplementary Note 1.

## Supplementary Table 5 | Fisher information of the detection schemes

| Scheme | Fisher matrix | What limits it | Relative to $F_Q$ |
|---|---|---|---|
| Quantum limit | $F_Q = 4\Phi \mathrm{d}x\ \Sigma\ \mathrm{Re}(G_{pm}^{\dagger} G_{pm})$ | distinguishability of the coherent states; reached only if Im $D = 0$ | 1 |
| Attainable (Holevo) optimum | cost within a factor $1 + R$ of the quantum bound; $R \geq 0.9996$ | incompatibility of the best measurements for different parts of the object | between 1/2 and 1 |
| Off-axis holography, heterodyne | $F_Q/2$ | both quadratures measured, with one unit of added noise | 1/2 |
| Bright-field camera, shot noise only | $F_{\mathrm{cam}} = 4\Phi \mathrm{d}x\ \Sigma\ r^{\mathrm{T}} r$, $r = \mathrm{Re}(e^{-i\varphi} g)$ | the quadrature out of phase with the lens background is lost | 1/2.6 to 1/2.1 on the channels that set resolution |
| Camera with systematic floor η | $\Sigma\ 4\Phi \mathrm{d}x(1 + \eta^2 n_b)^{-1} r^{\mathrm{T}} r$ | saturates at $4/(\eta^2\lvert B\rvert^2)$ per pixel | ceiling (Fig. 7c,d) |
| Camera after background rejection ε, transmission $\eta_s$ | $\Sigma\ 4\eta_s \Phi \mathrm{d}x(1 + \eta^2 \eta_s \varepsilon n_b)^{-1} r^{\mathrm{T}} r$ | effective floor $\eta\varepsilon^{1/2}$; needs $\varepsilon^{1/2}\lvert B\rvert > \lvert g\cdot c\rvert$ | ceiling raised by 1/ε; photons × $1/\eta_s$ |
| Static $K$-port sorter | $F_K = 4\Phi \mathrm{d}x\ \Sigma\ \mathrm{Re}[(P_K G_{pm})^{\dagger}(P_K G_{pm})]$ | only $K$ image-plane modes are detected | 24, 42, 64 and 88% of tr $F_Q$ for $K$ = 4, 8, 16, 32 |

Symbols as in Supplementary Notes 4 and 5: Φ, fluence per exposure; d$x$, pixel width; $G_{pm}$, image-plane operator of the object for lens position $p$ and illumination $m$, with rows $g$; $B$, background field, with phase φ; $n_b = \Phi \mathrm{d}x\lvert B\rvert^2$; $D = 4\Phi \mathrm{d}x\ \Sigma\ G_{pm}^{\dagger} G_{pm}$; $R$, incompatibility; $P_K$, projector onto the $K$ sorter modes. Sums run over pixels, illuminations and lens positions.

**Supplementary Table 6 | Photon price of field-wide resolution.** Photons per $\lambda_0^2$ summed over the 92 exposures, $N$, from which each line-pair spacing $s$ and all coarser ones stay resolved at all seven positions in all 20 noise realizations (lens $6\lambda_0$ wide, $n = 1.6$, with 17 contacts of radius $0.1\lambda_0$ unless stated).

| Detection or contact face | $s = 0.16\lambda_0$ | $s = 0.12\lambda_0$ | $s = 0.10\lambda_0$ | $s = 0.08\lambda_0$ | Finest within validated range |
|---|---|---|---|---|---|
| Ideal sorter (quantum bound) | $2.9 \times 10^{12}$ | $5.2 \times 10^{13}$ | $2.9 \times 10^{14}$ | $9.2 \times 10^{15}$ | $\lambda_0/12.5$ |
| Off-axis holography | $5.2 \times 10^{12}$ | $1.6 \times 10^{14}$ | $5.2 \times 10^{14}$ | $1.6 \times 10^{16}$ | $\lambda_0/12.5$ |
| Bright-field camera | $9.2 \times 10^{12}$ | $1.6 \times 10^{14}$ | $9.2 \times 10^{14}$ | $2.9 \times 10^{16}$ | $\lambda_0/12.5$ |
| Realistic sorter (3 dB loss) | $1.6 \times 10^{13}$ | $9.2 \times 10^{14}$ | $9.2 \times 10^{14}$ | $9.2 \times 10^{16}$* | $\lambda_0/10$ |
| Static sorter, 16 ports | $2.9 \times 10^{12}$ | $5.2 \times 10^{13}$ | $2.9 \times 10^{14}$ | $9.2 \times 10^{15}$ | $\lambda_0/12.5$ |
| Static sorter, 8 ports | $2.9 \times 10^{12}$ | $9.2 \times 10^{13}$ | $5.2 \times 10^{14}$ | $1.6 \times 10^{16}$ | $\lambda_0/12.5$ |
| Static sorter, 4 ports | $5.2 \times 10^{12}$ | $5.2 \times 10^{14}$ | $9.2 \times 10^{14}$ | $5.2 \times 10^{16}$ | $\lambda_0/12.5$ |
| Camera, $\eta = 10^{-5}$ | $9.2 \times 10^{12}$ | – | – | – | $\lambda_0/6$ |
| Realistic sorter, $\eta = 10^{-5}$, $\varepsilon = -20$ dB | $1.6 \times 10^{13}$ | $2.9 \times 10^{14}$ | $1.6 \times 10^{15}$ | – | $\lambda_0/10$ |
| Realistic sorter, $\eta = 10^{-5}$, $\varepsilon = -30$ dB | $1.6 \times 10^{13}$ | $2.9 \times 10^{14}$ | $9.2 \times 10^{14}$ | $9.2 \times 10^{18}$* | $\lambda_0/10$ |
| Camera, $\eta = 10^{-4}$ | – | – | – | – | $\lambda_0/3.3$ |
| Realistic sorter, $\eta = 10^{-4}$, $\varepsilon = -20$ or $-30$ dB | $1.6 \times 10^{13}$ | – | – | – | $\lambda_0/6$ |
| 17 contacts, $r = 0.05\lambda_0$, quantum bound | $9.2 \times 10^{12}$ | $9.2 \times 10^{13}$ | $1.6 \times 10^{14}$ | $9.2 \times 10^{14}$ | $\lambda_0/12.5$ |
| 17 contacts, $r = 0.05\lambda_0$, camera | $1.6 \times 10^{13}$ | $2.9 \times 10^{14}$ | $5.2 \times 10^{14}$ | $2.9 \times 10^{15}$ | $\lambda_0/12.5$ |
| Nine contacts, quantum bound | $9.2 \times 10^{12}$ | $1.6 \times 10^{14}$ | $1.6 \times 10^{15}$ | $9.2 \times 10^{16}$* | $\lambda_0/10$ |
| Flat face, quantum bound | – | – | – | – | $\lambda_0/3.3$ |
| Bare microscope, quantum bound | – | – | – | – | none ($s \leq 0.30\lambda_0$) |

Thresholds lie on a quarter-decade grid and are uncertain by one grid step (a factor of 1.8). *, above the 60 dB-equivalent budget of $6.9 \times 10^{16}$, outside the validated range of the linear model. –, not reached up to $9.2 \times 10^{19}$. Exposure at 1 kW cm$^{-2}$ and $\lambda_0 = 500$ nm: $t = N/(6.3 \times 10^{12}\ \mathrm{s}^{-1})$. With contacts of radius $0.05\lambda_0$, the linear model also resolves $\lambda_0/16.7$ at $9.2 \times 10^{18}$ and $\lambda_0/20$ at $2.9 \times 10^{19}$ (quantum bound).

## Supplementary Note 1 | A simple example: half-disk and half-sphere contacts

**Aim.** The main text counts channels with an operator built for a microscope: object waves, a lens and an objective of NA 0.9. This note asks the same question in the simplest setting that we could devise, with separate solvers written for it, to show the principle before any detailed analysis. Only three things exist in the calculation: a piece of glass ($n$ = 1.6) in air; a row of small light sources just below it, which plays the role of the object; and a detector that collects all the light radiated into the air. Three pieces of glass are compared (Supplementary Fig. 11a): a flat block $5\lambda_0$ wide and $\lambda_0$ thick, as a control; a half-disk of radius $R = \lambda_0$ resting on its flat face, dome up (case 1); and the same half-disk turned over, so that only the lowest point of its dome touches the source plane (case 2). In 3D the half-disk becomes a half-sphere and the block a flat disk.

**Counting in four steps.** (i) One source is switched on at a time and Maxwell's equations are solved on a grid (finite differences in the frequency domain, grid $\lambda_0/40$, absorbing layers at the boundaries). (ii) The radiated field is expanded in orthogonal outgoing waves: angular harmonics on a circle in 2D, vector spherical multipoles on a sphere in 3D. These amplitudes are the detector reading $d$. (iii) The readings of all sources form a table $M$, so that any object pattern $c$, a list of source strengths, gives

$$d = Mc, \qquad M = USV^{\dagger}, \qquad g_i = 20\log_{10}(s_i/s_1) \tag{S1}$$

The singular-value decomposition sorts the object patterns into independent channels with gains $s_1 \geq s_2 \geq \ldots$, quoted in decibels relative to the strongest channel of the system. (iv) Only fine detail is kept. For sources at positions $x_j$ spaced $p = 0.1\lambda_0$ apart, the fraction of the spectral weight of a pattern that lies beyond $nk_0$, its purity $\mu$, is

$$\mu = \frac{c^{\dagger}K_{\text{fine}}\,c}{c^{\dagger}c}, \qquad K_{\text{fine}} = I - K_{\text{low}}, \qquad (K_{\text{low}})_{jl} = \frac{p}{\pi}\ \frac{\sin[nk_0(x_j - x_l)]}{x_j - x_l} \tag{S2}$$

and the fine patterns are those spanned by the eigenvectors of $K_{\text{fine}}$ with $\mu \geq 0.999$. The channels beyond $nk_0$ are the singular values of $M$ restricted to these patterns, and those above −30 dB are counted. No contact size, patch or gap tolerance is assumed anywhere: the count follows from the singular values alone.

**Why purity matters.** A pattern confined to a finite window always contains a little coarse detail, and coarse detail is bright. The flat block measures this leakage: flat glass leaves $k_{\parallel}$ unchanged and cannot pass detail beyond $nk_0$, so any fine channel that it shows is spurious. With $\mu \geq 0.999$ its strongest fine channel lies at −40 dB in 2D, 10 dB below the counting threshold; with $\mu \geq 0.9999$ it lies at −48 dB in 2D and at −46 dB or lower in 3D (Supplementary Fig. 12c,d).

**Results in 2D** (sources $0.05\lambda_0$ below the glass, window $|x| \leq 1.5\lambda_0$, three source orientations; Supplementary Fig. 11b–d and Supplementary Table 3). The flat block passes no channel beyond $nk_0$. Case 1 passes five above −30 dB (eight above −40 dB): one for sources polarized along the edge (TE) and four for sources polarized in the plane (TM), read at the two corners where the flat face ends. Case 2 passes three (four above −40 dB), read at its tip. The channels of case 1 carry root-mean-square spatial frequencies of $2.5$–$2.9k_0$ (half-periods of $0.17$–$0.20\lambda_0$), up to 1.8 times finer than the solid-immersion limit, and one above −40 dB reaches $3.3k_0$; those of case 2 lie near $2.2$–$2.3k_0$. Grids of $\lambda_0/40$ and $\lambda_0/60$ give the same gains within 0.2 dB and the same counts; the fifth channel of case 1 lies on the threshold (−29.9 and −30.0 dB on the two grids).

**The gap decides.** Each further $0.025\lambda_0$ between sources and glass lowers the fine-detail gains by about 2 dB, as the decay of an evanescent wave across the gap predicts (equation (S6)). At a gap of $0.10\lambda_0$ each case keeps one channel above −30 dB, and at $0.15\lambda_0$ neither keeps any (Supplementary Fig. 11c,d). A count therefore belongs to a shape and a gap together, which is why the main text reports the gap sensitivity of every lens (Fig. 4d and Table 1).

**Results in 3D** (Supplementary Fig. 12a,b and Supplementary Table 4). In 3D the two corners of case 1 become a circular rim of radius $\lambda_0$. Each azimuthal order $l$, the number of times a pattern repeats around the axis, is solved separately, with source rings of all three orientations. With rings out to $1.45\lambda_0$ the half-sphere on its flat face passes 26 channels beyond $nk_0$ above −30 dB, and 24–30 with rings out to $1.95\lambda_0$ (for $\mu \geq 0.9999$ and 0.999), in orders up to $|l| = 7$. Turned over, it passes 5 (7 with the wider window), in orders up to $|l| = 2$–3. The flat disk passes none. The rim reads many orders because it is a circle; the tip reads few because it is nearly a point. Supplementary Note 2 turns this observation into a counting rule, and it is why rings are the natural contacts in 3D.

**Checks and scope.** The 2D solver reproduces the exact free-space field of a point source within 1.0% at $\lambda_0/40$ and 0.5% at $\lambda_0/60$ and its radiated power within 0.6%, and gives the same singular values for detector radii of $2\lambda_0$ and $2.5\lambda_0$. In the 3D solver the radiated multipole power equals the power supplied by the source within 2% in free space (grid $\lambda_0/20$; azimuthal orders 0–2, all source orientations) and within 0.2% with the glass present ($\lambda_0/40$), and the leading channels beyond $nk_0$ of two orders tested on the grid $\lambda_0/60$ agree with those at $\lambda_0/40$ within 0.1 dB. The counts of this note use point sources, a detector that collects all angles in air and a fixed gap, so they are to be compared with one another and not with the counts of the main text, which use evanescent object waves, an NA 0.9 objective and lenses in contact. Light trapped in the glass by total internal reflection is not counted; a single frequency and lossless glass are assumed. The conclusions are those of the main calculations: flat glass reads nothing beyond $nk_0$; an edge or a curved tip close to the object reads a few channels; a circular rim reads many; and a gap of tens of nanometres removes them.

## Supplementary Note 2 | A counting rule for contacts in three dimensions

Picture the fine patterns of the object as ripples running around the axis of the lens, like the petals of a flower: the order $l$ is the number of petals. A lens of a given size can carry only so many petals, about as many as wavelengths fit around its edge. A contact on the axis feels only the pattern without petals; a ring feels every pattern whose petals reach it. Counting channels is then bookkeeping: how many orders the lens carries, how many contacts each order meets, and how many polarizations survive the gap. The rest of this note makes the bookkeeping exact.

**The rule.** The 3D channel counts of the main text (Fig. 6a,b) follow from one rule,

$$N \approx \text{(orders delivered)} \times \text{(contacts met per order)} \times \text{(polarizations above threshold)} \quad \text{(S3)}$$

Supplementary Table 2 and Supplementary Fig. 13 give the breakdown for the lens $4\lambda_0$ across ($n$ = 1.6) whose textured face carries a central half-sphere bump and three half-round rings, each half of a torus, all of radius $0.1\lambda_0$, at radii of 0.6, 1.2 and $1.8\lambda_0$; the flat face between them stays $0.1\lambda_0$ above the object (40 nm at $\lambda_0$ = 400 nm).

**Orders that the lens delivers.** Object patterns sort by azimuthal order $l$. A lens carries orders only up to about the number of wavelengths that fit around it,

$$|l| \lesssim \frac{\pi D}{\lambda_0} \quad \text{(S4)}$$

which is 12.6 for $D = 4\lambda_0$, 8.8 for the sphere $2.8\lambda_0$ across and 3.1 for the sphere $\lambda_0$ across; the highest orders read in the full calculation are 13, 7 and 3. Beyond them the strongest channel falls below the threshold (to −31 to −34 dB for the $4\lambda_0$ lens). Contacts add channels within these orders; they add no new orders.

**Contacts that meet each order.** A pattern of order $l$ and spatial frequency $k_\parallel$ is dark within a distance of about $|l|/k_\parallel$ of the axis, so a contact at radius ρ meets only the orders

$$|l| \lesssim k_\parallel \rho \quad \text{(S5)}$$

with $k_\parallel \approx 1.6$–$2k_0$ for the detail just beyond $nk_0$. A bump on the axis ($\rho \leq 0.1\lambda_0$) meets only $l = 0$. Rings at 0.6, 1.2 and $1.8\lambda_0$ meet orders up to about 6–7, 12–15 and 18–22, the last two capped at 13 by the lens.

**Polarizations above the threshold.** Crossing a gap $g$ costs an evanescent wave the power factor

$$\frac{P(g)}{P(0)} = \exp(-2\alpha g), \qquad \alpha = \sqrt{k_\parallel^2 - k_0^2} \quad \text{(S6)}$$

that is, 6.8, 9.5 and 12.5 dB for $g = 0.1\lambda_0$ at $k_\parallel$ = 1.6, 2 and $2.5k_0$. This decides whether the second polarization of an order clears the 30 dB line.

**Each number.** *Smooth face in contact, 51.* The flat area reads nothing beyond $nk_0$; the rim, a circle of radius $2\lambda_0$ where the glass leaves the object abruptly, meets all 27 orders ($|l| \leq 13$), with two channels in most orders (at −17 to −24 dB and −25 to −29 dB) and one at $|l|$ = 11–13. *Face $0.1\lambda_0$ above the object, 33.* Relative to the brightest channel of the lens, which weakens too, the fine channels lose about 4 dB, and the second channel of most orders slips just below the line (−29 to −34 dB). *Central bump, +1.* It lifts one more $l = 0$ channel from −34 to −25 dB: a small contact on the axis is one pixel.

*Ring at 0.6$\lambda_0$, +24.* Both polarizations in the orders $|l| \leq 5$ and one at $|l| = 6$. *Rings at 1.2 and 1.8$\lambda_0$, +50.* Two to three channels per order for $|l| \leq 10$, about 25 per ring; at $|l| = 11$–$13$ their channels lie at −30 to −36 dB, just below the line, and the count rises to 167 at 40 dB. *Best smooth sphere, 28.* One contact whose patch, about half a wavelength wide, meets orders up to 7, with two nearly equal channels per order (the polarizations) and the next 30 dB weaker.

**Reading the rule.** The sphere is one wide contact in a window of 15 orders; the smooth face is one contact line in a window of 27; the textured face stacks up to four contact lines into the same 27. That gives 108 channels, 2.1 times the smooth face in contact and 3.9 times the best sphere. Channels are combined patterns, so the contribution of each ring is obtained by comparing counts with and without it. Only half-round bumps and rings of radius 0.1$\lambda_0$ were computed in 3D. The rule suggests what to try next: only the part of a contact within about 0.1$\lambda_0$ of the object matters, so sharper tips or edges should read finer detail per contact, a flat-topped disk should read through its rim as a ring does, and a raised lip at the rim of the textured face should restore the full contact of the rim. These are predictions, not results.

## Supplementary Note 3 | Resolution study, reconstruction and photon thresholds

Resolution is tested as an optician would test it: pairs of lines are moved closer together until they blur into one. Here the test is made at seven places across the field and repeated twenty times with fresh noise, and a spacing counts only if every pair is seen every time. The photon threshold of a spacing is the smallest exposure from which that remains true, so each number in Supplementary Table 6 is an exposure that an experimenter could set on a lamp.

**Lenses, illumination and test objects.** The faces compared are the flat face, the nine-bump face and faces with 17 bumps of radius $0.1\lambda_0$ or $0.05\lambda_0$ at a pitch of $0.3\lambda_0$, on lenses $6\lambda_0$ wide ($n$ = 1.6; Supplementary Fig. 9a). The 13 transmission angles lie within NA 0.9. The ten evanescent illuminations have in-plane wavenumbers ±1.0, 1.1, 1.2, 1.3 and $1.4k_0$ and enter from below by total internal reflection in the glass substrate (internal angles 43–74°). The object is read at 1–10 lens positions spaced evenly over one bump pitch, in whole steps of the $0.008\lambda_0$ object grid. Each test object consists of nine pairs of lines $0.024\lambda_0$ wide ($\Delta\varepsilon$ = 0.05) with spacing $s$ between $0.30\lambda_0$ and $0.06\lambda_0$, centred at $x = -2.70, -2.03, \ldots, 2.63\lambda_0$ so that together they sample every position relative to the contacts; the seven inner pairs lie under the contact array.

**Field-wide test at a given dynamic range.** Data are generated with the linear operator, with complex Gaussian noise 30–60 dB below the root-mean-square signal (20 realizations, drawn in the space of the normal equations, which is exact in distribution), and reconstructed by truncated singular-value inversion at the noise level. A pair is resolved when the two strongest peaks near it lie on either side of its centre, the reconstruction dips by at least 25% between them and both peaks reach at least half the largest value within $0.5\lambda_0$. A spacing counts as resolved when all seven inner pairs pass in every realization, as do all larger spacings. For the 17-bump face of radius $0.1\lambda_0$ with evanescent illumination and four positions the test was repeated with complete full-wave data (model errors 0.03–0.05%) and gives the same finest spacings (Supplementary Fig. 9d,e). Kernel widths are medians over $|x| \le 2.1\lambda_0$.

**Photon units.** The far fields are mapped to image-plane fields at unit magnification with the expression in 'Intensity detection' (Methods), sampled every $\mathrm{d}x = 0.1\lambda_0$ for $|x| \le 9\lambda_0$ (181 pixels), for each illumination $m$ and lens position $p$. The object vector is $c = \Delta\varepsilon\, w^{1/2}$ on the $0.008\lambda_0$ grid for $|x| \le 6.5\lambda_0$ ($N_\mathrm{c}$ = 1,625 real values). A fluence of $\Phi$ photons per $\lambda_0^2$ per exposure delivers $\Phi \mathrm{d}x|E|^2$ photons to a pixel (per unit length along the lines), and the budget is $N = 92\Phi$. At $\lambda_0$ = 500 nm, 1 kW cm$^{-2}$ corresponds to $6.29 \times 10^{12}$ photons per $\lambda_0^2$ per second, and one photon per $\lambda_0^2$ to $1.59 \times 10^{-10}$ J cm$^{-2}$; at $\lambda_0$ = 400 nm the same irradiance delivers $3.22 \times 10^{12}$ photons per $\lambda_0^2$ per second, so that exposure times are 1.95 times longer for an object of the same size in wavelengths. Obliquity and Fresnel factors are not included.

**Signal-matched reconstruction.** For every detection scheme the object is estimated from the Fisher matrix of that scheme, $F = V\Lambda V^\mathrm{T}$. In the space of the normal equations the data are $Fc + n$, which is exact in distribution for the linear Gaussian model, and

$$\hat{c} = V_k \Lambda_k^{-1} V_k^\mathrm{T}(Fc + n), \qquad n \sim \mathcal{N}(0, F), \qquad \lambda_i \ge \frac{N_\mathrm{c}}{\|c\|^2} \tag{S7}$$

keeps the $k$ eigenchannels whose expected signal exceeds the noise. The rule is the same for every scheme and uses the expected object energy as prior knowledge[29]. Scaling its threshold by 0.1 or 10 changes the finest spacing by about one step at fixed $N$ and the photon price of $\lambda_0/10$ and $\lambda_0/12.5$ by up to an order of magnitude; at these steps the camera stays within a factor of 1–3.3 of the quantum bound for every rule (Supplementary Fig. 10b).

**Photon thresholds.** Φ is stepped on a quarter-decade grid from $10^9$ to $10^{18}$ ($N = 9.2 \times 10^{10}$ to $9.2 \times 10^{19}$), and every spacing is tested independently at every Φ with fresh noise. The threshold of a spacing is the smallest Φ from which it stays resolved at every larger Φ, an isolated failure at a single grid point between two successes being treated as a noise dropout; the resolution at budget $N$ is the finest spacing whose threshold, and those of all coarser spacings, lie below $N$. Steps reached only at the last grid point are not reported, and thresholds are uncertain by one grid step (a factor of 1.8).

**Dynamic range in photons.** Complex Gaussian noise $X$ dB below the root-mean-square far-field signal $d$ of the object with $s = 0.12\lambda_0$ corresponds, in the same field units, to the fluence

$$\Phi_X = \frac{10^{X/10}}{\langle |d|^2 \rangle} \tag{S8}$$

that is, $N = 6.9 \times 10^{14}$, $6.9 \times 10^{15}$ and $6.9 \times 10^{16}$ for 40, 50 and 60 dB. With the far-field Fisher matrix of that noise model, the rule of equation (S7) resolves $\lambda_0/10$, $\lambda_0/10$ and $\lambda_0/12.5$ at 40, 50 and 60 dB, against $\lambda_0/6$, $\lambda_0/8$ and $\lambda_0/10$ with truncation at the noise level (Fig. 6f).

**Validity.** The 60 dB-equivalent budget is the largest at which the linear model was validated against complete full-wave data. Above it the neglected multiple-scattering terms (−68 dB) approach the noise, and results there (open symbols in Fig. 7, asterisks in Supplementary Table 6) indicate trends that require full-wave inversion.

## Supplementary Note 4 | Fisher information of the detection schemes

Fisher information measures how much a measurement tells us about the object: with twice the information, the best possible estimate has half the variance. It is the quantity that explains why blinking molecules can be located to a few nanometres[44], and it flows through an optical system much as energy does[35,45]. Light carries a fixed maximum, the quantum Fisher information, and each detector keeps a share of it. A camera keeps a little less than half, for a simple reason: the bright light of the lens acts as the reference beam of a hologram, and a hologram made with one reference reads only the part of the object wave that oscillates in step with it.

**What is computed.** The Fisher matrix $F$ of a measurement states how sharply it pins down the object: the covariance of any unbiased estimate $\hat{c}$ of the real object vector $c$ obeys $\mathrm{Cov}(\hat{c}) \geq F^{-1}$ (refs[42,43]). The largest $F$ that any measurement allowed by quantum mechanics can reach is the quantum Fisher matrix $F_{\mathrm{Q}}$. This note derives $F$ for every detection scheme of Fig. 7 and shows how close each comes to $F_{\mathrm{Q}}$; Supplementary Table 5 collects the results.

**Coherent image field.** The object is real (lossless). In the exposure with illumination $m$ and lens position $p$, the field in pixel $x$ and the amplitude of the coherent state that the pixel holds at fluence $\Phi$ are

$$E_{pm}(x) = B_m(x) + g_{pm}(x) \cdot c, \qquad \alpha_{pm}(x) = (\Phi\, \mathrm{d}x)^{1/2} E_{pm}(x) \tag{S9}$$

with $B_m$ the known background, the light of the lens alone, and $g_{pm}$ the row of $G_{pm}$ for that pixel (Methods). For a product of coherent states that depends on real parameters, the quantum (symmetric-logarithmic-derivative) Fisher matrix[42] is

$$F_{\mathrm{Q}} = 4 \textstyle\sum_{p,m,x} \mathrm{Re}\,(\partial_c \alpha^\dagger\ \partial_c \alpha) = 4\Phi\, \mathrm{d}x \textstyle\sum_{p,m} \mathrm{Re}\left(G_{pm}^\dagger G_{pm}\right) \tag{S10}$$

It does not depend on $B$, because a known background only displaces the state, and it adds over pixels, illuminations and lens positions. Off-axis holography and heterodyne detection measure both quadratures of every mode with one unit of added noise and reach $F_{\mathrm{Q}}/2$.

**Camera.** A bright-field camera counts photons, with mean

$$n(x) = \Phi\, \mathrm{d}x |B + g \cdot c|^2 = \Phi\, \mathrm{d}x[|B|^2 + 2\,\mathrm{Re}(B^*\, g \cdot c) + |g \cdot c|^2] \tag{S11}$$

The derivative of the count at $c = 0$ is $2\Phi \mathrm{d}x\ \mathrm{Re}(B^* g)$, and its Poisson variance is $n_{\mathrm{b}} = \Phi \mathrm{d}x|B|^2$, so that

$$F_{\mathrm{cam}} = \textstyle\sum_{p,m,x} \frac{(\partial_c n)^{\mathrm{T}}(\partial_c n)}{n_{\mathrm{b}}} = 4\Phi\, \mathrm{d}x \textstyle\sum_{p,m,x} r^{\mathrm{T}}\, r, \qquad r = \mathrm{Re}(e^{-i\varphi} g), \qquad \varphi = \arg B \tag{S12}$$

The amplitude of the background cancels: the interference signal and the shot noise both grow in proportion to $|B|$, so a brighter background neither helps nor hurts. Because $r^{\mathrm{T}} r + q^{\mathrm{T}} q = \mathrm{Re}(g^\dagger g)$ for the two quadratures of $e^{-i\varphi} g$,

$$F_{\mathrm{Q}} = F_{\mathrm{cam}} + F_{\mathrm{quad}}, \qquad F_{\mathrm{quad}} = 4\Phi\, \mathrm{d}x \textstyle\sum_{p,m,x} q^{\mathrm{T}}\, q \geq 0, \qquad q = \mathrm{Im}(e^{-i\varphi} g) \tag{S13}$$

Equation (S13) is exact. The camera is a homodyne detector whose reference is the lens background: it keeps all the information carried by the part of the object light that is in phase with $B$ and none of the part in quadrature. Because the phase between the two varies over pixels, illuminations and lens positions, the loss is partial: on the channels that set resolution the eigenvalues of $F_{\mathrm{cam}}$ are 2.1–2.6 times smaller than those of $F_{\mathrm{Q}}$ (Supplementary Fig. 10c), close to holography, without a reference beam.

**Static sorter.** Projecting every exposure onto the same $K$ orthonormal image-plane modes, with projector $P_K$, and detecting them at the quantum limit gives

$$F_K = 4\Phi\,\mathrm{d}x \textstyle\sum_{p,m} \mathrm{Re}\left[\left(P_K G_{pm}\right)^{\dagger}\left(P_K G_{pm}\right)\right] \leq F_{\mathrm{Q}} \quad \text{(S14)}$$

The modes are the $K$ leading eigenvectors of $\Sigma\, G_{pm} U U^{\mathrm{T}} G_{pm}{}^{\dagger}$, where the columns of $U$ span windowed object gratings with $|q| = 3.2$–$10k_0$ confined to $|x| \leq 2.4\lambda_0$ (80 dimensions); they collect 24%, 42%, 64% and 88% of the trace of $F_{\mathrm{Q}}$ for $K = 4$, 8, 16 and 32.

**Can the quantum bound be reached?** For a single parameter, yes. For several, only if the best measurements for the different parameters are compatible, which for coherent states requires $\mathrm{Im}\, D = 0$ on the estimated subspace. The incompatibility[52]

$$R = \left\|F_{\mathrm{Q}}^{-1}\,\mathrm{Im}\,D\right\|_{\infty}, \qquad D = 4\Phi\,\mathrm{d}x \textstyle\sum_{p,m} G_{pm}^{\dagger}\, G_{pm}, \qquad F_{\mathrm{Q}} = \mathrm{Re}\, D \quad \text{(S15)}$$

lies between 0 and 1 and bounds the attainable (Holevo) cost $C_{\mathrm{H}}$ for every positive weight matrix $W$:

$$\mathrm{Tr}\left(W F_{\mathrm{Q}}^{-1}\right) \leq C_{\mathrm{H}}(W) \leq (1 + R)\mathrm{Tr}\left(W F_{\mathrm{Q}}^{-1}\right) \quad \text{(S16)}$$

Physically, $\mathrm{Im}\, D \neq 0$ means that changes in two different parts of the object move the field in conjugate quadratures, which cannot both be measured without added noise. Here $R \geq 0.9996$ on the 133–206 channels retained for $N = 9.2 \times 10^{13}$–$9.2 \times 10^{16}$, so the bound leaves the attainable optimum anywhere between $F_{\mathrm{Q}}/2$ and $F_{\mathrm{Q}}$, the band shaded in Fig. 7b. Off-axis holography, at $F_{\mathrm{Q}}/2$, is therefore within a factor of two of the optimum, and on the channels that set resolution the camera is within a factor of 1.05–1.3 of holography.

## Supplementary Note 5 | Systematic floor, background rejection, detector load and shaped illumination

More photons average away random noise, but not the fixed errors of a detector. Weighing a feather on a truck scale is limited by the calibration of the scale, not by patience; the remedy is to take the truck off the scale first. The bright background of the lens is the truck, the light scattered by the object is the feather, and a mode sorter or any interferometric nulling removes the background before the light is detected.

**Systematic floor.** The floor η represents errors that do not decrease with more photons, such as fixed-pattern noise, drift and imperfect knowledge of $B$, expressed relative to the detected background in each pixel; it was set to $10^{-4}$ and $10^{-5}$. It adds a variance $\eta^2 n_b^2$ to the shot noise $n_b$, so the Fisher weight of a pixel in equation (S12) becomes

$$w(x) = \frac{4\Phi\,\mathrm{d}x}{1+\eta^2 n_\mathrm{b}(x)}, \qquad n_\mathrm{b}(x) = \Phi\,\mathrm{d}x|B(x)|^2 \tag{S17}$$

When $n_b \gg \eta^{-2}$ the fluence cancels and the information saturates at $4/(\eta^2|B|^2)$ per pixel, whatever the exposure. This produces the ceilings of Fig. 7c,d.

**Background rejection.** Attenuating the background power by ε before detection ($B \rightarrow \varepsilon^{1/2}B$) leaves the shot-noise information of equation (S12) unchanged and multiplies $n_b$ by ε. The saturated value becomes $4/(\eta^2\varepsilon|B|^2)$, that of a camera with the smaller floor $\eta\varepsilon^{1/2}$. An insertion loss, a transmission $\eta_s$ of all light, multiplies Φ; the realistic sorter has $\eta_s = 0.5$. The residual background serves as the reference, which requires it to exceed the object signal, $\varepsilon^{1/2}|B| > |g\cdot c|$. This holds here down to ε = −30 dB: $|Gc|/|B|$ is $1.1 \times 10^{-4}$ in median and at most $5.6 \times 10^{-3}$ for the $\lambda_0/10$ object.

**Detector load.** The brightest background pixel ($\max|B|^2 = 4.13$; mean 0.88) sets saturation. The number of pixels needed to image a field of area $A$ at a photon rate Γ per $\lambda_0^2$ without saturating wells of depth $W$ at a frame rate $f$, the detector-limited time for a budget $N$ with $n_\mathrm{pix}$ pixels, and the illumination-limited time are

$$n_\mathrm{req} = \frac{A\,\Gamma\,\varepsilon\max|B|^2}{Wf}, \qquad t_\mathrm{det} = \frac{NA\,\varepsilon\max|B|^2}{n_\mathrm{pix}Wf}, \qquad t_\mathrm{ill} = \frac{N}{\Gamma} \tag{S18}$$

with $A = 36\lambda_0^2$, $W = 3 \times 10^4$ electrons, $\Gamma = 6.29 \times 10^{12}$ s$^{-1}$ per $\lambda_0^2$ at 1 kW cm$^{-2}$ ($\lambda_0$ = 500 nm), $f$ = 1 kHz for the pixel count, and $n_\mathrm{pix} = 4.2 \times 10^6$ with $f$ = 100 s$^{-1}$ for the times. For $\lambda_0/10$ with a camera ($N = 9.2 \times 10^{14}$) the detector-limited time is about 3 h, against 2.4 min of illumination. After −20 dB rejection the detector load falls a hundredfold and the measurement is limited by the illumination, at about 4 min ($N = 1.6 \times 10^{15}$ for the realistic sorter at $\eta = 10^{-5}$).

**Shaped illumination.** The coupling of combinations of the 23 illuminating waves to fine detail at each lens position is

$$Q_{mm'} = \mathrm{tr}\big(U^\mathrm{T} G_m^\dagger G_{m'} U\big) \tag{S19}$$

with $U$ as in Supplementary Note 4. Its $K$ leading eigenvectors give $K$ wavefronts (weights over the 23 waves, unit norm) that share the photon budget of the 23 plane waves used in turn; resolution was computed as in Supplementary Note 3 for $K$ = 1, 2, 4, 8 and 23. The single best wavefront couples 10.9 times more strongly to fine detail than an average plane wave, yet it needs about thirty times more photons for $\lambda_0/10$, and eight shaped wavefronts do no better than the 23 plane waves (Supplementary Fig. 10a): field-wide imaging needs many independent views. With $K = 23$, a unitary recombination of the plane waves, the plane-wave thresholds are reproduced exactly.

## Supplementary Note 6 | Robustness of the simulated images

**Tests.** For Supplementary Fig. 8 each reconstruction of the five line pairs of Fig. 6e was repeated for 20 noise realizations; for noise 20 and 40 dB below the signal with truncation at the same level; for truncation at 25 and 35 dB; for dip criteria of 20 and 30%; and for data from mismatched lenses, inverted with the nominal operator: a gap of $0.01\lambda_0$ under the lens, a lens index of 1.61 instead of 1.60, or a second position displaced by $0.32\lambda_0$ instead of $0.30\lambda_0$. Pairs are judged by one test throughout (module nanolens.resolve): the two strongest peaks near the pair lie on either side of its centre, the reconstruction dips by at least 25% between them, and both reach at least half the largest value within $0.5\lambda_0$, which rejects ringing that a dip test alone would accept.

**Results.** For the nine-bump lens read at two positions, the finest resolved pair of Fig. 6e ($0.16\lambda_0$) is the same in all 20 noise realizations, for dip criteria of 20–30%, with a gap error of $0.01\lambda_0$ and with a position error of $0.02\lambda_0$. Noise 20 dB below the signal or truncation at 25 dB coarsens it to $0.30\lambda_0$, and an index error of 0.01 to $0.22\lambda_0$. From intensity images alone it is $0.22\lambda_0$ with one image per illumination angle, $0.30\lambda_0$ with one partially coherent image per position and $0.16\lambda_0$ with partially coherent focus stacks. The flat face resolves $0.30\lambda_0$ in every case (Supplementary Fig. 8).

## Supplementary Note 7 | Code, data and numerical checks

**Code and data.** All calculations, figures and quoted numbers are produced by the Python package nanolens-transfer, one archive containing the exact multipole solutions for spheres and cylinders and the 2D single-pass full-wave model (Figs. 2–5, Table 1 and Supplementary Figs. 1–4); the full-wave 3D model for rotationally symmetric lenses, the 2D complete-imaging model, intensity detection, image robustness and the resolution study (Fig. 6, Supplementary Figs. 5–9 and 13 and Supplementary Table 2); and the photon-budget analysis (Fig. 7, Supplementary Table 6 and Supplementary Fig. 10). The simple example of Supplementary Note 1 (Supplementary Figs. 11 and 12, Supplementary Tables 3 and 4) was computed with separate 2D and 3D finite-difference solvers.

**Threshold convention.** Channel powers $\lambda_i$ are eigenvalues of $H^{\dagger}H$; a threshold of $X$ dB counts channels with $\lambda_i \geq 10^{-X/10}\lambda_{max}$ ($10^{-2}$, $10^{-3}$ and $10^{-4}$ for 20, 30 and 40 dB), where $\lambda_{max}$ is the brightest channel of the same lens and measurement in either evanescent band, the larger of the leading eigenvalues of the immersion band and the band beyond it. This reference is used for every model; scanning and gap counts, labelled as such, are referred to a single position and to contact, as for a fixed detection floor.

**Spheres and cylinders.** Exact solutions reproduce vector waves to $10^{-13}$ and Mie efficiencies to ten digits. Halving the $k$-grid step changes the counts by at most 1 of 30 beyond $nk_0$ and 4 of 89 below it, and raising the upper band limit from $5k_0$ to $10k_0$ leaves the sphere counts unchanged at 20, 30 and 40 dB.

**2D full-wave model.** The collected response agrees with the exact cylinder solution within 0.4% for $D \leq \lambda_0$; full-wave and exact channel operators agree within 0.2% ($D = \lambda_0$) and 1% ($D = 6\lambda_0$) in their Gram matrices; refining the grids by 25% leaves every channel count of the smooth and textured lenses unchanged.

**3D full-wave model.** The model for rotationally symmetric lenses reproduces the exact Mie eigenvalues of spheres within 0.5 dB with the same counts, and refining its grid by 25% leaves every count of the smooth and textured lenses unchanged at 20, 30 and 40 dB.

**Complete imaging.** The linear imaging operator reproduces complete full-wave solutions with the object included with a relative error of about $0.01\Delta\varepsilon$ (1% for the test objects). A coarser grid for the largest lenses changes their reach by at most $0.02k_0$; raising the band limit to $10k_0$ changes the counts beyond $2nk_0$ by at most one channel; and intensity detection reproduces the reach of complex-field detection within $0.16k_0$.

**Resolution study.** For the field-wide test (Supplementary Fig. 9) the linear operator reproduces complete full-wave solutions to 0.04% (−68 dB) for the weak test objects ($\Delta\varepsilon = 0.05$), so the data are generated with it; for the 17-bump face of radius $0.1\lambda_0$ with evanescent illumination and four positions, complete solutions for every spacing from $0.30\lambda_0$ to $0.08\lambda_0$ give the same finest resolved spacing as the linear calculation at 30, 40, 50 and 60 dB (0.22, 0.16, 0.12 and $0.10\lambda_0$). Lens positions are whole steps of the $0.008\lambda_0$ object grid; with positions off the grid, the discretized object and the shifted operator differ by a fraction of a line width, which limits the model accuracy to about 2% and must be avoided at high dynamic range.

**Photon budgets.** The first-pass sweeps on a half-decade grid agree with the quarter-decade maps within one grid step. Of the 8,820 interior points of the 252 success maps, 31 (0.35%) are isolated failures between two successes and are treated as noise dropouts. Image-plane and far-field normalizations of the photon units differ by 26% in the trace of the quantum Fisher matrix. $|Gc|/|B|$ is at most $5.6 \times 10^{-3}$ for the $\lambda_0/10$ object, so a background attenuated by up to 30 dB still exceeds the object signal and the homodyne linearization holds.

## Supplementary Note 8 | Fundamental questions and answers

**1. Why can an ordinary microscope not see very small things?** Light leaving an object carries its details as ripples. Coarse ripples travel to the lens; ripples finer than about half a wavelength do not travel at all. They cling to the object and fade within a few tens of nanometres, as the treble of a distant band dies out long before the bass. A microscope far from the object therefore never receives the fine detail, however perfect its lenses (Fig. 1a).

**2. What does a small transparent bead change?** A bead resting on the object sits inside the zone where the fine ripples still exist. Its curved surface turns part of them into ordinary travelling light, which the microscope collects, as a fingertip reads Braille dots that the eye cannot see from across a room (Fig. 1b). This is the idea behind the nanolenses of 2009 and the microsphere nanoscopes that followed[7,9].

**3. What is a channel, and why count channels?** An image is a list of numbers about the object. A channel is one independent number that the lens delivers to the detector, like one pixel of a camera or one lane of a road. A lens with few lanes for fine detail cannot deliver a sharp image, whatever is done afterwards. We count the lanes that are no more than a thousand times weaker than the brightest one, a limit of 30 dB (Fig. 1c and Fig. 2c). Counting channels gives every lens, of any shape, the same yardstick.

**4. Why does only the contact matter?** Because the fine ripples fade within tens of nanometres, only glass that close to the object can catch them. The simplest demonstration uses three pieces of glass on a plane of light sources (Supplementary Note 1). A flat block delivers no fine detail at all. A half-disk delivers a few channels, and only from its corners or its tip, where the glass leaves the object. Lift the glass by a further twentieth of a wavelength and most of them are gone.

**5. If the contact is so small, how can a bead show an image of a whole field?** A bead is two instruments at once (Fig. 2a). For moderately fine detail it is an immersion lens with a wide field, and light that bounces between bead and object, the re-illumination identified by Maslov and Astratov[28], doubles its reach (Fig. 6c). For the finest detail it is a single nanoscale pixel under its contact, which is why scanning the bead adds information (Fig. 4c). This also explains an old puzzle: a larger bead or a higher index does not see finer, because a larger bead touches more gently and traps the fine-detail light inside itself (Fig. 4a,b).

**6. How can one lens read many places at once?** By giving it many contacts. Small bumps on a flat lens face each act as a pixel: nine bumps raise the count from 2 to 12 in two dimensions (Fig. 5), and in three dimensions a bump and three rings raise it to 108, against 28 for the best bead (Fig. 6a,b). A counting rule predicts these numbers before any simulation is run (Supplementary Note 2).

**7. What resolution does this give?** A face with 17 contacts, read at four positions under evanescent illumination, separates lines a tenth of a wavelength apart everywhere in a field four wavelengths wide, and lines 12.5 times finer than the wavelength with more light (Fig. 6f and Fig. 7). For violet light of 400 nm these are 40 and 32 nm across 1.6 μm, with an ordinary objective, no labels and no scanning tip. These are predictions of a 2D model, stated with the conditions under which they were validated (Supplementary Note 3).

**8. How much light does that take, and is an exotic detector needed?** Detecting light is like counting raindrops: ten times better precision needs a hundred times more drops. At the limit set by quantum mechanics, a tenth of a wavelength takes 46 s at an irradiance of 1 kW $cm^{-2}$ for a weak test object, and an everyday camera needs only three times longer, because the bright light of the lens serves as a built-in reference beam (Fig. 7b and Supplementary Note 4). Each further step of resolution costs about ten times more light. A more elaborate detector, a mode sorter, pays off only where a camera is stopped by its own fixed errors or by saturation (Fig. 7c,d and Supplementary Note 5).

**9. How does this differ from other label-free routes to fine detail?** Superoscillatory lenses shape travelling light into hotspots smaller than the wavelength without touching the object, and pay with energy sent into sidebands[49]. Limited-size-object microscopy uses prior knowledge of the size of an isolated object[31]. Photonic chips illuminate unlabelled specimens with evanescent light and exploit its fluctuations[48]. A lens in contact converts the object's own evanescent waves on the spot and needs no prior knowledge of the object. The routes are complementary, and they share one currency: finer detail is always bought with signal-to-noise, which Fisher information counts[35,44,45]. The contribution of this paper is to state that price in photons for a lens in contact, and to show that the contacts, not the detector, set the ceiling (Fig. 7e).

**10. What should an experimentalist build and test first?** Three predictions follow (Discussion). Large beads should resolve periods down to about $\lambda_0/(n + 1)$ and no finer. A textured face should show high-frequency contrast under each contact. And at 400 nm, contacts 40 nm in radius and 120 nm apart, read at four positions under evanescent illumination, should resolve line pairs 40 nm apart across 1.6 μm within minutes with a camera. The design advice is as short: do not reach for a larger bead or a higher index; keep the lens within a few tens of nanometres of the object, and give it many sharp contacts.